\documentclass[11pt]{article}
\pdfoutput=1
\usepackage{amsmath}
\usepackage{amssymb}
\usepackage{graphicx,bbm,mathrsfs}
\usepackage{nicefrac}
\usepackage{bbm}
\usepackage{geometry}       
\usepackage{jheppub}       

\def\ie{{\it i.e.}}

\newcommand{\be}{\begin{equation}}  
\newcommand{\ee}{\end{equation}}  
\newcommand{\bea}{\begin{eqnarray}}  
\newcommand{\eea}{\end{eqnarray}}

\newcommand\lsim{\mathrel{\rlap{\lower4pt\hbox{\hskip1pt$\sim$}}
    \raise1pt\hbox{$<$}}}
\newcommand\gsim{\mathrel{\rlap{\lower4pt\hbox{\hskip1pt$\sim$}}
    \raise1pt\hbox{$>$}}}

\newcommand{\captionfonts}{\small}

\newcommand{\approptoinn}[2]{\mathrel{\vcenter{
  \offinterlineskip\halign{\hfil$##$\cr
    #1\propto\cr\noalign{\kern2pt}#1\sim\cr\noalign{\kern-2pt}}}}}

\makeatletter  
\long\def\@makecaption#1#2{%
  \vskip\abovecaptionskip
  \sbox\@tempboxa{{\captionfonts #1: #2}}%
  \ifdim \wd\@tempboxa >\hsize
    {\captionfonts #1: #2\par}
  \else
    \hbox to\hsize{\hfil\box\@tempboxa\hfil}%
  \fi
  \vskip\belowcaptionskip}
\makeatother   

\begin{document}

\begin{flushright}LPT-Orsay-15-62\end{flushright} 

\vspace*{.5cm}

\begin{center}

\thispagestyle{empty}

{\Large\bf 
Anatomy of the Higgs fits: a first guide to statistical treatments of the theoretical uncertainties
}\\[10mm]

\renewcommand{\thefootnote}{\fnsymbol{footnote}}

{\large Sylvain~Fichet$^{\,a,b}$\footnote{sylvain@ift.unesp.br}, 
\large Gr\'egory~Moreau$^{\,c}$\footnote{moreau@th.u-psud.fr}}\\[10mm]

\addtocounter{footnote}{-3}

{\it
$^a$ ICTP South American Institute for Fundamental Research, Instituto de Fisica Teorica 
Sao Paulo State University, Brazil \\
$^{b}$~ International Institute of Physics, UFRN, 
Av. Odilon Gomes de Lima, 1722 - Capim~Macio - 59078-400 - Natal-RN, Brazil \\
$^{c}$~ Laboratoire de Physique Th\'eorique, B\^at. 210, CNRS,
Universit\'e Paris-sud 11 \\  F-91405 Orsay Cedex, France \\
}

\vspace*{12mm}

{  \bf  Abstract }
\end{center}

\noindent 
The studies of the Higgs boson couplings based on the recent and upcoming LHC data open up a new window on physics beyond the Standard Model.
 In this paper, we propose a statistical guide to the consistent treatment of the theoretical uncertainties entering the Higgs rate fits. Both the Bayesian and frequentist approaches are systematically analysed in a unified formalism.
We present  analytical expressions for the marginal likelihoods, useful  
to implement simultaneously the experimental and theoretical uncertainties. 
We review the various origins of the theoretical errors (QCD, EFT, PDF, production mode contamination\dots).
All these individual uncertainties are thoroughly combined with the help of moment-based considerations. The theoretical correlations among Higgs detection channels appear to affect 
the location and size of  the best-fit regions in the space of Higgs couplings.
We discuss the recurrent question of the shape of the prior distributions for the individual  theoretical errors and find that a nearly Gaussian prior arises from the error combinations.
We also develop the bias approach, which is an alternative to marginalisation providing more conservative results.   The statistical framework to apply the bias principle is introduced and two realisations of the bias are proposed.
Finally, depending on the statistical treatment, the Standard Model prediction for the Higgs signal strengths is found to lie within either the $68\%$ or $95\%$ confidence level region 
obtained from the latest analyses of the $7$ and $8$ TeV LHC datasets.

\clearpage

\tableofcontents

\newpage

\section{Introduction and summary}  \label{se:intro}

Besides the historical discovery of a resonance around $125$~GeV~\cite{ATLAS-disc,CMS-disc} that is most probably the Brout-Englert-Higgs 
boson responsible for the 
ElectroWeak (EW) symmetry breaking~\cite{BEHiggs}, the ATLAS and CMS Collaborations have provided a set of 88 rate measurements~-- 
based on the full dataset collected so far with luminosities of $\sim 5$~fb$^{-1}$ at the center of mass energy $\sqrt s=7$~TeV and $\sim 20$~fb$^{-1}$ at 
$\sqrt s=8$~TeV~\cite{ATLASfit,CMSfit} (see also Ref.~\cite{ATLASweb,CMSweb})~-- 
that constitutes a new and precious source of indirect information on physics beyond the Standard Model (SM). 
Indeed, observing deviations of the Higgs boson rates with respect to their SM predictions would reveal the presence of an underlying theory 
while the absence of such deviations allows one to strongly constrain new models (see for example Ref.~\cite{RSfit} for higher-dimensional models,
Ref.~\cite{COMPfit} for composite Higgs theories and Ref.~\cite{SUSYfit} for supersymmetric scenarios).
So far, no signs from an unknown world have came out from the data,
but this is only the beginning of a long exploration,
 given the
expected LHC upgrades~\cite{EurStrat}.

The fits of the Higgs rates ({\it c.f.} Ref.~\cite{Fits} for the first set of analyses, Ref.~\cite{pMfits,Strumia,FirstBias,BaySylvain} 
for the results after the Moriond 2013 winter conference and Ref.~\cite{ATLASfit,CMSfit} for the latest official ATLAS and CMS analyses) 
are thus obviously important. Now certain aspects of these analyses remain to be worked out in order to obtain the final fits for testing new physics.
First, the precise likelihood functions associated to the experimental rates (in particular their specific shapes and the complete correlations 
between channels) are not provided in the present public papers, although they might be expected at some point. Second, a major part of the
theoretical uncertainties is due to QCD calculations of the Higgs production rates~\cite{LHCHWGweb,LHCHWG1,LHCHWG2,LHCHWG3}  
and their treatments in the fits raise questions in the Higgs physics community (see Ref.~\cite{Kyle,Cacciari} for recent discussions). 
Taking carefully into account these theoretical uncertainties is crucial for the Higgs fits due to the following reasons.

First, theoretical uncertainties can be sizeable with respect to the experimental ones.  
The QCD uncertainty on the gluon-gluon fusion mechanism dominantly involved in most of the Higgs discovery channels induces 
typically an error of $\sim 10\%$ on signal strengths (see Section~\ref{se:marg}), that is already comparable 
to the experimental error bars in several Higgs channels which reach values down to $\sim 20\%$~\cite{ATLASfit,CMSfit,ATLASweb,CMSweb}. 
Besides, considering for instance the CMS prospectives at $\sqrt s=14$~TeV with a luminosity of $300$~fb$^{-1}$, the 
experimental error bars are around $\sim 5\%$ (with same systematic errors as today)
for the diphoton final state and less than $\sim 10\%$ for the $\tau$-lepton, Z and W boson channels~\cite{EurStrat} so that the theoretical error might even become 
the dominant one in some channels. 
\\
Second, theoretical uncertainties might be of the same magnitude as the main potential deviations due to new physics. For instance 
 the maximal corrections to Higgs couplings estimated  in Ref.~\cite{Gupta} for characteristic composite Higgs and 
supersymmetric models~\footnote{In the case of no new states, related to the EW symmetry breaking, directly observed at the LHC.} 
lead typically to deviations of the signal strengths between $\sim 2\%$ and tens of percent compared to SM. This is of the same order as the theoretical error
mentioned  above, so that one is precisely in the situation where the theoretical error deserves a careful treatment to test new physics scenarios.~\footnote{This 
intermediate situation is to be contrasted with the two extreme cases of expected signal strength deviations much higher than the theoretical error (which can then be neglected) 
or deviations well smaller (no hope to detect them). In both of these cases, a detailed treatment of the theoretical error would not be really needed to test new physic scenarios.}

\vspace{0.3cm}

Therefore, in this paper, our primarily goal is to answer precisely the question~:
{\it what is the correct treatment of the theoretical uncertainties in the fits of the Higgs boson rates?}~\footnote{Throughout this paper, we use generically the
expression ``theoretical error'' to denote any error on the SM prediction for the Higgs rates. This is a slight wording abuse, because certain of these errors like the ones from the PDF determination  have a partial experimental origin.}  
This seemingly simple question has lead us to  several new developments, summarized in the
 three lines of work described in the paragraphs below. 
\\
First, we present a systematic survey of the various statistical treatments of the theoretical error and their applications to the Higgs fits within a unified formalism.
We confront the frequentist and 
Bayesian frameworks,~\footnote{Sometimes in the literature, there are inconsistencies in the sense that errors are combined in a frequentist
way (combination depending on the prior shape) while the priors are convoluted in a Bayesian way (convolution via integrations).}~\footnote{A pure Bayesian fit of the Higgs rates  has been carried out in Ref.~\cite{BaySylvain}.}
 that prove to exhibit a certain degree of convergence at the level of accuracy of the present LHC data.
\footnote{To be contrasted with the preliminary study of
Ref.~\cite{Dirk} based on simulated Higgs data.} We also compare the marginalisation and bias treatments.
In  the former, we consider the  representative cases of Gaussian and flat combined priors because of 
the lack of knowledge inherent to the distribution of theoretical uncertainties.~\footnote{To the best of our knowledge, a flat prior for the theoretical uncertainty is for the first time applied to 
the  Higgs fits. 
Notice also that the combination in quadrature of the theoretical and experimental errors, sometimes made in the literature, is  equivalent to a marginalisation assuming Gaussian distributions for both  sources of errors and neglecting the correlations.
 This is true in both  frequentist 
and Bayesian cases.} 
We find the Gaussian prior to be well motivated by  the full combination of each indi\-vidual theoretical uncertainty. 
It turns out that the choice of one among all these statistical approaches may affect significantly the determination of the Higgs properties.
It is thus important to  understand precisely the conceptual differences between these approaches. Finally, this survey is the opportunity to provide useful analytical expressions for the 
 marginalised likelihood functions, including the theoretical correlations among the Higgs channels.
\\
Second, we explain precisely the principle of bias \footnote{A bias  has been  applied once in Ref.~\cite{FirstBias}.
The analysis developed here improves  
the bias performed in Ref.~\cite{FirstBias} by including more effects like the production contamination, the individual scale/EFT/PDF errors, the branching 
fraction uncertainties, the correlations between Higgs channels and the Bayesian/frequentist cases.}
 and its fundamental differences with the marginalisation 
principle. 
The bias principle is  more conservative than the marginalisation principle by construction and does not depend on  the shape of the priors of the nuisance parameters.
This thorough examination of the bias principle leads naturally to introduce a statistical framework for biasing.
 We propose two realisations of the  bias, 
referred to as  the extremal bias and the envelope method, that apply in both frequentist and Bayesian contexts. Regarding the error combinations, important 
differences arise between the marginalisation and bias frameworks.~\footnote{For example,   the PDF and amplitude uncertainties for the ggF mechanism are summed in quadrature  in the Bayesian marginalisation, whereas they are linearly summed in the bias approach.}
\\
Third, we discuss and implement several improvements in the treatment of the theoretical uncertainties. {\it (i)} For the cross sections, the combinations of all the individual uncertainties are discussed exhaustively, including in particular the several errors constituting the parton PDF uncertainty.
The so-called leading moment approximation is developed to facilitate the combination of  such a high number of errors. 
{\it (ii)} The error contamination by various production modes and the errors on the Higgs branching ratios  are taken into account. 
{\it (iii)} The correlations between the theoretical errors on the various Higgs detection channels are included.~\footnote{We notice that such 
correlations were included {\it e.g.} in Ref.~\cite{Strumia} for the specific assumption of errors with Gaussian priors and neglecting the correlations among different Higgs production modes.}
We show that these theoretical correlations induce significant shifts of the best-fit regions in the Higgs coupling parameter space.
{\it (iv)} A Higgs fit with more conservative theoretical errors is shown to illustrate the potential impact from the imperfect knowledge of the magnitude of these errors. 
\\
For each of the statistical approaches developed along these three lines of work, we provide the up-to-date Higgs fit results based on the latest available data
from the $7$ and $8$~TeV LHC, that can be readily used for new physics tests. From the theory side, we have updated the major gluon-gluon 
Fusion mechanism by using its reduced perturbative QCD error, issued from the recent calculation up to N$^3$LO~\cite{Duhr}. We have also included the theoretical uncertainty on this production mode due to the use of an Effective Field Theory in the amplitude 
calculation~\cite{Duhr,bottomEFT,QCDEWnnlo}, so that the whole error on the cross section remains at $\sim 10\%$.

\section{Statistical preliminaries}  \label{se:STAT}

This section condenses the basic elements of frequentist and Bayesian statistics that will be used along the paper. 
In addition to statistical basics, the principle of bias is also presented. 

\subsection{Need-to-know frequentist and Bayesian statistics}  \label{se:STATbasics}

In order to extract some information about a new physics model from a set of data, the central quantity to study is the likelihood function \cite{LPberger}.~\footnote{Note this is an abuse of language, the likelihood function is actually a distribution.}
The likelihood function is equal to the conditional probability density for  obtaining the observed data, taken as a function of the hypothesis. In the case of predictions made in a given hypothesis $H$ with $n$ parameters $ \{\theta_n\} \equiv \theta $, the likelihood function reads
\be
L(\theta)\equiv p(d|H,\theta)\,,
\ee 
where $d$ represents the set of data. Note that the likelihood is defined up to an overall factor.
In the present work, the data we will consider are the set of signal strength measurements from LHC and Tevatron, described in Section~\ref{sec:ExpSide}.

In particle physics, the likelihood function encloses a \textit{statistical} uncertainty associated with the data. This is the uncertainty coming from the fluctuations inherent to the observation of a quantum process. This statistical uncertainty tends to zero in the limit of a large amount of data. However, other sources of uncertainty can be present, both on the experimental or the theoretical side. For example, uncertainties arise from the finite  resolution of a detector, or from the finite accuracy of a computation. These \textit{systematic} uncertainties do not depend on the amount of data, and need to be taken into carefully. In this paper, we are going to have a close look at the \textit{theoretical} systematic uncertainties.

The  starting point for modeling a systematic  uncertainty is  to explicitly parametrize it. Namely, one introduces a set of new parameters, $\delta \equiv \{\delta_i\}$, which explicitly modifies the likelihood, 
 \be
 L(\theta,\delta)\,.
 \ee
These new parameters are named \textit{nuisance} parameters, as opposite to the $\theta$'s which are considered as the parameters of interest. 
This  step of  parametrisation is common to the frequentist and Bayesian frameworks, and is fairly universal. Discrepancies will  
appear in the way the $\delta$'s are treated, and will be at the center of our attention in the rest of the paper.
Two fundamentally different points of view on how to treat the nuisance parameters, denoted as \textit{marginalisation} and \textit{bias}, will be 
further identified (in both the frequentist and Bayesian contexts).

\vspace{0.5cm}

In Bayesian statistics, model parameters are genuine random variables.
 They are associated with a so-called prior distribution, noted 
$\pi(\theta)$. In order to carry out a process of inference (for example, setting exclusion bounds), the relevant object to study is the posterior distribution, 
\be
p(H,\theta|d)\propto L(\theta) \, \pi(\theta)\,.  \label{eq:PostBay}
\ee
In this framework, a so-called $1-\alpha$ Bayesian credible region is defined by the domain  
$\Omega_\alpha=\{\theta\,|\,p(H,\theta|d)> p_\alpha \}$, where $p_\alpha$ is determined by the fraction of integrated posterior 
\be
\frac{\int_{\Omega_\alpha}\, d\theta\,p(H,\theta|d)}{ \int_{\Omega}\, d\theta\,p(H,\theta|d)} = 1-\alpha\, ,
\label{eq:bayes_contours}
\ee
$\Omega$ being the whole parameter space.
The $1-\alpha$ Bayesian Credible (BC) contour  is the boundary of $\Omega_\alpha$ and it corresponds to the contour level  
defined as $\{\theta\,|\,p(H,\theta|d)= p_\alpha \}$.
 In what follows we will use the BC contours at
\be
1-\alpha=\{ 68.27\% \, , \, 95.45\% \, , \, 99.73\%  \}\,.  \label{eq:CLBay}
\ee

\vspace{0.5cm}

In frequentist statistics,
 the likelihood function is employed to build a statistical test, like the likelihood ratio~\footnote{In classical frequentist statistics, 
hypotheses  and parameters are not associated with probabilities. In this paper, for the frequentist side,  we adopt the more general framework of hybrid Bayesian-frequentist statistics, in which a distribution can be attributed to a nuisance parameter. Conceptually, such distribution cannot be seen as a prior \textit{pdf}, but corresponds to the likelihood for a real or imaginary measurement constraining the nuisance parameter (see Ref.~\cite{CMS-NOTE-2011-005}, p.~4). However, by  abuse of language, we will sometimes use the term ``prior'' in frequentist statistics as well. 
Classical frequentist statistics are recovered by giving a flat shape to these frequentist ``prior'' distributions.
} 
\be q(\theta)=-2\log \left[\frac{ L(\theta)\pi(\theta)}{\max\limits_{\theta\in\Omega} L(\theta)\pi(\theta)}\right]
\label{eq:freq_contours}
 \,.\ee 
The probability density function ($pdf$) of this test is then computed by simulation (typically, using Monte-Carlo pseudo-data). 
The $pdf$  of $q(\theta)$, noted $f_q$, can then be used to evaluate a $\mathbf{p}$-value, typically of the form
\be
\mathbf{p}(\theta)=\int^\infty_{q_{d}}\,f_q(q'|\theta)\,dq'\,,  \label{eq:pValFreq}
\ee
where $q_d$ is the value given by the actual data.  The $1-\alpha$ confidence regions  are then obtained by solving $\mathbf{p}(\theta)=\alpha$, \ie~the confidence regions are given by $\Omega_\alpha=\{\theta|\mathbf{p}(\theta)>\alpha\}$.

Whenever the likelihood is Gaussian, $q$ follows a $\chi^2$ distribution. One has then $1-\alpha=F_{\chi^2}^{(n)}(q_\alpha)$, where $F_{\chi^2}^{(n)}$ is the $\chi^2$ cumulative function with $n$ degrees of freedoms. Confidence regions can thus be obtained by plotting $q(\theta)=q_\alpha$. This simpler procedure is commonly used in the literature, even when the likelihood is not Gaussian. 
We adopt this procedure throughout this paper. In the case where the likelihoods are bivariate (which will be the case of our example of Higgs fit), we adopt the threshold values 
\be
q=\{ 2.30,\,6.18,\,11.83  \}\,. \label{eq:Chi2val}   
\ee
In the Gaussian limit, these values match exactly the confidence levels $1-\alpha=\{ 68.27\%$, $95.45\%$,  $99.73\%\}$.

\subsection{Treatment of nuisance parameters}

\subsubsection{Marginalisation principle} \label{sec:MargPrin}

Having introduced the nuisance parameters $\delta$~\footnote{Recall 
that we have defined $\delta$ as a set of nuisance parameters, $\delta\equiv\{\delta_i\}$. The subsequent integrations and maximisations
will thus be multidimensional. } in the likelihood  $L(\theta,\delta)$, the next step is to eliminate them. 
This will effectively deform the likelihood, enlarging the preferred regions, and possibly shift their central values.
In the Bayesian framework, this is naturally done by integrating over $\delta$, so that 
\be
L_{\rm B}(\theta) = \int_{\cal D} d\delta\, L(\theta,\delta)\pi(\delta)\,, \label{eq:marg}
\ee
where $\pi(\delta)$ is the prior distribution for the  $\delta$ parameters.
This operation is  named \textit{marginalisation}.
In the frequentist framework, the likelihood is instead maximized,
\be
 L_{\rm F}(\theta)=  \max\limits_{\delta\in{\cal D}} \left[ L(\theta,\delta)\pi(\delta)\right]\,.
 \label{eq:marg_freq}
\ee
This operation is usually named \textit{profiling}. Here however, in order to emphasize the parallel between Bayesian and frequentist cases, we also refer to it as ``marginalisation''. The outcome of
Bayesian and frequentist  marginalisation gives respectively the marginal likelihoods  $L_{\rm B}$ and $L_{\rm F}$. The best-fit regions are then obtained by using $L_{\rm B}$ and $L_{\rm F}$ in  Eqs.~\eqref{eq:bayes_contours} and \eqref{eq:freq_contours}, respectively. 
Finally, let us notice that in the frequentist case, it is clear that  the marginalisation operation has the effect of selecting the  values of $\delta$ preferred by the data.

\subsubsection{Bias principle} \label{sec:BiasPrin}

The common feature of Bayesian and frequentist marginalisations  is that nuisance para\-meters contribute to goodness-of-fit. This implies that the nuisance parameters can relax a tension among various measurements, which in turn induces a shift of the best-fit regions. 
In the context of the search for new physics, such a shift could also be characteristic of the presence of a new physics signal. It is thus of highest importance to correctly understand the effects of nuisance parameters, in order not to confuse  systematic uncertainties with the presence of new physics! 

In order to explicitly expose the  shifts induced by nuisance parameters, and ultimately obtain more conservative results, a useful approach is to define a new operation, alternative to marginalising,  with the   requirement that the nuisance para\-meters \textit{do not} contribute to goodness-of-fit. 
We will refer to this principle as \textit{bias}, as opposite to the marginalisation principle. 
We will see that the bias principle provides results that are independent of the shape of the prior of the nuisance parameters.

The bias principle can be intuitively grasped as follows. Consider the likelihood $L(\theta,\delta)$ with a single nuisance parameter on the interval $\delta\in[\delta_a,\delta_b]$. Instead of marginalising over $\delta$, one can look at the contours of the likelihood for various
\textit{discrete} values of $\delta$, say $\delta=\delta_a,\delta_b$. 
For each value of $\delta$, the contours are given by  Eq.~\eqref{eq:bayes_contours} (Bayesian) or Eq.~\eqref{eq:freq_contours} (frequentist). 
To obtain the contours, we can see that the likelihood is separately normalised for $\delta_a$ and $\delta_b$. This normalisation is in general not the same for $\delta_a$ and $\delta_b$. 
Because of this normalisation factor,  no particular value of $\delta$ is preferred by the fit. 
It is this normalisation factor that concretely realises  the bias principle.

In Bayesian statistics, the bias principle finds a general realisation as follows.
The requirement one wants to implement is that the nuisance parameters $\delta$ do not contribute to goodness-of-fit. This is equivalent to ask that the $\delta$ do not have a preferred region once data are taken into account. To translate formally this condition, the relevant quantity to involve is the marginal posterior of $\delta$, $p(\delta|d)$. 
To implement the bias principle, one should thus require $p(\delta|d)$ to  be constant,  which translates into the condition
\be
\frac{\partial}{\partial \delta} p(\delta|d)=0\,,\label{eq:p_constraint}
\ee
with \be
p(\delta|d)=\int_\Omega \,d\theta\, L(\theta,\delta)\, \pi(\delta)\,\pi(\theta)\,.\label{eq:pdelta}  
\ee
We see that the condition \eqref{eq:p_constraint} fixes  the $\pi(\delta)$ prior  to be
\be
\pi(\delta)=\frac{1}{ \int_\Omega \,d\theta\, L(\theta,\delta)\,\pi(\theta)}\,.  \label{eq:DIVc}
\ee
This peculiar prior is not independent on data, and is  thus not orthodox with respect to the usual Bayesian philosophy. This is an expected consequence of biasing and all quantities are  nevertheless well defined. It follows that the posterior for $\theta$ and $\delta$ has the form $L(\theta,\delta)\pi(\theta)/\int d\theta\, [ L(\theta,\delta)\pi(\theta) ]$.
The Bayesian bias likelihood is then given by marginalising this particular posterior with respect to the nuisance parameters,
\be
 \bar L_{\rm B}(\theta)=  \int_{\mathcal{D}} d\delta \left[ \frac{L(\theta,\delta)}{ \int_\Omega d\theta  L(\theta,\delta) \pi(\theta) }\right]\,. \label{eq:bias_bayes}
\ee

In frequentist statistics, the bias principle is realized in a very similar way to the Bayesian case. The quantity telling how $\delta$ is constrained by the data is the marginal likelihood for $\delta$ (with its associated ``prior''), $\max\limits_{\theta\in\Omega} \left[ L(\theta,\delta)\pi(\theta)\pi(\delta)\right]$, which selects the preferred $\theta$ for a given $\delta$. One requires this marginal likelihood to be constant,
\be
\frac{\partial}{\partial \delta} \max\limits_{\theta\in\Omega}\left[  L(\theta,\delta)\pi(\theta)\pi(\delta)\right]=0\,.\label{eq:p_constraint_freq}
\ee
This implies that the $\pi(\delta)$ ``prior'' satisfies
\be
\pi(\delta)=\frac{1}{\max\limits_{\theta\in\Omega} L(\theta,\delta)\pi(\theta)}\,.\label{eq:DIVc_freq}
\ee
The marginal likelihood of $\theta$  is then given by
\be
 \bar L_{\rm F}(\theta)=  \max\limits_{\delta\in{\cal D}} \left[ \frac{L(\theta,\delta)}{ \max\limits_{\theta\in\Omega} [ L(\theta,\delta)\pi(\theta) ] }\right]\,. \label{eq:bias_freq}
\ee
 This operation is sometimes referred to as the envelope method. This is because, for a continuous domain ${\cal D}$, it
 draws continuous regions which are wider than the ones obtained by marginalising.
  \footnote{
Using  $L=e^{-\chi^2/2}$, one has the equivalent formulation of the envelope method in terms of $\chi^2$,
\be \bar \chi^2(\theta)=  \min_\delta \left[ \chi^2(\theta,\delta)-2\log \pi(\theta) - \min_\theta [ \chi^2(\theta,\delta) -2\log \pi(\theta)] \right]\,. \ee
In case of classical frequentist statistics, $\pi(\theta)$ is a constant, so that the two $\log\pi(\theta)$ terms cancel.
}

Comparing the Bayesian and frequentist realisations of the bias principle, Eq.~\eqref{eq:bias_bayes} and Eq.~\eqref{eq:bias_freq}, it appears that the resulting bias operations are fully similar: the expressions Eq.~\eqref{eq:bias_bayes} and Eq.~\eqref{eq:bias_freq} are identical up to interchanging  maximisation and integration.

Let us finally comment about the best-fit regions for the bias likelihoods. 
 The Bayesian bias is a particular case of Bayesian marginalisation with a well-chosen prior. The contours are thus obtained by integration, using $\bar L_{\rm B}$ in Eq.~\eqref{eq:bayes_contours}. For the frequentist bias, the bias likelihood $\bar L_{\rm F}$ can be treated using the usual likelihood ratio test and computing the associated p-value, as described in Eq.~\eqref{eq:freq_contours}. 
We conclude that the best-fit regions for both the Bayesian and frequentist bias are well-defined.

Let us make an important comment which will turn useful for the frequentist treatments in Section~\ref{se:bias}. For a single $\delta$ in the discrete domain ${\cal D}=\{\delta_{a},\delta_b\}$, the best-fit regions 
obtained by inserting the likelihood~(\ref{eq:bias_freq}) in Eq.~\eqref{eq:freq_contours} reproduce exactly the ones in the discrete version of the bias described earlier in this subsection. 
Indeed, the normalized likelihood~(\ref{eq:bias_freq}) will lead to a denominator equal to one in Eq.~\eqref{eq:freq_contours} and the role of this denominator in the contour definition will be played instead by
the  denominator of Eq.~(\ref{eq:bias_freq}).

In this paper, we will refer to the general realisations of the bias principle given by Eq.~\eqref{eq:bias_bayes}, \eqref{eq:bias_freq} as the \textit{envelope method}, for both the Bayesian and frequentist versions.
In contrast, the discrete version of the bias previously introduced  can be seen as a minimal realisation of this principle. In this paper, we will refer to it as the \textit{extremal bias}, for both the Bayesian and frequentist versions.

\section{Combinations of theoretical uncertainties}  \label{se:Combinations}

This section applies to any systematic uncertainties. Nevertheless, since in this paper our main focus is  on theoretical uncertainties, we will readily use this term. 
In the previous section, we have seen that the correct procedure to incorporate theoretical uncertainties into the likelihood is to model these uncertainties using nuisance parameters and 
treat them using either  the marginalisation or the bias approach.
From the practical point of view, this step of marginalisation can be computationally heavy to carry out, both in the  Bayesian and frequentist cases. Indeed,  for each point in the space of parameters of interest, for $n$ nuisance parameters, either a $n$-dimensional integration or a $n$-dimensional maximisation has to be done,
whose complexity typically grows exponentially with $n$.

Because of the cost of  exact marginalisation, it is a common practice in the high-energy physics community to combine certain uncertainties in a preliminary step, before carrying out the operation of marginalising. This approach of ``preliminary combinations'' should be followed with some care, because it can be approximative and may contain implicit assumptions.
In this section, we revisit and develop the various operations of preliminary combination on a firm statistical ground.

\subsection{Error modelisation}\label{se:conventions}

Let $Q$ be an arbitrary quantity entering into a base likelihood $L[Q]$.
The uncertainty about  $Q$  can be modelled via a dependence of the form
\be
Q\mapsto Q\times(1+\delta\, \Delta)\,, \label{eq:deltaA}
\ee
where $\delta$ is the nuisance parameter, associated with a distribution $\pi(\delta)$, defined over the domain $\mathcal{D}$. 
Here and throughout this paper, without loss of generality,  we let all the $\delta$ follow a ``standard distribution'', such that  all the information about the magnitude of the uncertainty will be contained in the coefficient $\Delta$. 
With this parametrisation, $\Delta$ represents the \textit{relative} uncertainty associated with $Q$. This linear model \eqref{eq:deltaA} is valid for any $\pi$ distribution, provided that  the magnitude of the relative error is small, $\Delta\ll 1$.
The actual definition  of  $\pi$  depends on the statistical approach adopted. 
In the Bayesian case,  $\delta$ is  a random variable, so that one chooses ${\rm E}[\delta]=0$, ${\rm V}[\delta]=1$.~\footnote{${\rm E}$ and ${\rm V}$ respectively denote the expected value and variance operators, ${\rm E}[\delta]=\int_{\mathcal{D}}d\delta\,\delta\, \pi(\delta)$ and
${\rm V}[\delta]=\int_{\mathcal{D}}d\delta\,\delta^2\, \pi(\delta)-({\rm E}[\delta])^2$.
} Note that the domain of $\delta$ can be either finite or infinite. In the hybrid frequentist case, one can follow the same conventions as for the Bayesian case.  The classical frequentist case is equivalent to have a flat $\pi$, and one sets the domain to be  $\mathcal{D} \equiv [-1,1]$ in that case.  For the errors we will consider, $\pi$ will always be centred on zero.

\subsection{Bayesian combination of theoretical uncertainties} \label{se:combTU}

In the Bayesian framework, a nuisance parameter $\delta$ is rigorously taken as  a random variable with prior distribution $\pi$. 
In presence of various nuisance parameters, one may wish to combine various sources of error, say $\delta_A$ and $\delta_B$.
A combination of these sources can be done if they appear systematically into a single combination inside the likelihood, $L[\delta_A\Delta_A+\delta_B\Delta_B]$. One can then define the combined error $\delta_C\Delta_C=\delta_A\Delta_A+\delta_B\Delta_B$, so that
\be\label{eq:comb_marg}
L[\delta_C\Delta_C]\,\pi_C(\delta_C)\propto \int d\delta_A \ d\delta_B\,\, \mathbb{\delta}[\delta_A\Delta_A+\delta_B\Delta_B-\delta_C\Delta_C]\,L[\delta_A\Delta_A+\delta_B\Delta_B] \ \pi_{A,B}(\delta_A,\delta_B)\,,
\ee
where $\delta[x]$ is the Dirac distribution. 
Here $\pi_{A,B}$ is the common prior of $\delta_A$, $\delta_B$. If these are independent, one has $\pi_{A,B}(\delta_A,\delta_B)=\pi_A(\delta_A)\pi_B(\delta_B)$.
Note that the integration over $\delta_C$ of the left-hand side of this equation recovers Eq.~\eqref{eq:marg}.

When $\delta_A$ and $\delta_B$ are independent,  Eq.~\eqref{eq:marg} implies that the distribution of $\delta_C$ is exactly given by a convolution product,  
\be
\pi_C\left(\frac{x_C}{\Delta_C}\right)\,=\, \int d x\ \pi_A\left( \frac{x}{\Delta_A} \right)\,\pi_B\left( \frac{x_C-x}{\Delta_B} \right)\,.\label{eq:conv_bayes}
\ee 
The variable $x$ can be seen as $\delta\Delta$. It is convenient to define $\bar \pi_C (x)=\pi_C \left(\frac{x}{\Delta_C}\right)$, so that the width of  $\bar \pi_C$ is given by $\Delta_C$.  In contrast, recall that the width of $\pi_C$ is always normalized to one by convention. 
Using the $\bar \pi$ definition, the  convolution~\eqref{eq:conv_bayes}  can simply be written as 
\be
\bar\pi_C\left(x_C\right)\,=\, \int d x\ \bar\pi_A\left( x \right)\,\bar\pi_B\left( x_C-x \right)\,,  \label{eq:conv_bayesII}
\ee
or more shortly
\be\bar \pi_C=\bar \pi_A\star \bar \pi_B\,.  \label{eq:conv_bayesIII}  \ee
The resulting distribution $\pi_C$ has in general a non trivial shape, except
for example
 when both $\pi_A$and $\pi_B$ are Gaussian, in which case $\pi_C$ is Gaussian as well.
In contrast, Eq.~\eqref{eq:conv_bayes} implies that the magnitudes of the errors $\Delta_A$, $\Delta_B$   are  combined following
\be
 \Delta_C^2=\Delta_A^2+\Delta_B^2\,, \label{eq:comb_Bayes_1D}
\ee
irrespective of the shape of the distributions. That is, the errors are always combined in quadrature, \ie~the variances always add-up. Note the $\Delta^2$'s correspond to the variance of the $\bar \pi$ distributions.

In case of two independent sets of several correlated variables $\delta_{A,i}$, $\delta_{B,i}$ with respective covariance matrices $\mathcal{C}_A$, $\mathcal{C}_B$, combined as
 $\delta_{C,i}=\delta_{A,i}+\delta_{B,i}$,~\footnote{
Note that in this case, for simplicity, we used  a different convention from the one-variable case: we do not
factor out the magnitude of the uncertainties ($\Delta_i$) in front of the $\delta_i$.
} 
 the combination is naturally generalized to
\be
\mathcal{C}_C=\mathcal{C}_A+\mathcal{C}_B\,.\label{eq:comb_Bayes_ND}
\ee
Again, this is independent of the prior shapes. The distribution of $\delta_{C,i}$ is again obtained using Eq.~\eqref{eq:comb_marg}.

Finally, one may wish to combine nuisance parameters that are themselves correlated. In the case  of two nuisance parameters $\delta_A$, $\delta_B$ with a correlation coefficient $\rho$, one gets
\be
 \Delta_C^2=\Delta_A^2+\Delta_B^2+2\rho\Delta_A\Delta_B\,, \label{eq:comb_Bayes_corr_1D}
\ee
giving rise to a linear combination in the fully (anti-)correlated case $\rho=\pm 1$, and to Eq.~\eqref{eq:comb_Bayes_1D} in the de-correlated case $\rho= 0$.
The combination~\eqref{eq:comb_Bayes_corr_1D} is still independent of the prior shapes.
Note that in this case $\pi_C$ is still obtained from Eq.~\eqref{eq:comb_marg}, but is not given anymore by a convolution product because $\pi_A$ and $\pi_B$ are not factorised anymore.

Finally, in the case of two sets of nuisance parameters $\delta_{A, i}$, $\delta_{B, i}$ with a relative correlation matrix $\mathcal {C}_{AB}$, one gets
\be
\mathcal{C}_C=\mathcal{C}_A+\mathcal{C}_B+2\mathcal{C}_{AB}\,.
\ee
All the results of this subsection are straightforward to derive using characteristic functions (see Appendix~\ref{app:approx}).

In the limit $\Delta_A\gg \Delta_B$, it appears that $\pi_C\sim\pi_A$, \ie~the combined prior has mainly the shape of the leading uncertainty. In Section~\ref{se:LMA}, we demonstrate that it is well justified to use Eq.~\eqref{eq:comb_Bayes_1D}, which is exact, together with  the approximation $\pi_C \approx\pi_A$.
Beyond the $\Delta_A\gg \Delta_B$ limit, if one wishes to care about the shape of $\pi_C$, a conservative approach is to consider both extreme cases $\pi_C=\pi_A$ and $\pi_C=\pi_B$. This is because the actual shape of $\pi_C$ is always an intermediate distribution between $\pi_A$ and $\pi_B$, as dictated by the convolution product.

\subsection{Frequentist combination of theoretical uncertainties} \label{se:combTUfreq}

Let us start again with the nuisance parameters $\delta_A$, $\delta_B$ and  their associated ``prior'' distribution $\pi_{A,B}$. If the nuisance parameters enter as a single combination in the likelihood, $L[\delta_A\Delta_A+\delta_B\Delta_B]$, one can define the nuisance parameter $\delta_C$ as above, and write
\be
L[\delta_C\Delta_C] \ \pi_C(\delta_C) \ \propto \ \max_{\delta_A,\delta_B}  \, \bigg[\mathbb{\delta}[\delta_A\Delta_A+\delta_B\Delta_B-\delta_C\Delta_C]\,L[\delta_A\Delta_A+\delta_B\Delta_B] \ \pi_{A,B}(\delta_A,\delta_B)\bigg]\,,\label{eq:comb_freq}
\ee
where again $\delta[x]$ is the Dirac distribution.~\footnote{Here $\delta[x]$ can be taken as the regularised Dirac peak.} We emphasis that this formula is exactly similar to the Bayesian one, Eq.~\eqref{eq:comb_marg}, with integration replaced by marginalisation. 
When $\pi_{A,B}(\delta_A,\delta_B)=\pi_{A}(\delta_A)\pi_{B}(\delta_B)$, it appears then that the distribution of $\delta_C$ is given by
\be
\pi_C\left(\frac{x_C}{\Delta_C}\right)\propto\max_{x}\bigg[ \pi_A\left(\frac{x}{\Delta_A}\right) \pi_B\left(\frac{x_C-x}{\Delta_B}\right)\bigg]\,. \label{eq:freq_prod}
\ee
This formula has  a convolution product structure, where the integration has been replaced by a maximisation. From that point, it is then possible to compute the frequentist correlation matrix, $\mathcal{C}^{-1}_{ij}=-\partial^2 \log L/\partial \theta_i\partial \theta_j$. The general formula for the combination of $\mathcal{C}_A$, $\mathcal{C}_B$ is straightforward but tedious to compute. 
In sharp contrast with the Bayesian case, it appears in the frequentist case that the combination of the correlation matrices $\mathcal{C}_A$, $\mathcal{C}_B$ 
accordingly to Eq.~\eqref{eq:freq_prod} depends on the shape of the $\pi_A$, $\pi_B$ distributions. 

In the particular case where both $\pi_A$, $\pi_B$ are Gaussian, the combination appears to be in quadrature, as in the Bayesian case. The combination formulas then match exactly the Bayesian ones, Eqs.~\eqref{eq:comb_Bayes_1D} and \eqref{eq:comb_Bayes_ND}. Moreover $\pi_C$ is also Gaussian.
Another important particular case is the one of flat priors. In that case, $\pi_C$ appears to be flat, and the combination is \textit{linear}, 
\be
\Delta_C=\Delta_A+\Delta_B\,. \label{eq:comb_freq_lin}
\ee
Note that no correlation matrix can be defined in the flat case. 
\footnote{In the multivariate case, $\delta_{A,i}$ and $\delta_{B,i}$ have in general a non-trivial domain  $\mathcal{D}_A$, $\mathcal{D}_B$. The combined domain $\mathcal{D}_C$ is given by the distance $||\delta_{C,i}||$ for which the centers of $\mathcal{D}_A$ and $\mathcal{D}_B$ are aligned with $\delta_{C,i}$ and 
 the domain $\mathcal{D}_A$ and $\mathcal{D}_B$ share a single point. For example if $\mathcal{D}_A$, $\mathcal{D}_B$ are ``hyper-rectangles" with size $\Delta_{A,i}$, $\Delta_{B,i}$, the sizes simply add up just like in the one-dimensional case, $\Delta_{C,i}=\Delta_{A,i}+\Delta_{B,i}$.}

In the case where $\delta_A$ and $\delta_B$ are correlated, they should be  treated with a common ``prior'' as in the Bayesian case.

\subsection{The leading moment approximation} \label{se:LMA}

Consider again the Bayesian case of a combination of two nuisance parameters, $\delta_C\Delta_C\equiv \delta_A\Delta_A+\delta_B\Delta_B$. Recall that the $\delta$ parameters have zero mean and have a standard distribution so that ${\rm E}[\delta]=0 $, ${\rm V}[\delta]=1 $. Assume further that the magnitude of the uncertainty $B$ is small with respect to the uncertainty $A$, 
\be
\Delta_A\gg \Delta_B\,.
\ee
When this condition is satisfied, the source of uncertainty $B$ can be treated as a perturbation to the source of uncertainty $A$. Starting from this observation, one can obtain $\pi_C$   up to $\Delta_B/\Delta_A$ corrections (see Eq.~\eqref{eq:LMA_piC}). This is demonstrated in Appendix~\ref{app:approx} using characteristic functions.
In particular, for independent variables, at the first non-trivial order in the expansion, one obtains that
\be
\pi_C\approx\pi_A  \label{eq:piCpiA}
\ee
\be
\Delta_C^2=\Delta_A^2+\Delta_B^2\,. \label{eq:quad2}
\ee
Recall that $\pi_C$ is determined by the convolution product $\bar \pi_C = \bar \pi_A \star \bar \pi_B$. Hence for $\Delta_A\gg\Delta_B$, one can intuitively expect that the shape of $\bar \pi_A$ and $\bar \pi_C$ are similar (see Eq.~\eqref{eq:piCpiA}), even though their widths are different (according to Eq.~\eqref{eq:quad2}).
In case $\delta_A$ and $\delta_B$ are correlated, Eq.~\eqref{eq:quad2} has to be replaced be Eq.~\eqref{eq:comb_Bayes_corr_1D}.

This ``leading moment'' approximation is useful in presence of a hierarchy between the magnitude of the various  uncertainties.
It dictates how to consistently capture the main effects of the uncertainties into the likelihood.
This in turn  allows one to obtain  an approximate form for the combined priors, which opens up the possibility of obtaining analytical expressions for the marginal likelihoods.

The leading moment approximation also applies when $\delta_A$ and $\delta_B$ appear in various linear combinations within the likelihood. This situation typically happens when various observables are affected by the same source of uncertainty.
 The case of two nuisance parameters and two combinations is discussed in Appendix~\ref{app:approx}.  
 One considers two combinations $\delta_{C_1}\Delta_{C_1}=\delta_{A}\Delta_{A_1}+\delta_{B}\Delta_{B_1}$, $\delta_{C_2}\Delta_{C_2}=\delta_{A}\Delta_{A_2}+\delta_{B}\Delta_{B_2}$. 
It is found that the  $\Delta_{C_{1,2}}$ are obtained as in the one-combination case discussed above. 
The  correlation coefficient between $\delta_{C_1}$ and $\delta_{C_2}$ requires more attention. 
If $\Delta_{A_1}\gg \Delta_{B_1}$, $\Delta_{A_2}\gg \Delta_{B_2}$, it is found to be  approximately equal to one. 
 This implies that the shapes of the distributions of $\delta_{C_1}$, $\delta_{C_2}$ and $\delta_{A}$ are the same up to $\Delta_{B_{1,2}}/\Delta_{A_{1,2}}$ corrections (see Eq.~\eqref{eq:LMA_comb_app1}), that is  
  \be\pi_{C_1 C_2}(\delta_{C_1},\delta_{C_2}) \approx \pi_A(\delta_{C_1})\,\delta[\delta_{C_1}-\delta_{C_2}] \,. \label{eq:LMA_comb_12} \ee
From Eq.~\eqref{eq:LMA_comb_12}, it appears that the leading moment approximation reduces the number of nuisance parameters in the likelihood.
In the case where $\Delta_{A_1}\gg \Delta_{B_1}$, $\Delta_{A_2}\ll \Delta_{B_2}$, it appears that the correlation coefficient between $\delta_{C_1}$ and $\delta_{C_2}$ is approximately equal to the correlation coefficient between $\delta_A$ and $\delta_B$ (see Eq.~\eqref{eq:LMA_comb_app2}), so that 
\be
\pi_{C_1C_2}\approx \pi_{AB}\,.
\ee
In the particular case where $\delta_A$ and $\delta_B$ are independent, one has
\be
\pi_{C_1C_2}\approx\pi_{C_1}\pi_{C_2}\,,\quad \pi_{C_1}\approx \pi_A\,,\quad\pi_{C_2}\approx \pi_B\,. \label{eq:LMA_comb_3a}
\ee
In the other particular case where $\delta_A$ and $\delta_B$ are $100\%$ correlated or anti-correlated, one has 
 \be\pi_{C_1 C_2}(\delta_{C_1},\delta_{C_2}) \approx \pi_A(\delta_{C_1})\,\delta[\delta_{C_1}\pm\delta_{C_2}] \,. \label{eq:LMA_comb_3} \ee
 All the cases with more variables or more combinations can be deduced recursively from the case with two parameters and two combinations studied here.~\footnote{This 
 leading moment approximation will be applied to the theoretical uncertainties on the Higgs rates in Sections~\ref{CTerr} and \ref{BRerr}.}

\subsection{Combining uncertainties in the bias approach}
\label{se:comb_bias}
We now analyse how the combination of uncertainties arises in the case of the method of bias. 
We still consider a combination of nuisance parameters $\delta_{A,B}$ entering in the likelihood as $L[\delta_A\Delta_A+\delta_B\Delta_B]$. Recall that in our conventions, $\delta$ is a random variable with a fixed domain, while $\Delta$ is a number representing the magnitude of the uncertainty.  In the bias approach, by definition, the shape of the distribution of $\delta$ is set so that $\delta$ does not participate to the fit. The information about  the uncertainty is thus encoded only in the \textit{domain} of the variable $\delta \Delta$. 
The choice of this domain has some degree of arbitrariness. This choice depends on how conservative one wants the results to be. In the following we choose to let  $\delta$ vary in the interval $[-1,1]$ and we identify $\Delta$ as a $1\sigma$ error, \ie~the same way it is defined for the marginalisation.

The operation of Bayesian bias can be seen as a special case  of marginalisation, where the prior is set by Eq.~\eqref{eq:DIVc}. As the likelihood we consider in this section depends only on the combination $\delta_A\Delta_A+\delta_B\Delta_B$, this peculiar prior depends only on the combination $\delta_A\Delta_A+\delta_B\Delta_B$ by construction. Let us denote it as $\pi_{\rm bias}^B(\delta_A\Delta_A+\delta_B\Delta_B)$. In order to get the combination $\delta_C\Delta_C=\delta_A\Delta_A+\delta_B\Delta_B$, one applies the definition of Eq.~\eqref{eq:comb_marg} using  the $\pi_{\rm bias}^B$ prior. It turns out that $\pi_C(\delta_C)=\pi_{\rm bias}^B(\delta_C \Delta_C)$. This means that the domain of $\delta_C\Delta_C$ is given by the domain of $\delta_A\Delta_A+\delta_B\Delta_B$,
\be
\mathcal{D}_{\delta_C\Delta_C}=\mathcal{D}_{\delta_A\Delta_A+\delta_B\Delta_B}\,.
\ee
When $\delta_A$ and $\delta_B$ are independent, one has simply
\be
\Delta_C=\Delta_A+\Delta_B\,. \label{eq:comb_bias_0}
\ee
When $\delta_A$ and $\delta_B$ are  $100\%$ correlated positively (\ie~$\delta_A=\delta_B$), it turns out that one has again the combination
\be
\Delta_C=\Delta_A+\Delta_B\,.\label{eq:comb_bias_+1}
\ee
When $\delta_A$ and $\delta_B$ are $100\%$  correlated negatively  (\ie~$\delta_A=-\delta_B$), the combination reads
\be
\Delta_C=|\Delta_A-\Delta_B|\,.\label{eq:comb_bias_-1}
\ee
Let us stress that the correlation between $\delta_A$ and $\delta_B$ is determined by their common domain $\mathcal{D}_{\delta_A\Delta_A,\delta_B\Delta_B}$. The above extreme cases are easily determined.  
The case of an intermediate correlation is trickier as it requires a precise definition of the domain. The case of an arbitrary correlation will not be needed throughout this paper. 
We see that the uncertainties are automatically combined \textit{linearly} in the Bayesian bias method.

These results above can be applied recursively to more complex  combinations. For example if $\delta_D\Delta_D=\delta_A\Delta_A+\delta_B\Delta_B+\delta_C\Delta_C$, with $\delta_A$ and $\delta_B$ $100\%$ anti-correlated and $\delta_c$ independent from the two others, the bias combination gives 
\be
\Delta_D=|\Delta_A-\Delta_B|+\Delta_C\,.\label{eq:comb_bias_3}
\ee
Also, the bias combination  applies in presence of various linear combinations (labelled by $i$) of the same nuisance parameters. In that case, the result of the combination  is a common nuisance parameter $\delta$, coming with  different magnitudes $\Delta_i$ for each combination.

The frequentist bias has the same structure as the Bayesian bias. The starting point to determine the error combination is to use the frequentist version of the bias prior of Eq.\eqref{eq:DIVc_freq} in Eq.~\eqref{eq:comb_freq}.
It follows that the frequentist combinations are the same as in the Bayesian case.
We can thus conclude that in the bias approach, the preliminary combinations of uncertainties are done linearly, in both the frequentist and Bayesian cases. 
One should remark that such a combination is systematically more conservative than the combinations from both the Bayesian and frequentist marginalisations, as can be seen comparing Eqs.~\eqref{eq:comb_bias_0}, \eqref{eq:comb_bias_+1}, \eqref{eq:comb_bias_-1} with for example  Eq.~\eqref{eq:comb_Bayes_corr_1D}. Note that the  combination in the frequentist marginalisation  with flat prior (see {\it e.g.} Eq.~\eqref{eq:comb_freq_lin}) is the same as the bias combination.
 Therefore the bias method is also  more conservative than the standard marginalisation at the level of error combinations.

\section{The Higgs boson rates }  \label{se:MU}

The couplings of the Higgs boson $h$ are all predicted in the Standard Model, so that any deviation from the SM predictions would constitute a sign of the existence of physics beyond the SM. 
The  Higgs couplings can be probed by collider experiments, which can produce the Higgs on-shell and observe its decays. This process of Higgs production followed by its decay is parametrised as
\be pp\,(p\bar p)  \xrightarrow{X} h \rightarrow Y \,.\ee
The SM Higgs production mechanisms accessible at the LHC (and Tevatron) are {\it i)} gluon-gluon fusion (ggF), {\it ii)} vector boson
fusion (VBF), {\it iii)} associated production with an electroweak gauge
boson $V=W,Z$ (VH), and {\it iv)} associated production with a $t\bar{t}$ pair (ttH). 
The main SM Higgs decays observed at the colliders are decays into gauge bosons, $h\rightarrow \gamma\gamma$, $ZZ$, $W^+W^-$,  and into heavy fermions,  $h\rightarrow b\bar b$, $\tau\bar\tau$. The production modes $X$ and final states $Y$ will be therefore taken in the following list,
\be
X=\{\textrm{ggF, VBF, VH, ttH} \}\,,
\ee
\be
Y=\{ \gamma\gamma, ZZ,  WW,  b\bar b, \tau\bar \tau \}\,.
\ee

\subsection{The data}
\label{sec:ExpSide}

The Higgs searches at ATLAS, CMS and the Tevatron are focussed on a specific final state $Y$. For each final state, various channels are defined using mutually exclusive cuts.  Throughout this paper, these experimental channels will be labelled by lower case latin indices $(i,j \dots)$.  We will consider all the $88$ channels. A given $i$ contains the information on the final state and the specific channel. 
In the following, it will be sometimes useful to refer to the final state $Y$ corresponding to a given channel $i$. We will use the short notation $Y_i $, meaning that $Y$ is taken as a  function of the variable $i$, \ie~ 
$Y_i\equiv Y(i) $.

The results from Higgs searches at the LHC and the Tevatron are reported in terms  of signal strengths $\mu_i^{\rm ex}$. A signal strength is defined as the ratio of the observed event number with the expected SM event number,  \be\mu_i^{\rm ex}=\frac{N_i^{\rm ex}}{N_i^{\rm SM}}\,.\ee

The predicted SM event rate of a process $pp\,(p\bar p)  \xrightarrow{X} h \rightarrow Y$ is given, in the narrow width approximation,  by $\mathscr{L}\sigma_X^{\rm SM} B_Y^{\rm SM}$. Here $\sigma_X^{\rm SM}$ is the production rate, $B_Y^{\rm SM}$ is the branching ratio $B_Y^{\rm SM}=\Gamma_Y^{\rm SM}/\sum_{Y'} \Gamma_{Y'}^{\rm SM}$ and $\mathscr{L}$ is the integrated luminosity. 
However, from the experimental viewpoint, 
 all the production processes contribute to a given final state. Hence the Higgs production cross sections have to be weighted by a selection efficiency  $\epsilon_{X,i }^{\textrm{SM}}$ encoding the effects of kinematical cuts. 
    The actual expected event rates are thus given by \be N^{\rm SM}_i=\mathscr{L}\,\sum_X  \epsilon_{X,i }^{\textrm{SM}} \sigma_X^{\textrm{SM}} B_i^{\textrm{SM}}\,\,,
\label{eq:N_SM}    
    \ee
where the notation $B_i^{\rm SM}$ is a shortcut for $B^{\rm SM}_{Y(i)}$, \ie~the index $i$ selects the final state $Y$.
 The experimental Higgs signal strengths have thus the form 
 \be
 \mu_{i}^{\rm ex}    \ = \  \frac{N_{i}^{\rm ex}}{\mathscr{ L}\,\sum_X  \epsilon_{X,i }^{\textrm{SM}} \sigma_X^\textrm{SM}\, B_i^\textrm{SM}\,   } \,. \label{eq:REALmu}
 \ee
Note that the kinematical cuts have been to some extent designed to disentangle the production modes, so that  often one of the efficiencies will dominate over the others.

The experimental central values of the $\mu_{i}^{\rm ex}$, the associated statistical errors, the experimental systematic errors,    and the selection efficiencies $\epsilon_{X,i}^{\textrm{SM}}$ that we will exploit in our analysis are taken from the following references.
The statistical and experimental systematic errors are often combined within these references and will be denoted here as $ \Delta\mu_i^{\rm ex}$. 
\\
Regarding the ATLAS data, 
the diphoton final state results are taken from Ref.~\cite{A-diphoton}, 
the $ZZ$ channel is from Ref.~\cite{A-ZZ}, 
the $WW$ channel from Ref.~\cite{A-WW}, 
the $b \bar b$ from Ref.~\cite{A-bb} and 
the $\tau\bar \tau$ from Ref.~\cite{A-tautau}.
Results are presented as well in Ref.~\cite{ATLASweb} and the combined channels are studied in Ref.~\cite{ATLASfit}.
\\ 
As for the CMS results,
the diphoton final state has been presented in Ref.~\cite{C-diphoton}, 
the $ZZ$ channel measurements are provided in Ref.~\cite{C-ZZ}, 
the $WW$ ones in Ref.~\cite{C-WW},
the $b \bar b$ in Ref.~\cite{C-bb} and 
the $\tau\bar \tau$ in Ref.~\cite{C-tautau} (see also Ref.~\cite{CMSweb} and the combined channel analyses~\cite{CMSfit}).
\\
Finally, the latest results from the Tevatron (D0 and CDF Collaborations) can be found
in Ref.~\cite{TevatWEB,muTevatron}.

Apart from statistical and experimental systematic errors, certain theoretical errors on $\mu_{i}^{\rm ex}$ are included 
in the public results. To the best of our knowledge, the combination between these experimental and theoretical uncertainties  is often made in quadrature.
We thus subtract in quadrature these theoretical errors from the provided total uncertainties. 
How to properly (re)introduce  the theoretical errors constitutes the main topic of this paper, and will be discussed at length in the upcoming sections.

Finally, we mention that we do not include in our fits more challenging observables related to the Higgs pair production~\cite{Djouadi:2005gi}, off-shell effects, 
loop-induced $Z\gamma$ final state, electron/muon pair final states, final states induced by flavour-changing Higgs couplings, nor exotic or invisible final states. 
Some of those would require to introduce new parameters in the Lagrangian that we will consider in Eq.~(\ref{Eq:LagEff}). The motivation is to keep a simple physical 
framework in order to discuss easily the statistical aspects. In any case, the present experimental limits on such Higgs observables are still not stringent enough to affect 
drastically the Higgs fits. Moreover, all the 
statistical concepts discussed throughout the paper can be simply extended to new Higgs observables.

\subsection{New physics parametrisation}
\label{sec:EffLag}
 
The new physics possibly lying beyond the SM may induce a distortion of the SM Higgs couplings. The correct way of dealing with the low-energy manifestation of heavy new physics is through the use of an effective Lagrangian (see {\it e.g.} Ref.~\cite{BaySylvain} for global fits of the Higgs effective Lagrangian). The leading effects on the Higgs sector appear through dimension-6 operators.
The effective Lagrangian then induces anomalous couplings between the Higgs and the SM particles. The anomalous couplings to weak bosons and to  heavy fermions can be parametrised as   
\begin{eqnarray} 
{\cal L}_H  & = &  
 \ c_W \ g_{hWW} \ h \ W_{\mu}^+ W^{-\mu} + \ c_Z \ g_{hZZ} \ h \  Z_{\mu}^0 Z^{0 \mu} \nonumber \\
  & & - \ c_t \ y_t \ h \ \bar t_L t_R \  - c_b \ y_b \ h \ \bar b_L b_R \ - c_c \ y_c \ h \ \bar c_L c_R \ - c_\tau \ y_\tau \ h \ \bar \tau_L \tau_R + \ {\rm h.c.}  
\label{Eq:LagEff}
\end{eqnarray} 
where $y_{t,b,c,\tau}$ are the SM Yukawa coupling constants (in mass eigenbasis), 
the subscript $L/R$ indicates the fermion chirality,  
$v$ is the Higgs vacuum expectation value, 
$g_{hWW} = 2M^2_W/v$ and $g_{hZZ} = M^2_Z/v$ are the EW gauge boson couplings.
 The $c_{W,Z,t,b,c,\tau}$ parameters are defined  such that the limiting case $c_{W,Z,t,b,c,\tau}\to 1$ corresponds to the SM. 
New tensor structures are also generated by the effective Lagrangian but are not taken into account here.


Our focus being on theoretical uncertainties,  we adopt a fairly simple parametrisation of the new physics effects. 
We assume universal deviations for fermion couplings, $c_f\equiv c_{t}=c_{b}=c_{c}=c_{\tau}$, and for weak bosons, $c_V\equiv c_W=c_Z$. The $c_f$ are assumed to be real.
Clearly, this simplified description of the new physics effects represents only a piece (operators with no extra derivatives) of  the full dimension-6 effective Lagrangian.
Having $c_W\approx c_Z$ and $c_f$ universality is  however approximately compatible with certain new physics scenarios, like for a warped extra-dimension with bulk custodial  
symmetry vanishing IR brane kinetic terms for EW gauge bosons \cite{NeubertWED,SylvainWED}.~\footnote{Note 
that contrary to a widespread belief, $c_W=c_Z$ is not entirely justified by custodial symmetry \cite{SylvainWED}. }
Having only two parameters in this simplified framework, the results of our fits will systematically be presented in the $c_V-c_f$ plane.

In the hypothesis of the existence of a physics Beyond the SM (BSM) parametrised by $c_V-c_f$, the
 expected signal strength  is given by 
\begin{eqnarray}
\mu_{i}^{\rm th} [c_V,c_f]\ = \frac{N_i^{\rm BSM}[c_V,c_f]}{N_i^{\rm SM}} = \frac{\sum_X  \epsilon_{X,i }^{\textrm{BSM}} \sigma_X^\textrm{BSM}\, B_i^\textrm{BSM}\,}{\sum_X  \epsilon_{X,i }^{\textrm{SM}} \sigma_X^\textrm{SM}\, B_i^\textrm{SM}\,  } \, ,
\label{eq:THmu}
\end{eqnarray}
$N_i^{\rm SM}$ being defined in Eq.~\eqref{eq:N_SM}.
This is the theoretical prediction of the experimental signal strength defined in Eq.~\eqref{eq:REALmu}.
Both  BSM cross sections and branching ratios $\sigma_X^{\rm BSM}$, $B_i^{\rm BSM}$ can be expressed in terms  of the SM amplitudes and of $c_V,c_f$. The expressions can for example be found in Ref.~\cite{EFIT}, whose procedure is closely followed here. 
In all generality, the BSM efficiencies are not the same as the ones of the SM either. However, this happens when couplings with new tensors structures are generated by new physics. In our simplified framework, this does not happen, such that one can safely take $\epsilon_{X,i}^{\rm BSM}= \epsilon_{X,i}^{\rm SM}\equiv \epsilon_{X}^i$.

The SM production cross sections and partial decay widths for the Higgs boson are taken, respectively, from the LHC Higgs cross section Working Group 
(LHCHWG) Ref.~\cite{LHCHWGweb} (see also Ref.~\cite{LHCHWG1,LHCHWG2,LHCHWG3} as well as the recent N$^3$LO ggF computation~\cite{Duhr}) and Ref.~\cite{LHCHWGweb,LHCHWG3}. 
These numerical results correspond to the rates calculated at the highest orders of EW and QCD corrections known so far (mixed EW-QCD 
at NNLO for the ggF mechanism~\cite{QCDEWnnlo} and at NLO for other Higgs production modes).

\section{The Higgs likelihood}  \label{se:LikeUncer}

\subsection{The base likelihood} \label{se:BaseLike}

Having introduced the statistical framework and the Higgs data  in Sections~\ref{se:STAT}  to  \ref{se:MU}, we can proceed with building the Higgs likelihood function. We define the  \textit{base} likelihood $L_0$ as the likelihood containing the central values of Higgs signal strengths, and the experimental uncertainties.  
The theoretical  uncertainties are kept apart from now. Their inclusion into the base likelihood will be discussed at length in the next sections and is the central topic of this paper.

In absence of any experimental systematic errors, a signal strength variable follows a Poisson statistics, and the associated likelihood is thus a Poisson distribution.  Whenever the event number is large enough, about $O(10)$ in practice, the likelihood can be approximated by a Gaussian. 
In contrast, in presence of systematic uncertainties, this approximation generally does not  hold. In practice however, 
the complete likelihood resulting from the combination of statistical and experimental systematic errors is not provided in the experimental public results. We will therefore model the base likelihood using Gaussian distributions, just as if the shape came out only from the statistical error. Such an approximation is expected to be good as long as the systematic error is small with respect to the statistical error, as shown in Section~\ref{se:LMA} and Appendix~\ref{app:approx}.

The observed rates in the current 88 channels (labelled by $i,j$) are potentially correlated, for example because of the experimental error on the luminosity.  The base likelihood  follows therefore a multivariate normal distribution,
\be
L_\mu(\mu^{\rm th}_i;\mu^{\rm ex}_i)=\exp\left[{-\frac{1}{2}\sum_{i,j}(\mu_{i}^{\rm th}-\mu_{i}^{\rm ex})\, \mathcal{C}^{{\rm ex}\,-1}_{ij}\, (\mu_{j}^{\rm th}-\mu^{\rm ex}_{j}) }\right]\,,
\label{pdfGAUSS}
\ee
where  $\mathcal{C}^{\rm ex}_{ij}$ is the correlation matrix among all channels.

Ideally,  each individual observed channel $i$ must be considered in order to take into account all the experimental information available on the signal strengths.
In practice,  few elements of this correlation matrix have been provided by the Collaborations up to now. 
Therefore in the following, we will  include only the diagonal elements of $\mathcal{C}_{ij}^{\rm ex}$, given by $\mathcal{C}_{ii}^{\rm ex}=(\Delta \mu_i ^{\rm ex})^2$, where $\Delta \mu_i ^{\rm ex}$ is the experimental uncertainty extracted from the public experimental results.
For future releases, we encourage the experimental Collaborations to provide  as many elements as possible for the correlation matrix    of the individual signal strengths.
\footnote{ Also, we suggest that both the magnitudes of the uncertainties $\Delta \mu_i ^{\rm ex}$ and the correlations should be presented without ambiguities, 
 so that the people exterior to the Collaborations be able to properly reconstruct the likelihood function.}
\\  Alternatively, to perform the Higgs fits one could think of using the correlations between
the combined observed rates,  that are currently provided by the LHC Collaborations. 
Although  instructive, these combined rates do not keep track of all information since they are grouping together different Higgs production modes (which were originally measured independently), like $\mu^{\rm ex}_{\rm VBF,VH}$ and $\mu^{\rm ex}_{\rm ggF,ttH}$ for each Higgs decay 
channel~\cite{ATLASweb,CMSweb}.
Notice that such combined signal strengths also hide some information in the sense that they can result from summations over various exclusive selection cut categories.

\subsection{The uncertainty on the  signal strengths }

The Higgs theoretical uncertainties we will refer to are the theoretical uncertainties associated with the \textit{expected} event rates $ N^{\rm SM}_i$ defined in Eq.~\eqref{eq:N_SM}, that are obtained through analytical and numerical computations in quantum field theory. These uncertainties will propagate both into the experimental signal strengths $\mu^{\rm ex}_i$ and into the theoretical strengths $\mu^{\rm th}_i$, defined in Eqs.~\eqref{eq:REALmu}, \eqref{eq:THmu}. 
Following our conventions (see Section~\ref{se:Combinations}, Eq.~\eqref{eq:deltaA}), the theoretical uncertainty on the Standard Model expected rate in a channel $i$ is written under the form
\be
 N^{\rm SM}_i(1+\delta^N_i \Delta^N_i)\,,
\ee
where  $\delta^N_i$ is the nuisance parameter with ${\rm E}[\delta^N_i]=0$, ${\rm V}[\delta^N_i]=1$, and $\Delta^N_i$ represents the relative magnitude of the uncertainty.

The theoretical uncertainty on $N_i^{\rm SM}$ propagates to the experimental signal strength as
\be
\mu_i^{\rm ex}(1+\delta^{\mu}_i\Delta^{\mu}_i)=\mu_i^{\rm ex}(1-\delta^N_i \Delta^N_i)\,. \label{eq:mu_ex_error }
\ee
The case of the theoretical signal strength $\mu^{\rm th}_i= N^{\rm BSM}_i/ N^{\rm SM}_i$ is slightly trickier. Here we focus on the most realistic case where the deviations induced by new physics are small, so that the anomalous couplings $c_a$ (with $a=(W,Z,t,b,c,\tau)$) are close to one, \ie~$|c_a-1|\ll1$. The contributions from new physics can be linearised with respect to the small parameters $c_a-1$, so that the BSM event rate in the channel $i$ can be written as
\be
 N^{\rm BSM}_i= N^{\rm SM}_i+\sum_a (c_a-1) N^{\rm BSM}_{a,i}+O((c_a-1)^2)\,.
\ee 
In this expression, 
it appears that the leading source of uncertainty comes from the SM event rate uncertainty $\Delta^N_i$.  In the expression of $\mu^{\rm th}_i$, it turns out that this uncertainty cancels out at first order between the numerator ($N_i^{\rm BSM}$) and the denominator ($N_i^{\rm SM}$). 
The subleading uncertainties would then come from a term quadratic  in $\Delta^N_i$ and from the relative uncertainty $(c_a-1) \frac{\Delta N^{\rm BSM}_{a,i}}{N^{\rm BSM}_{a,i}}$  
on the components $N^{\rm BSM}_{a,i}$. Notice that one can reasonably expect similar QCD errors in the SM and BSM predictions so that $\frac{\Delta N^{\rm BSM}_{a,i}}{N^{\rm BSM}_{a,i}} \sim \Delta^N_i$. 
These higher-order contributions are subleading compared to the error on the experimental signal strength, given in Eq.~\eqref{eq:mu_ex_error }, which is of order $ \Delta^N_i$. In the following, we will thus focus only on the  
uncertainty of the experimental signal strength $\mu_i^{\rm ex}(1+\delta^{\mu}_i\Delta^{\mu}_i)$.

\subsection{The structure of the Higgs theoretical uncertainties }

The theoretical uncertainty on $N^{\rm SM}_i$ comes from the errors on the Higgs cross sections $\sigma^{\rm SM}_X$ and partial decay widths  
$\Gamma_Y^\textrm{SM}$. Still following our conventions, these relative uncertainties are written as
\be
\sigma_X^{\rm SM}(1+\delta^\sigma_X\Delta^\sigma_X)\,,
\ee
\be
\Gamma_Y^\textrm{SM}(1+\delta^\Gamma_{Y} \Delta^\Gamma_{Y})\,. 
\ee
The exact content of these errors  will be discussed in details in the next section. 

The uncertainty on the partial decay width propagates to the branching ratios. Defi\-ning the relative error on the branching ratios as $B_Y^\textrm{SM}(1+\delta^B_{Y} \Delta^B_{Y})$, 
one has~\footnote{$\delta_{YY'}$ represents the Kronecker symbol.}
\begin{eqnarray}
\delta^B_{Y} \Delta^B_{Y}  = \sum_{Y'} \delta^{\Gamma}_{Y'} \Delta_{Y'}^{\Gamma} \bigg ( B^{\rm SM}_{Y'} -  \delta_{{Y}Y'} \bigg )\,.
\label{eq:comb_B}
\end{eqnarray}
The uncertainty from the cross sections and branching ratios then propagates to the signal strength~(\ref{eq:REALmu}) and is thus 
encoded in a factor $\mu_i^{\rm ex}(1+\delta^\mu_i \Delta^\mu_i)$ where
\be
\delta^\mu_i \Delta^\mu_i=-\delta^N_i \Delta^N_i=
-\frac{\sum_X  \epsilon_{X}^{i} \sigma_X^\textrm{SM}\,\delta_X^\sigma\Delta_X^\sigma}{\sum_{X'}\epsilon^{i}_{\rm X'} \sigma^{\rm SM}_{\rm X'}}
-\delta^B_{Y_i}\Delta^B_{Y_i}\, ,
\label{eq:combination_NSM}
\ee 
$Y_i=Y(i)$ being the $Y$ decay mode of the Higgs channel detection $i$. Note that 
the sign after the first equal symbol is just a convention if the errors are symmetric.

Finally, the errors on cross sections and partial widths come from several sources. One can write those generically as
\begin{eqnarray}
\delta^\sigma_X\Delta^\sigma_X=\sum_n \delta^{n}_X\Delta^{n}_X\,,  \label{eq:subPROD}
\\ 
\delta^\Gamma_{Y} \Delta^\Gamma_{Y} \ = \ \sum_{n'} \delta_Y^{n'}  \Delta_{Y}^{n'}   \label{eq:subDECAY}
\end{eqnarray}
with the relative errors $\Delta^{n}_X$, $\Delta^{n'}_Y$ to be detailed in the following. \footnote{
Throughout the paper, we will systematically denote the values of $\Delta^{n}_X$, $\Delta^{n'}_Y$ taken from the literature by $\Delta \ \vert_0$ or $\Delta \ ^0$. The possible ambiguities in the interpretation of these numbers  will be discussed case by case.}

Knowing the base likelihood of Eq.~\eqref{pdfGAUSS}, and knowing where exactly the theoretical uncertainties enter, we have the complete Higgs likelihood 
as a function of all the quantities that will have to be treated statistically, namely the nuisance parameters and the effective BSM parameters, \footnote{In the following, to adopt compact notation, we will omit the $c_V,c_f$ arguments of the likelihood function when no ambiguity is possible. }
\be 
L_\mu\left(\mu_i^{\rm th}[c_V,c_f] ;\,  \mu_i^{\rm ex} (1+\delta^\mu_i \Delta^\mu_i ) \right) \ = \ 
L_0\left(c_V, c_f ; \delta^{n}_X, \delta^{n'}_Y  \right) \,. \label{eq:L0}
\ee
Rigorously, the next step is to eliminate the nuisance parameters, $\delta^{n}_X$, $\delta^{n'}_Y$, applying either the marginalisation or the bias method. 
In general these steps should be performed numerically, and are computationally heavy. 
Here however, we will use the methods of preliminary combinations advocated in Section~\ref{se:Combinations}.
Then it will appear that the subsequent Higgs likelihoods are much lighter to treat.

\section{Combining the Higgs rate uncertainties }  \label{se:marg}

In this section we shall combine the Higgs rate uncertainties that will be used in  the marginal
likelihood   studied in Section~\ref{se:MargBF}. 
The most clear and rigorous statistical context for the marginalisation procedure is arguably the one of Bayesian statistics. In particular, the nuisance parameters   are treated on the same ground as the variables of interest and are thus automatically given a probability distribution (see for instance Ref.~\cite{Trotta:2008qt}). 
For that reason we focus in this section on the error combinations within the \textit{Bayesian} context. The resulting likelihood involving the  combined errors will be formally treated within both the Bayesian and frequentist marginalisations in Section~\ref{se:MargBF}.

As we have described in Section~\ref{sec:MargPrin}, 
the Bayesian marginalisation procedure eliminates the dependence of the likelihood on the  nuisance parameters through an integration. For the Higgs likelihood Eq.~\eqref{eq:L0}, this integration reads
\begin{eqnarray}
L(c_V,c_f) \ = \  \int \bigg(\prod_{n,n', X, Y}  d \delta^n_X \ d \delta^{n'}_Y\bigg)\,\pi_0(\delta^n_X,\delta^{n'}_Y)\, \ L_0 (c_V,c_f;\delta^n_X,\delta^{n'}_Y)\,,
\end{eqnarray}
where $\pi_0$ is the joint prior of all the nuisance parameters. Recall that this prior factorises when parameters are independent.
More explicitly, this marginal likelihood reads
\begin{eqnarray} \label{GenLprior}  
L(c_V,c_f)  & = & \int \bigg(\prod_{n,n', X, Y}  d \delta^n_X \ d \delta^{n'}_Y\bigg)\,\pi_0(\delta^n_X,\delta^{n'}_Y)\, \times
\\
&& \exp\left[{-\frac{1}{2}\sum_{i,j}(\mu_{i}^{\rm th}[c_V,c_f]-\mu_{i}^{\rm ex}(1+\delta^\mu_i\Delta^\mu_i))\, 
\mathcal{C}^{{\rm ex}\,-1}_{ij}\, (\mu_{j}^{\rm th}[c_V,c_f]-\mu^{\rm ex}_{j}(1+\delta^\mu_j\Delta^\mu_j)) }\right] . 
\nonumber 
\end{eqnarray}
The theoretical uncertainties $\delta^\mu_i\Delta^\mu_i$ on each signal strength $\mu_i$ are expressed in terms of the uncertainties on cross section $\delta^n_X\Delta^n_X$  and partial decay width $\delta^{n'}_Y\Delta^{n'}_Y$ through Eqs.~\eqref{eq:comb_B} to \eqref{eq:subDECAY}.

In the following subsections, starting from Eq.~\eqref{GenLprior}, we will combine all the sources of uncertainty step-by-step, following the combination formalism established in Section \ref{se:Combinations}.
The aim of this section is to provide a clear and exhaustive treatment of all the Higgs theoretical uncertainties.

\subsection{Combining the PDF and $\alpha_s$ uncertainties}  \label{PDFsole}

Let us first discuss the errors on QCD predictions for the Higgs production cross sections at the proton level. 
Those are induced by the uncertainties on the parton Probability Density Functions (PDF) inside the proton. 
First, one may distinguish between two distinct origins to the PDF uncertainties: an experimental source --~as the PDF are reconstructed from collider data~-- 
and the choice of a specific PDF set (MSTW, CT/CTEQ, NNPDF\dots). 
\\
Second, we consider simultaneously the parametric uncertainty coming from the strong coupling constant, $\alpha_s$. 
We consider both PDF and $\alpha_s$ uncertainties simultaneously because  they contribute in an intricate way to the cross section, as $\alpha_s$ enters both in the hard process matrix element and the PDF themselves.
\\ \\ 
$\bullet$ {\sc Modeling the uncertainties:} \\ 
The uncertainties from $\alpha_s$ and the collider data are modeled by the nuisance parameters
$\delta^{\alpha_s}$, $\delta^{\rm data}$ and constitute independent sources of uncertainty (hence with factorisable priors). 
 The relative uncertainties on $\alpha_s$ and the PDF data can be parametrised as
\be
\alpha_s(1+\delta^{\alpha_s}\Delta^{\alpha_s})\,,\quad
{\rm data} (1+\delta^{\rm data}\Delta^{\rm data})\,.
\ee
The $\alpha_s$ error enters in the cross section in two different ways. 
On one hand, $\alpha_s$ is used in the fit of the data aimed at determining the PDF themselves. 
On the other hand, $\alpha_s$ is also involved in the hard subprocess that is convoluted with the PDF
to obtain the final cross section. 
These two contributions to the cross section uncertainty, named here as $\Delta^{ \alpha_s,{ \rm fit}}$ and $\Delta^{ \alpha_s,{ \rm hard}}$, are not available in the literature. However, we will show that the knowledge 
of these two separate contributions  is not necessary either. Rather, provided that the relative errors $\Delta^{ \alpha_s,{ \rm fit}}$  and $\Delta^{ \rm data}$   are small enough 
 to be linearised, only the sum $\Delta^{ \alpha_s,{ \rm hard}}+\Delta^{ \alpha_s,{ \rm fit}}$ is needed. This sum can typically be inferred from the literature. 

In order to understand the interplay among the $\alpha_s$ and the data uncertainties, it is instructive to write explicitly  how they enter into the cross 
section. One should start with the form
\be
\sigma^{\rm SM}_X[f_{\rm PDF}[\alpha_s,{\rm data}],\alpha_s  ]\,,
\ee
where the first argument corresponds to the PDF input, while the second argument re\-presents the $\alpha_s$-dependence coming from the partonic process. From this general form,  one then introduces the $\delta^{\alpha_s}$ and $\delta^{\rm data}$ nuisance parameters, and expand the expression at first order, \footnote{
The $\partial_{1,2}$ represents derivative with respect to the first and second argument of the function respectively, $\partial_1 f =\partial f(x,y)/ \partial x$, $\partial_2 f =\partial f(x,y)/ \partial y$.} 
\be
\begin{split} 
\label{eq:sigma_propa}
&\sigma^{\rm SM}_X\big[f_{\rm PDF}[\alpha_s(1+\delta^{\alpha_s}\Delta^{\alpha_s}),{\rm data}(1+\delta^{\rm data}\Delta^{\rm data})],\alpha_s(1+\delta^{\alpha_s}\Delta^{\alpha_s})  \big]=\\
& \  \   \  \   \  \   \  \  \sigma^{\rm SM}_X\big[f_{\rm PDF}\,[\alpha_s,{\rm data}],\alpha_s  \big]\bigg(1+
\delta^{\alpha_s}(\partial_1 f_{\rm PDF}\, \partial_1\sigma^{\rm SM}_X\, \Delta^{\alpha_s}) + \delta^{\rm data}(\partial_2 f_{\rm PDF}\, \partial_1\sigma^{\rm SM}_X\, \Delta^{\rm data})
\\& \  \   \  \   \   \  \  + \delta^{\alpha_s}(\partial_1 f_{\rm PDF}\, \partial_2\sigma^{\rm SM}_X\, \Delta^{\alpha_s}) +O(\Delta^2)\bigg)\,.  
\end{split}
\ee
The terms in the last two lines represent the errors propagated to the cross section at first order in $\Delta$, expressed as partial derivatives of $\sigma^{\rm SM}_X$, and correspond precisely to the
relative errors on the cross section, \footnote{Note that the $\Delta$'s in Eq.~\eqref{eq:sigma_propaBIS} can be negative as they are identified from the partial derivatives in Eq.~\eqref{eq:sigma_propa}. In the rest of the paper however, the $\Delta$'s are taken positive by convention. Different signs for the $\Delta$'s would correspond to a negative correlation, that is instead included at the level of the $\delta$'s in the rest of the paper. }
\be
\delta^{\alpha_s} \Delta^{\alpha_s,{\rm fit}}_X+
\delta^{\rm data}_X \Delta^{\rm data}_X+
\delta^{\alpha_s} \Delta^{\alpha_s,{\rm hard}}_X\,.  \label{eq:sigma_propaBIS}
\ee
It appears clearly that only the sum $\Delta_X^{ \alpha_s,{ \rm hard}}+\Delta_X^{ \alpha_s,{ \rm fit}}$ is needed. Fortunately, this is what is provided in the literature. 
This sum $\Delta_X^{\alpha_s}\equiv\Delta_X^{ \alpha_s,{ \rm hard}}+\Delta_X^{ \alpha_s,{ \rm fit}}$  can be read for example from Ref.~\cite{LHCHWG3}. 
 Note also that the nuisance parameter $\delta^{\alpha_s}$ is common to any production mode,
\ie~it does not carry the index $X$. In contrast, the nuisance parameter $\delta^{\rm data}_X$ carries an index $X$ because each production mode potentially involves different initial states. 
These initial states correspond to  different PDF, which are fitted from different data sets.


Finally, one should check the validity of the error propagation at linear order in the cross sections  (\ie~that the $O(\Delta^2)$ in Eq.~\eqref{eq:sigma_propa} is well negligible). 
From Eq.~\eqref{eq:sigma_propa}-\eqref{eq:sigma_propaBIS}, one can see that at linear order, for any fixed value of $\alpha_s$ (\ie~fixed value of $\delta^{\alpha_s}$), the error bar on $\sigma^{\rm SM}_X$ induced by the data uncertainty (obtained  from varying $\delta^{\rm data}$, e.g. in [-1,1]) should have the same size.
A change with $\alpha_s$ of this bar size could thus come only from higher order terms such like $$\delta^{\alpha_s}\delta^{\rm data}\,(\partial_1 \partial_2 f_{\rm PDF}
\  \partial^2_1\sigma^{\rm SM}_X\, \Delta^{\alpha_s}\,\Delta^{\rm data}).$$
On the Fig.~(57)-(58)-(59) of Ref.~\cite{LHCHWG3} for the various Higgs production reactions at the $8$~TeV LHC, 
we see that the change of this bar size (vertical bar there) is small with respect to the shift (\ie~$\Delta_X^{\alpha_s}$) of the bar central values. We conclude that one can restrict the expansion Eq.~\eqref{eq:sigma_propa} to linear order in a  good approximation.

Notice that a customary way to write  these uncertainties is by splitting between the overall PDF error and the hard subprocess error, 
$\delta^{\rm PDF}_X\Delta_X^{\rm PDF}+\delta^{ \alpha_s}\Delta_X^{ \alpha_s,{ \rm hard}}$, with  $\delta^{\rm PDF}_X\Delta_X^{\rm PDF}= \delta^{ \alpha_s}\Delta_X^{ \alpha_s,{ \rm fit}}
+\delta^{\rm data}_X\Delta_X^{\rm data}\,.$
The trouble when using this form is that the $\delta_X^{\rm PDF}$ and $\delta^{ \alpha_s,{ \rm hard}}$ contributions are correlated via $\alpha_s$. 
Combining these uncertainties then requires to know such a correlation coefficient, which is fixed by $\Delta^{ \alpha_s,{ \rm fit}}$, as well as  $\Delta^{ \alpha_s,{ \rm hard}}$.
We emphasize that the use of this intermediate parametrisation brings unnecessary complications, and we recommend thus to avoid it.

Hence according to Eq.~\eqref{eq:sigma_propaBIS}, the parametric uncertainties from $\alpha_s$ 
are cast into a single error $\Delta^{\alpha_s}_X$, and add up with 
the statistical error from the data as
\begin{eqnarray}  
\delta^{\rm data}_{X} \Delta^{\rm data}_{X} + \delta^{\alpha_s} \Delta^{\alpha_s}_{X} \ . 
\label{ChangeVar0}
\end{eqnarray} 
Using this approach, one deals directly with the elementary sources of uncertainty. These two sources of error have no intrinsic relation and are thus independent, meaning that $\delta^{\rm data}$ and $\delta^{\alpha_s}$ have factorisable priors.

 Similarly, the uncertainty from the choice of a specific PDF set, modeled by $\delta^{\rm set}$, 
can be added up linearly to the errors of Eq.~\eqref{ChangeVar0} in a good approximation. 
The linear approximation can be justified from  Fig.~(57) in Ref.~\cite{LHCHWG3}. 
There one can see that the size of the data error bars as well as the shifts induced by $\alpha_s$ depend only weakly on the PDF set choice.
The $\delta^{\rm set}$ error is also independent from the $\delta^{\rm data}_{X}$,
$\delta^\alpha_s$ errors and in turn possesses its own prior distribution. 
All those errors induce three terms in the sum of theoretical errors entering Eq.~(\ref{eq:subPROD}). These terms can be cast into a global PDF uncertainty,   
\begin{eqnarray}
\delta^{\rm PDF+\alpha_s}_{X} \Delta^{\rm PDF+\alpha_s}_{X} \ \hat =\
\delta^{\rm set} \Delta^{\rm set}_{X} + \delta^{\rm data}_{X} \Delta^{\rm data}_{X} + \delta^{\alpha_s} \Delta^{\alpha_s}_{X} \ .
\label{ChangeVar1}
\end{eqnarray} 
We recall that $X = \{ {\rm ggF},{\rm VBF},{\rm VH},{\rm ttH} \}$ and that the 
$\Delta$'s are relative errors, which are chosen by convention to correspond to one standard deviation. Those are related to the $1\sigma$ absolute errors on the SM Higgs cross section through \textit{e.g.}
 $$\Delta^{\rm data}_{X} \ \hat = \ \frac{
\Delta\sigma^{\rm data}_{X} }{ \sigma^{\rm SM}_{X}} \ .$$
\\  
$\bullet$ {\sc Combining the three uncertainties:} \\ 
Here we combine the three sources of theoretical uncertainty described in Eq.~(\ref{ChangeVar1}).
We will add up more and more errors progressively in the following subsections.
These three independent sources of error are associated with three priors $\pi^{\alpha_s}$, $\pi^{{\rm data}}_X$, $\pi^{\rm set}$. These nuisance parameters appear in Eq.~(\ref{GenLprior}), where they are integrated over.
We now proceed to combine these errors following the analysis of Section~\ref{se:Combinations}, starting from Eq.~(\ref{eq:comb_marg}). In practice, for the discussion,  it will be convenient to combine only two errors at a time.
One then finds a likelihood of the type~(\ref{GenLprior}) depending only on the nuisance parameter $\delta^{\rm PDF+\alpha_s}_{X}$. 
The distribution of this nuisance parameter comes with a   $1\sigma$ width $\Delta^{\rm PDF+\alpha_s}_{X}$ given by 
\begin{eqnarray}
(\Delta^{\rm PDF+\alpha_s}_{X})^2=(\Delta^{\rm set}_{X})^2+(\Delta^{\rm data}_{X})^2+(\Delta^{\alpha_s}_{X})^2 \ .
\label{PDFComb}
\end{eqnarray}
The nuisance parameter $\delta^{\rm PDF+\alpha_s}_{X}$ obeys a new  prior $\pi_{X}^{\rm PDF+\alpha_s}$, obtained via two successive convolutions of the initial priors (as in Eq.~(\ref{eq:conv_bayes})-(\ref{eq:conv_bayesII})-(\ref{eq:conv_bayesIII})),  
\begin{eqnarray}
\bar \pi_{X}^{\rm PDF+\alpha_s} \ = \ \bar \pi_{X}^{\rm set} \star   \bar \pi_{X}^{\rm data} \star \bar \pi_{X}^{\alpha_s}   \, , 
\label{PriorConv}
\end{eqnarray}
where $\bar \pi_{X}^{\rm PDF+\alpha_s} (x) = \pi_{X}^{\rm PDF+\alpha_s} (x/\Delta^{\rm PDF+\alpha_s}_{X})$ and the variable $x$ corresponds to the relative error 
$\delta^{\rm PDF+\alpha_s}_{X}\Delta^{\rm PDF+\alpha_s}_{X}$. 
For the initial priors one has for example $\bar \pi^{\alpha_s}_X(x)= \pi^{\alpha_s}(x/\Delta^{\alpha_s}_X) $.
The Eq.~(\ref{PDFComb}) and then (\ref{PriorConv}) are justified in details in the rest of this subsection. 
\\ \\
$\bullet$ {\sc Details on the data and $\alpha_s$ error combinations:} 
\\ We emphasize that the Bayesian combination of the $1\sigma$ widths, as here in Eq.~(\ref{PDFComb}), is \textit{independent} of the shapes of the prior distributions. 
This combination only depends on the possible correlations among individual errors [{\it c.f.} Section~\ref{se:combTU}]. 
In the present case, there is no correlation between the $\delta^{\rm data}_{X}$ and $\delta^{\alpha_s}_{X}$ parameters, as explained right below Eq.~(\ref{ChangeVar0}).
This leads to the sum in quadrature of the $1\sigma$ errors $(\Delta^{\rm data}_{X})^2+(\Delta^{\alpha_s}_{X})^2$ 
in Eq.~(\ref{PDFComb}).
\\
Let us comment about those uncertainties. First, the error associated to $\pi_{X}^{\rm data}$ originates mainly from measurements:  
it is mainly induced by the limited accuracy of data points used to perform the fit for recons\-tructing PDF. Hence this error is mostly of statistical nature. 
There exists of course systematic errors as well,  but it has been checked by several groups that the final $\pi_{X}^{\rm data}$ distribution can be reasonably taken as Gaussian~\cite{LHCHWG1}. 
\\
Second, the uncertainty on $\alpha_s$ 
originates mainly from lattice calculation errors (mainly theoretical) and especially from perturbative truncation errors~\cite{HPQCD}~\footnote{The only source of 
experimental error is, $m_{\eta_c}$,$m_{\eta_b}$, and is minor -- as can be read from the Table~IV of Ref.~\cite{HPQCD}.}. Indeed the $\alpha_s$ determination 
from lattice methods (most accurate one in Ref.~\cite{HPQCD}) represents today the most precise determination and hence essentially dictates the final world
average error~\cite{PDG}. The FLAG Working Group on lattice calculations has estimated a more conservative 
uncertainty on $\alpha_s$, which is increased by a new QCD perturbative error estimation~\cite{FLAG}, thus still leading to a dominant theoretical uncertainty.
\\
At this level, a comment is needed on the link between the $1\sigma$ errors and the uncertainty magnitudes provided in literature. 
To remain conservative we use $\Delta^{\alpha_s}_{X} = \Delta^{\alpha_s}_{X}\vert_0$ for the $1\sigma$ error, where $\Delta^{\alpha_s}_{X}\vert_0$ 
is the  error provided by Ref.~\cite{LHCHWGweb,LHCHWG3}. There is indeed a somewhat arbitrary choice for the 
relation between $\Delta^{\alpha_s}_{X}$ and $\Delta^{\alpha_s}_{X}\vert_0$, due to the theoretical (QCD) nature of the uncertainty. The origin of this arbitrariness is the fact that the QCD errors are just estimated by varying the renormalisation and factorisation 
scales on arbitrary intervals. We present a similar discussion in the beginning of next Section (\ref{EFTsca}) for $\Delta^{\rm scale}_{X}$. 
Concerning the $1\sigma$ error from data, one can adopt $\Delta^{\rm data}_{X}=\Delta^{\rm data}_{X}\vert_0$ ($\Delta^{\rm data}_{X}\vert_0$ 
being read from Ref.~\cite{LHCHWGweb,LHCHWG3}). Indeed, the probability distribution for the uncertainty induced by the experimental data can be safely 
described by a Gaussian, as described above, so that the errors provided by Ref.~\cite{LHCHWGweb,LHCHWG3} can reasonably be interpreted as $1\sigma$ errors.

Let us now discuss the convolution between $\bar \pi_{X}^{\rm data}$ and $\bar \pi_X^{\alpha_s}$ that appears in Eq.~(\ref{PriorConv}). 
For that purpose, we first need to discuss the form of the $\pi^{\alpha_s}$ distribution.
The shape of $\pi^{\alpha_s}$ can be taken as flat since the uncertainty on $\alpha_s$ 
originates mainly from theoretical uncertainty, as mentioned above. 
However, the choice of the prior for a theoretical uncertainty is often controversial, so that 
we will also consider the case of a non-flat $\pi^{\alpha_s}$ 
distribution.~\footnote{To be consistent throughout the paper, concerning the initial priors,  we will assume a flat shape for the distributions whose shape is unknown (uncertainties from QCD, 
parametrisation\dots).
\label{FootFlat}}
\\
Finally, the convolution of the Gaussian prior, $\bar\pi_{X}^{\rm data}$, with a flat prior, $\bar\pi_{X}^{\alpha_s}$, 
gives rise to a Gaussian distribution, $\bar \pi_{X}^{\rm data} \star \bar \pi_{X}^{\alpha_s}$, in a 
good approximation for the various Higgs production modes. The justification is that the $\bar\pi_{X}^{\alpha_s}$  width, $\Delta^{\alpha_s}_{X}$, is systematically smaller or 
of the same order as $\Delta^{\rm data}_{X}$,~\footnote{For the ggF example, our conservative treatment of the errors provided in Fig.~(59) of Ref.~\cite{LHCHWG3} gives 
an half absolute width, $W/2 \ \hat = \ \sqrt{3}\Delta\sigma^{\alpha_s}_{X} = \sqrt{3}\Delta\sigma^{\alpha_s}_X\vert_0 \simeq 0.5$~pb, which is indeed comparable to, 
$\Delta\sigma^{\rm data}_{X} = \Delta\sigma^{\rm data}_{X} \vert_0 \simeq 0.5$~pb. In the alternative case (see the analogous discussion at the start of Section~\ref{EFTsca}), 
one has instead, $W/2 \ \hat = \ \sqrt{3}\Delta\sigma^{\alpha_s}_{X} = \Delta\sigma^{\alpha_s}_{X}\vert_0 \simeq 0.3$~pb, which is clearly smaller than, 
$\Delta\sigma^{\rm data}_{X} \simeq 0.5$~pb, so that the Gaussian approximation for the final convolution would be even better because this case would tend to 
a situation where the non-Gaussian error becomes negligible.} in which case the convolution leads to an almost pure Gaussian prior. This will be demonstrated explicitly in Fig.~(\ref{fig:Conv3})  for other priors. 
\\ \\ 
$\bullet$ {\sc Details on the combination with the PDF set error:} \\ The various PDF estimations provided by the different fitting groups reflect several sources of 
error~\cite{hep1101.0538,PDF4LHCweb,hep1301.6754}. Indeed, these groups make different choices/hypotheses about the numbers of free parameters 
used to model the PDF~\footnote{The infinite-dimensional problem of representing a space of functions is reduced to a finite-dimensional form, in order to 
be manageable, by introducing a parametrisation of the PDF.}, the statistical methods adopted to fit the data~\footnote{There exist mainly two classes of 
methodology currently used to determine a confidence interval re\-presented in the space of functions: some variations of the Hessian approach 
(multi-Gaussian probability distributions) and the Monte Carlo approach. Both types of methods have their own limitations.}, the number of independently 
parameterized PDF (in particular regarding (anti-) strangeness), the collider results exploited,   the matching methods applied to include heavy-quark 
mass effects in the flavour number scheme and the variable- or fixed-flavour number scheme. All these sources of uncertainty are synthesized in the $1\sigma$ error on the Higgs production 
rates noted $\Delta^{\rm set}_{X}$. To remain conservative, we assume $\Delta^{\rm set}_{X}=\Delta^{\rm set}_{X}\vert_0$, where 
$\Delta^{\rm set}_{X}\vert_0$ is the error read from Fig.~(57)-(59) of Ref.~\cite{LHCHWG3}. $\Delta^{\rm set}_{X}\vert_0$ can be estimated by taking half 
the interval obtained by using the various PDF sets which lead to a finite number of predictions for the Higgs rate central values. Of course, this determination of $\Delta^{\rm set}_{X}\vert_0$ 
is probably underestimated as {\it (i)} the hypotheses made by the groups provide illustrative examples which do not necessarily indicate the extremal 
values of the PDF, and, {\it (ii)} the effects of the various sources of error listed above can potentially compensate each other.
We comment on this point
in the following paragraph.   
\\ 
In Eq.~(\ref{PDFComb}), the sum in quadrature between the $\Delta^{\rm set}_{X}$ error and the data and $\alpha_s$ errors is justified because these are independent uncertainties. Nevertheless, in practice, for our numerical applications,  we use the so-called envelope method~\footnote{This ``envelope method'' 
corresponds precisely to the uncertainty  combinations in the bias approach, see Section~\ref{se:comb_bias}. What we call envelope method in the present paper is rather described in Section~\ref{sec:BiasPrin}.}
 to determine $\Delta^{\rm PDF+\alpha_s}_{X}$ as done in Ref.~\cite{LHCHWG3,0905.3531}~\footnote{In the envelope method used in this reference, the whole uncertainty interval is found by searching at the minimum and maxi\-mum rates (considering the various PDF sets, $\alpha_s$ values and including 
the possibility to move  along the data-error bars). Then dividing by two this interval gives an estimation of the combined error  as well as a central value for the rate. }  
and calculated by the LHCHWG~\cite{LHCHWGweb}. 
Note that the envelope method overestimates the combined errors,  compensating somehow for the underestimation of the PDF set error. For the ggF me\-chanism, the $\Delta\sigma^{\rm PDF+\alpha_s}_{\rm ggF}$ error derived in this way has to be reduced by 
$\sim 40 \%$ to recover the quadrature summation of Eq.~(\ref{PDFComb}), and the decrease is smaller for the other Higgs production reactions. Hence, we conclude that the use of
the envelope method to determine the global PDF uncertainties gives rise  to a substantial overestimation of these errors.

We finally discuss the shape of the prior of the final combination $\pi^{{\rm PDF}+\alpha_s}_X$. Most of the sources of error taken into account in $\Delta^{\rm set}_{X}$ are of theoretical nature and all the errors  have unknown distributions. The shape of $\pi_{X}^{\rm set}$ is therefore assumed to be flat.
 The convolution of $\bar \pi_{X}^{\rm set}$ (see Eq.~(\ref{PriorConv}))
with the nearly Gaussian distribution $\bar \pi_{X}^{\rm data} \star \bar \pi_{X}^{\alpha_s}$ leads in a good approximation to a final Gaussian prior, $\pi_{X}^{\rm PDF+\alpha_s}$~\footnote{
Given that there are several sources of errors contained in the  PDF set uncertainty, one may expect  the $\pi_{X}^{\rm set}$ prior to be somehow peaked. This feature improves even more the Gaussian approximation of  $\pi_{X}^{\rm PDF+\alpha_s}$.
}. Once more, this is guaranteed by the fact that for any Higgs 
production mode at the LHC, $\Delta^{\rm set}_{X}$ is smaller or comparable to the combination of $\Delta^{\rm data}_{X}$ and $\Delta^{\alpha_s}_{X}$ (see for instance 
Ref.~\cite{LHCHWG3}).

\subsection{Scale and EFT errors: the amplitude uncertainties}  \label{EFTsca}

\vspace{0.3cm}
\noindent 
$\bullet$ {\sc Scale error:} \\ 
There exists another major type of error, this time at the parton level, on the QCD prediction for Higgs production cross sections. 
It originates from the lack of know\-ledge on the higher order contributions to the amplitude in the perturbative expansion, and can be recast 
into the dependence on the QCD renormalisation and factorisation scales. We note $\delta^{\rm scale}_X$ the nuisance parameter representing this ``scale uncertainty''.
\\
There are no strong arguments to choose the shape for $\pi_{X}^{\rm scale}$. As for many other theoretical uncertainties, the choice of the prior is typically a subject of controversy. Here we choose $\pi_{X}^{\rm scale}$ to be flat. 
 Concerning the magnitude of the scale uncertainty $\Delta^{\rm scale}_X$, it is also not 
clear to which width exactly corresponds the provided value, noted $\Delta^{0}_{X}$ here, that is found in Ref.~\cite{LHCHWGweb,LHCHWG1,Duhr}. It is reasonable to expect 
$ \Delta^{\rm scale}_{X}$   to be of order $\Delta^{0}_{X} $. 
 To be more precise, we could make the two different assumptions,  
$\Delta^{0}_{X} \hat = \Delta^{\rm scale}_{X}$ or $\Delta^{0}_{X} \hat = {\cal W}/2$ where ${\cal W}$  is defined as the support of the 
distribution,~\footnote{Recall that the support of a distribution is the domain where this distribution is not zero-valued.}
 with {\it e.g.} in the case of a flat distribution on an interval with size ${\cal W}$:
$2\Delta^{\rm scale}_{X} \hat= {\cal W}/\sqrt{3}=2\Delta^{0}_{X}/\sqrt{3}$. In order to be conservative in the choice of $\Delta^{\rm scale}_{X}$, we choose the former hypothesis throughout this paper: $\Delta^{\rm scale}_{X} = \Delta^{0}_{X}$.
\\
It is remarkable that recently~\cite{Duhr}, the  calculation for the ggF mechanism has been pushed up to the complete N$^3$LO order in perturbative QCD. This has allowed a reduction of the symmetrized~\footnote{Symmetrized over the positive and negative errors as, 
$\Delta=[(\Delta_+^2+\Delta_-^2)/2]^{1/2}$.} scale error from $\Delta^{0}_{\rm ggF} \simeq 7.51\%$ (with the renormalisation/factorisation scale $\mu_0 = m_H/2$  
to absorb some of the soft-gluon resummation corrections~\cite{Anastasiou})~\cite{LHCHWGweb,LHCHWG1}, down to $\Delta^{0}_{\rm ggF} \simeq 4.16\%$ (with 
$\mu_0 = m_H$~\footnote{Choosing instead, $\mu_0 = m_H/2$, could be motivated by a faster convergence of the perturbative series~\cite{Duhr}. However, since it 
would lead to a significantly smaller uncertainty, $\Delta^{0}_{\rm ggF} \simeq 2.13\%$, we stick to the  central choice, $\mu_0 = m_H$, in order to remain 
conservative.})~\cite{Duhr}. The error was obtained in both cases by spanning the interval $[\mu_0/2,2\mu_0]$, for the renormalisation/factorisation scale 
$\mu = \mu_{\rm R} = \mu_{\rm F}$, at an energy $\sqrt s=8$~TeV and for $m_H \simeq 125.2$~GeV.
\begin{figure}[h!]
\center
\includegraphics[width=7.5cm]{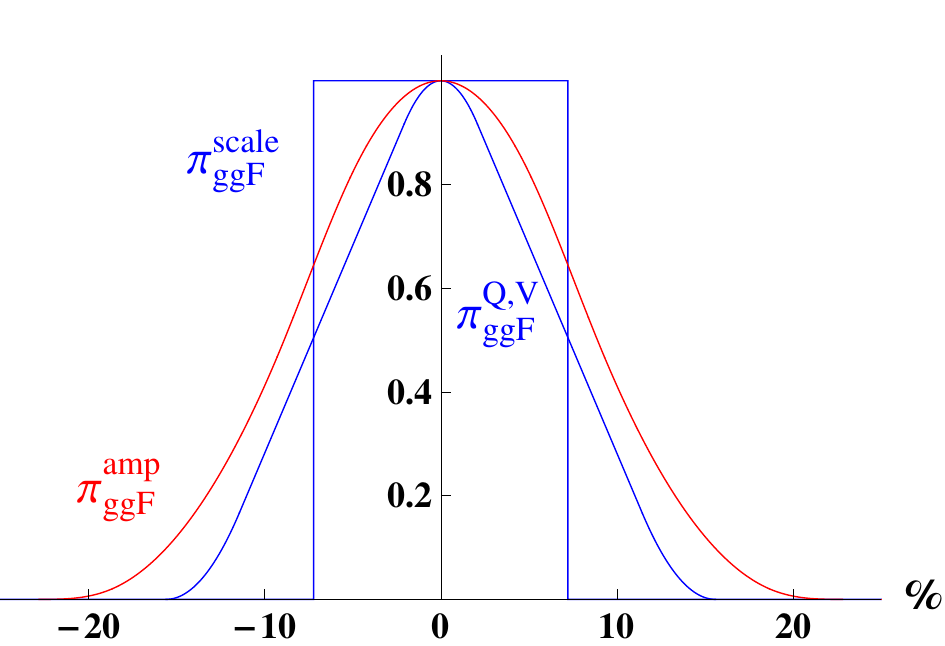}
\caption{
Probability density distribution, $\pi_{\rm ggF}^{\rm amp}(x/ \Delta_{\rm ggF}^{\rm amp})$ (in red), involving the relative error $x$ (in $\%$) of the ggF cross section,  
as derived through the convolution of the $\pi_{\rm ggF}^{\rm Q,V}$ and $\pi_{\rm ggF}^{\rm scale}$ priors (both in blue). 
The quantity $\Delta_{\rm ggF}^{\rm amp}$ represents the relative $1\sigma$ error on the Higgs production rate (see text). For  better comparison, the normalisation 
is chosen such that all the functions possess the same maximum, equal to unity at the origin.
}\label{fig:Conv1}
\end{figure}
\\ \\ \\ 
$\bullet$ {\sc EFT error:} \\
In the specific case of the ggF mechanism, another source of error arises in the amplitude of the Higgs production~\cite{EFTorigin}, that we describe now. The 
evaluation of this amplitude beyond the NLO level is possible within the Effective Field Theory (EFT) approach, where the particles running in the triangle loop are assumed 
to be much heavier than the produced Higgs boson to integrate out the heavy particles.
\\
 For the top quark exchange, the infinite mass assumption, $m_t \gg m_H$, induces a negligible error 
on the ggF amplitude~\cite{QCDEWnnlo,topEFT}. In contrast, the EFT approach is clearly not valid for the other significant ggF contribution: the bottom quark 
exchange~\cite{Duhr}. This inappropriate use of the EFT limit 
introduces some non-negligible error mainly through the interference between the bottom and dominant top quark loops (this error being smaller at the 
Tevatron than at the LHC)~\cite{bottomEFTevatron}. 
\\
A similar uncertainty originates from the mixed QCD-EW corrections to the ggF process~\cite{QCDEWnnlo}. Those have been calculated at NNLO via the EFT approach based 
on the simplifying but unrealistic assumption, $M_{W,Z} \gg m_H$. For all the EFT errors, some approximative estimations can be computed at NNLO (using $K$-factors obtained 
at NLO and NNLO for the top loop)~\cite{topEFT,bottomEFT}. 
\\
A related uncertainty comes from the freedom in the choice of a renormalisation scheme for the bottom quark mass, involved in the ggF amplitude (on-shell scheme, 
$\overline{\rm MS}$ scheme\dots). The error from the renormalisation scheme dependence can be approximately estimated at NLO~\cite{topEFT}.
\\ 
These three sources of theoretical uncertainty, namely the two kinds of EFT assumptions (on the heavy quark masses, $m_Q$ ($Q=b,t$), and vector boson masses, 
$M_V$ ($V=W,Z$)) and the $m_b$ scheme dependence, are independent and their respective priors are unknown. We assume these priors to be flat. 
To be conservative, we take the three $1\sigma$ errors to be equal to the numbers estimated in Ref.~\cite{topEFT,bottomEFT}, for the $8$~TeV LHC. 
Summing those in quadrature gives rise to the relative rate error, $\Delta_{\rm ggF}^{\rm Q,V} \hat = \Delta\sigma^{\rm Q,V}_{\rm ggF} / \sigma^{\rm SM}_{\rm ggF} \simeq 5.6\%$. 
The convolution of the three flat priors (accordingly to Eq.~(\ref{eq:conv_bayesIII})) leads to the blue distribution, $\pi_{\rm ggF}^{\rm Q,V}$, shown in Fig.~(\ref{fig:Conv1}),
which already resembles a Gaussian shape as predicted by the central limit theorem.
\\ \\ 
$\bullet$ {\sc Combining the $\Delta^{\rm scale}_{\rm ggF}$ and $\Delta_{\rm ggF}^{\rm Q,V}$ errors:} \\ The theoretical scale and 
EFT uncertainties on the ggF mechanism are of different nature and are thus independent. The combined ggF $1\sigma$ error is in turn given by 
\begin{eqnarray}
(\Delta_{\rm ggF}^{\rm amp})^2  \ = \  (\Delta^{\rm scale}_{\rm ggF})^2 + (\Delta_{\rm ggF}^{\rm Q,V})^2  \ . 
\label{eq:ggFamp}
\end{eqnarray}
This error constitutes the characteristic width of the $\pi_{\rm ggF}^{\rm amp}$ 
distribution obtained by convoluting the $\bar \pi_{\rm ggF}^{\rm scale}$ and $\bar \pi_{\rm ggF}^{\rm Q,V}$ priors, as performed in 
Fig.~(\ref{fig:Conv1}) (see the final red curve). Remar\-kably, this distribution, 
\begin{eqnarray}
\bar \pi_{\rm ggF}^{\rm amp} \ \equiv \ \bar \pi_{\rm ggF}^{\rm scale} \star \bar \pi_{\rm ggF}^{\rm Q,V} \, , 
\label{PriorConvAmp}
\end{eqnarray}
derived from four purely flat priors, is Gaussian in a good approximation. This can be also seen in Fig.~(\ref{fig:Conv2}) where $\pi_{\rm ggF}^{\rm amp}$ is plotted together with a pure Gaussian 
distribution (blue curves). 
Recall that $\bar \pi_{\rm ggF}^{\rm amp} (x) = \pi_{\rm ggF}^{\rm amp} (x/\Delta_{\rm ggF}^{\rm amp})$ and the variable $x$ corresponds to $\delta^{\rm amp}_{{\rm ggF}}\Delta_{\rm ggF}^{\rm amp}$.

\subsection{Combination of the PDF and amplitude errors}  \label{PDFsca}

For the various Higgs production modes -- except the ggF process that will be discussed separately below, one has to combine the PDF and scale errors to determine the final uncertainty on the whole cross section. 
The scale error adds up to the PDF error of Eq.~(\ref{ChangeVar1}), according to Eq.~(\ref{eq:subPROD}), defining the total uncertainty on the cross section,  
\begin{eqnarray}
\delta^\sigma_{X} \Delta_{X}^{\sigma}  
= \delta^{\rm PDF+\alpha_s}_{X} \Delta^{\rm PDF+\alpha_s}_{X} +  \delta^{\rm scale}_{X} \Delta^{\rm scale}_{X} \,.
\label{ChVarSig}
\end{eqnarray}
These errors being independent, the $1\sigma$ widths add-up in quadrature,
\begin{eqnarray}
(\Delta_{X}^{\sigma})^2 = (\Delta^{\rm PDF+\alpha_s}_{X})^2 + (\Delta^{\rm scale}_{X})^2 \ ,
\label{VarianComb}
\end{eqnarray}
as dictated by Section~\ref{se:combTU}, \ie~irrespective of the $\pi_{X}^{\rm PDF+\alpha_s}$ 
and $\pi_{X}^{\rm scale}$ shapes. Recall that $\Delta_{X}^{\sigma}$ is the $1\sigma$ width of the resulting $\bar \pi_{X}^\sigma$ distribution. 
The prior $\pi_{X}^\sigma$ of this total uncertainty is then given by (see Eq.~(\ref{eq:conv_bayesIII}))  
\begin{eqnarray}
\bar \pi_{X}^\sigma \ \equiv \ \bar \pi_{X}^{\rm PDF+\alpha_s}  \star \bar  \pi_{X}^{\rm scale}  \, ,
\label{PriorConvScale}
\end{eqnarray}
with $\bar \pi_{X}^\sigma (x) = \pi_{X}^\sigma (x/\Delta^\sigma_{X})$ and $x$  corresponding to $\delta^\sigma_{X}\Delta^\sigma_{X}$.

\begin{figure}[h!]
\center
\includegraphics[width=7.5cm]{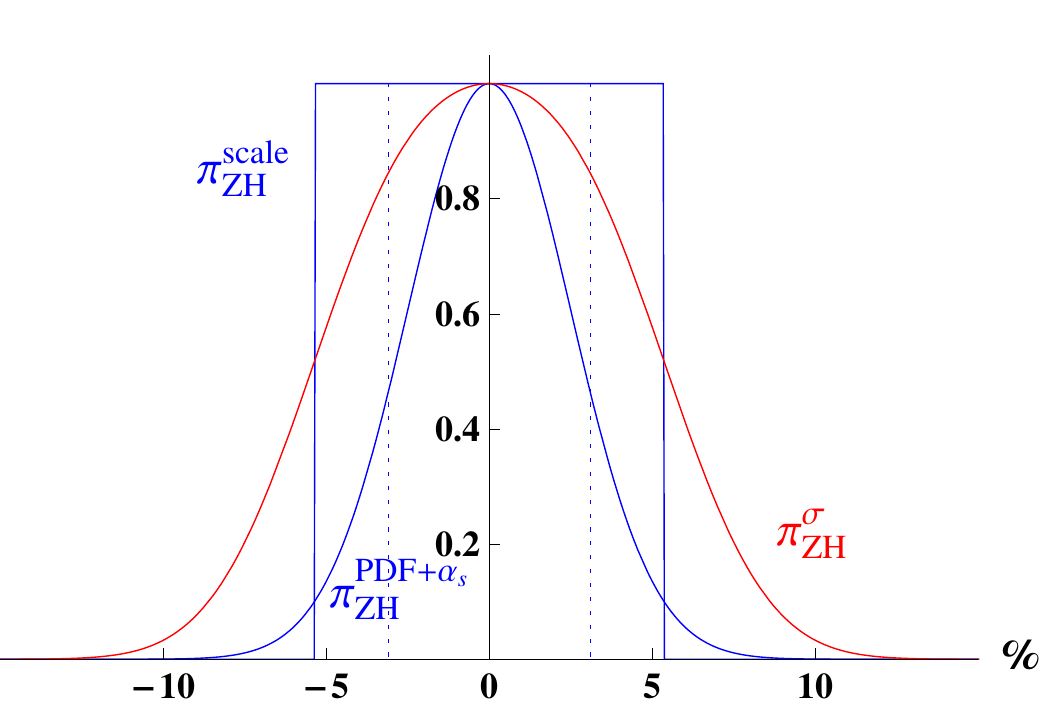}
\caption{
Probability density distribution, $\pi_{Z{\rm H}}^\sigma(x/\Delta_{Z{\rm H}}^{\sigma})$ (in red), involving the relative error $x$ (in $\%$) of the $Z$H production cross section, 
as derived through the convolution of a Gaussian $\pi_{Z{\rm H}}^{\rm PDF+\alpha_s}$ and a flat $\pi_{Z{\rm H}}^{\rm scale}$ priors (both in blue). 
The quantity, $\Delta_{Z{\rm H}}^{\sigma}$, represents the relative $1\sigma$ error on the Higgs production rate. The normalisation 
is chosen such that all the functions possess the same maximum, equal to unity at the origin. The $1\sigma$ band for the $\pi_{Z{\rm H}}^{\rm scale}$ distribution is 
indicated by the vertical dotted lines.
}\label{fig:Conv3}
\end{figure}

\begin{figure}[h!]
\center
\includegraphics[width=7.5cm]{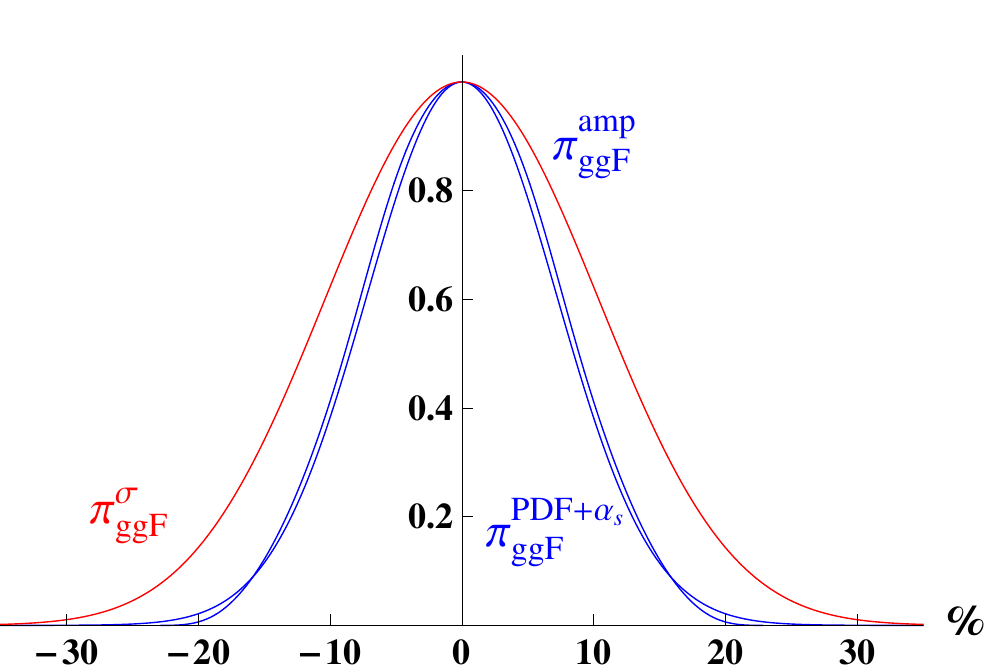}
\caption{
Probability density distribution, $\pi_{\rm ggF}^\sigma(x/\Delta_{\rm ggF}^{\sigma})$ (in red), involving the relative error $x$ (in $\%$) of the ggF
cross section, as derived through the convolution of a Gaussian $\pi_{\rm ggF}^{\rm PDF+\alpha_s}$ prior and the $\pi_{\rm ggF}^{\rm amp}$ 
distribution obtained in Fig.~(\ref{fig:Conv1}) (both in blue). The quantity, $\Delta_{\rm ggF}^{\sigma}$, represents the relative $1\sigma$ error for the ggF rate. 
}\label{fig:Conv2}
\end{figure}

Let us discuss  the form of the $\pi_{X}^\sigma$ function, as generated through Eq.~(\ref{PriorConvScale}). The shape of $\pi_{X}^{\rm scale}$ being unknown, we assume a flat $\pi_{X}^{\rm scale}$ distribution. Remind that this  error is simply obtained by
varying the QCD scale, so that no favoured value is predicted for the cross section. It is therefore a sensible choice to assign equal probabilities to all the 
values of $\delta^{\rm scale}_{X}$ (or equivalently of the Higgs cross section) inside a certain range. 
On the other hand, we have seen in Section~\ref{PDFsole} that $\pi^{\rm PDF+\alpha_s}_{X}$  is approximatively Gaussian.
Given the relative values 
of $\Delta^{\rm PDF+\alpha_s}_{X}$ and $\Delta^{\rm scale}_{X}$ for each process $X$ --~which are syste\-matically such that either 
$\Delta^{\rm PDF+\alpha_s}_{X} > \Delta^{\rm scale}_{X}$ or $\Delta^{\rm PDF+\alpha_s}_{X} \approx \Delta^{\rm scale}_{X}$~\footnote{whatever is the
prescription: $\Delta^{\rm scale}_{X} = \Delta^{0}_{X}$ or $\Delta^{\rm scale}_{X} = \Delta^{0}_{X}/\sqrt{3}$.}~-- a Gaussian 
$\pi_{X}^{\rm PDF+\alpha_s}$ and a flat $\pi_{X}^{\rm scale}$ lead in a good approximation to a final Gaussian $\pi_{X}^\sigma$.
This combination is shown in Fig.~(\ref{fig:Conv3}) for $Z$H production, for which 
$\Delta^{\rm PDF+\alpha_s}_{Z{\rm H}} \simeq 2.5\%$ and $\Delta^{\rm scale}_{Z{\rm H}} = \Delta_{Z{\rm H}}^0 \simeq 3.1\%$ (at $\sqrt s=8$~TeV 
with $m_H \simeq 125.2$~GeV)~\cite{LHCHWGweb}.
\\ \\ 
$\bullet$ {\sc The ${\rm ggF}$ reaction:} \\ 
In the case of Higgs production via the ggF mechanism, the PDF error has to be combined with the whole amplitude error studied previously 
in Section~\ref{EFTsca}. The resulting total  error on the cross section is
\begin{eqnarray}
\delta^\sigma_{{\rm ggF}} \Delta_{\rm ggF}^{\sigma}  
= \delta^{\rm PDF+\alpha_s}_{{\rm ggF}} \Delta^{\rm PDF+\alpha_s}_{\rm ggF} +  \delta^{\rm amp}_{{\rm ggF}} \Delta^{\rm amp}_{\rm ggF} .
\label{ChVarSiggF}
\end{eqnarray}
These two errors being independent, their widths add-up in quadrature, 
\begin{eqnarray}
(\Delta_{\rm ggF}^{\sigma})^2=(\Delta^{\rm PDF+\alpha_s}_{\rm ggF})^2 + (\Delta^{\rm amp}_{\rm ggF})^2 \ , 
\label{VarianCoggF}
\end{eqnarray}
and their priors are convoluted following 
\begin{eqnarray}
\bar \pi_{\rm ggF}^\sigma \ \equiv \ \bar \pi_{\rm ggF}^{\rm PDF+\alpha_s}  \star \bar \pi_{\rm ggF}^{\rm amp}  \, .
\label{PriorConvScggF}
\end{eqnarray}
This convolution~(\ref{PriorConvScggF}) is performed in Fig.~(\ref{fig:Conv2}), using the $\pi_{\rm ggF}^{\rm amp}$ distribution obtained 
in Fig.~(\ref{fig:Conv1}) 
and the value $\Delta^{\rm PDF+\alpha_s}_{\rm ggF} \simeq 7.20\%$ 
(at $\sqrt s=8$~TeV with $m_H \simeq 125.2$~GeV)~\cite{LHCHWGweb}. 
Both priors $ \pi_{\rm ggF}^{\rm amp}$, $\bar \pi_{\rm ggF}^{\rm PDF+\alpha_s}$  being nearly Gaussian, the final distribution is  almost Gaussian. \footnote{Recall the convolution of two Gaussian distributions gives rise to a Gaussian distribution.}

\subsection{The production contamination}  \label{CTerr}

There are several production mechanisms for the Higgs boson (recall that $X=\{$ggF, VBF, WH, ZH, ttH$\}$). The cross section for each of these production modes is associated with a theoretical uncertainty, that has been obtained through subsections \ref{PDFsole} to \ref{PDFsca}. In fact, one may note that the
uncertainties of these various cross sections are potentially correlated, as they partly arise from common sources like the $\alpha_s$ parametric error.  Therefore the $\delta^\sigma_X$ follow a common distribution $\pi^\sigma$, which does not necessarily factorise into $\pi^\sigma_{\rm ggF}\pi^\sigma_{\rm VBF}\times~\ldots$
The aspect of correlations among the cross section errors will be further discussed in Section~\ref{se:Corr}. Here we shall proceed using the most general prior $\pi^\sigma$, and we denote the resulting correlation matrix  as $\rho_{XX'}^\sigma$. \footnote{In
 Section~\ref{se:Corr}, the assumptions adopted for  $\rho_{XX'}^\sigma$  will allow us to express  $\pi^\sigma$  in terms of the $\pi^\sigma_X$.  }

The contribution from the cross sections errors in a given detection channel  can be read from 
Eq.~\eqref{eq:combination_NSM}. Let us first adopt a more compact notation,
\begin{eqnarray}  
\frac{\sum_X  \epsilon_{X}^{i} \sigma_X^\textrm{SM}\,\delta_X^\sigma\Delta_X^\sigma}{\sum_{X'}\epsilon^{i}_{\rm X'} \sigma^{\rm SM}_{\rm X'}}\ \hat= \ \sum_X  \delta^\sigma_{X} \Delta_{X,i}\,, 
\label{DmuCt}
\end{eqnarray}
where  the $\delta^\sigma_X\Delta^\sigma_X$ are defined in Eqs.~\eqref{ChVarSig},  \eqref{ChVarSiggF}.
The Higgs detection channels have been designed to select predominantly a certain mode of production. That is, for a given channel $i$, the experimental cuts are profiled so that  typically the  efficiency $\epsilon^i_X$  for one of the production modes $X$ (see Eq.~\eqref{eq:REALmu}) is much larger than for the others, implying a hierarchy among the $\Delta_{X,i}$.
We can therefore use the leading moment approximation, developed in Section~\ref{se:Combinations} and Appendix~\ref{app:approx}, to proceed to the combination of the errors. 
Applying the leading moment approximation amounts to treat the contaminations as a small perturbation of the uncertainty from the leading production mode.
The  cross section uncertainties propagate  in a given detection channel as ($P$ stands for production)
\begin{eqnarray}
\delta^P_{X_i}\Delta^{P}_i  \ = \  \delta^\sigma_{{\rm ggF}} \Delta_{{\rm ggF},i} + \delta^\sigma_{{\rm VBF}} \Delta_{{\rm VBF},i} 
+ \delta^\sigma_{{\rm ZH}} \Delta_{{\rm ZH},i} 
+ \delta^\sigma_{{\rm WH}} \Delta_{{\rm WH},i} 
+ \delta^\sigma_{{\rm ttH}} \Delta_{{\rm ttH},i} \,.
\label{ContRedef}
\end{eqnarray} 
 Here the label of the combined nuisance parameter $\delta^P_{X_i}$ is chosen to be the label of the dominant production mode  in the $i$ channel.
 Note that $X_i$ should be understood as $X(i)$.  
 This naming refers to the fact that the shape of the combined nuisance parameter prior corresponds approximatively to the shape for the dominant uncertainty, see Eq.~\eqref{eq:piCpiA}. 
  For example, if the production mode ggF dominates in the channel $i$, one has \be\delta_{X_i}^P=\delta_{\rm ggF}^P\,.\ee
The various nuisance parameters $\delta_{X}^P$ are potentially correlated. They should thus follow a joint prior distribution, $\pi^P$, generating a correlation matrix $\rho_{XX'}^P$. 

Assuming generic correlations $\rho_{XX'}^\sigma$ among the various cross section errors, the ma\-gnitude of the combined production uncertainty in a channel $i$ is given exactly by 
\begin{eqnarray}
(\Delta_{i}^{P})^2=\sum_{XX'} \rho_{XX'}^\sigma\Delta_{X,i}\Delta_{X',i} \ .
\label{VarianCombGen}
\end{eqnarray}
The leading moment approximation then dictates (see Eqs.~\eqref{eq:LMA_comb_12}--\eqref{eq:LMA_comb_3}) that 
\be
\pi^P\approx \pi^\sigma\,. \label{eq:piPpisig}
\ee
Equation~\eqref{eq:piPpisig} implies that the correlations among the $\delta^P_X$ are approximatively the same as the ones between  the $\delta^\sigma_X$, \ie~
\be\rho_{XX'}^P \approx \rho^\sigma_{XX'}\,.\label{eq:rhoprhosig}\ee This fact can be understood as follows. Consider only two detection channels, $i$ and $j$. If the same production mode $X \,\hat =\, X_i = X_j$ dominates in both channels, they are nearly $100\%$ correlated, so that they are described by a single nuisance parameter $\delta^P_X$, which is equivalent to say that $\rho_{XX}^P\approx1$. Note that one has $\rho_{XX}^\sigma=1$ by definition, so that $\rho^P_{XX}\approx \rho^\sigma_{XX}$. Besides, if two different production modes $X_i\neq X_j$ dominate respectively in the $i$ and $j$ channels, the uncertainties in both channels are respectively described by $\delta^P_{X_i}$ and $\delta^P_{X_j}$. These two nuisance parameters inherit the correlation from the leading production modes $X_i$ and $X_j$, which is given by $\rho^\sigma_{X_i X_j}$. Therefore one recovers Eq.~\eqref{eq:rhoprhosig}.

Finally, notice that for certain kinematical cuts selecting the $ttH$ mode in the diphoton decay channel~\cite{A-diphoton}, even additional production modes can slightly contribute, like the $bbH$, $tHW$ and $tHbq$ productions. These production modes participate in the contamination and 
 have thus been included in the combination of production modes in Eq.~\eqref{ContRedef}.

\subsection{The uncertainties on branching ratios}  \label{BRerr}

Two sources of error affect the Higgs signal strengths:  the  production and the decay rate uncertainties (see Eq.~(\ref{eq:REALmu})).
The latter is  often not considered in the Higgs fits. Still following our approach of step-by-step combinations, one should start with the signal strength error Eq.~\eqref{eq:combination_NSM}, where all uncertainties on production modes have been already combined (Eq.~\eqref{ContRedef}). 
The uncertainties on production and decay rates  combine thus as,  up to an irrelevant global sign, 
\begin{eqnarray}
\delta^\mu_{X_i} \Delta^{\mu}_i  = \delta^P_{X_i}\Delta^{P}_i  + \delta^B_{Y_i} \Delta^B_{Y_i}  \ \  \ \mbox{with} \ \ \
\Delta^B_{Y_i} = \frac{\Delta B^{\rm SM}_i}{B^{\rm SM}_i} \ , \  B^{\rm SM}_i=\frac{\Gamma^{\rm SM}_{Y_i}}{\Gamma_{\rm tot}}
\label{DmuB}
\end{eqnarray}
where $\Gamma^{\rm SM}_{Y_i}$ is the SM partial decay width for the detection channel $i$. 
In this equation,  we apply the leading moment approximation to treat the branching ratios errors as perturbations of the leading error from production modes. This is why    
the $\delta^\mu_{X_i}$ parameters carry the index $X_i$, which is the index of the dominant production mode in the channel $i$, as  in the previous subsection.   
For example, if the production mode ggF dominates in the channel $i$, one has \be\delta_{X_i}^\mu=\delta_{\rm ggF}^\mu\,. \label{eq:ExdmuXi} \ee

The relative error $\delta^B_{Y_i} \Delta^B_{Y_i}$ on the SM branching ratio is expressed as in Eq.~(\ref{eq:comb_B}), where the decay width uncertainty~\eqref{eq:subDECAY} can now be specified in terms of the various sources of 
error ({\it c.f.} Section~3 of Ref.~\cite{topEFT} for a recent overview, and references therein),
\begin{eqnarray}
\delta^{\Gamma}_{Y} \Delta^\Gamma_{Y} & = & 
\sum_{a} \delta^{{\rm pu}_a}_Y\Delta^{{\rm pu}_a}_Y + \delta^{\rm thu}_Y\Delta^{\rm thu}_Y  \    \  \    \mbox{where {\it e.g.}}  
\  \  \  \Delta^{\rm thu}_{Y} = \frac{\Delta \Gamma^{\rm thu}_Y }{ \Gamma^{\rm SM}_Y } \ .
\label{eq:DetailGa}
\end{eqnarray}
The partial decay width errors $\Delta\Gamma^{\rm thu/{\rm pu}_a}_Y$ are 
taken from the LHCHWG~\cite{LHCHWGweb,LHCHWG1,LHCHWG3}. The $\Delta\Gamma^{\rm thu}_Y$ denote the theoretical uncertainties due to the limitations of  QCD perturbative calculations.  The $\Delta\Gamma^{{\rm pu}_a}_Y$ represent the parametric uncertainties induced by the experimental errors on the input parameters, labelled by
$a\equiv \alpha_s,m_c,m_b,m_t$ (charm, bottom and top quark masses). Typically, one has $\Delta^{\rm thu/{\rm pu}_a}_{b\bar b} \gg 
\Delta^{\rm thu/{\rm pu}_a}_{VV^*}$, $\Delta^{\rm thu/{\rm pu}_a}_{\tau \bar \tau}$ since the QCD corrections to the $h\to VV^*$, $\tau \bar \tau$ 
decay channels arise only at orders higher or equal to $O(\alpha_s^2)$.

The $\Delta\Gamma^{{\rm pu}_a}_Y$ errors are associated to Gaussian distributions, and are thus identified without ambiguity with the errors defined in Ref.~\cite{LHCHWG3}.  
 The  $\Delta\Gamma^{\rm thu}_Y$ errors are purely theoretical,  so that one associates them with
 flat priors. To adopt a conservative prescription, as in Section~\ref{PDFsca}, we interpret the numbers given in \cite{LHCHWGweb} as $1\sigma$-widths. These numbers  are thus directly identified with the $\Delta \Gamma^{\rm thu}_{Y}$.
 
Now inserting Eq.~(\ref{eq:DetailGa}) into Eq.~(\ref{eq:comb_B}) provides the contributions of the theoretical and parametric uncertainties to the branching ratios,
\begin{eqnarray}
\delta^B_{Y_i} \Delta^B_{Y_i} & = & 
\sum_{Y,a} \delta^{{\rm pu}_a}_Y\Delta^{{\rm pu}_a}_Y \bigg ( B^{\rm SM}_Y -  \delta_{{Y_i}Y} \bigg )
+ \sum_{Y} \delta^{\rm thu}_Y\Delta^{\rm thu}_Y \bigg ( B^{\rm SM}_Y -  \delta_{{Y_i}Y} \bigg )
\nonumber \\ & \hat = & \sum_{Y,a} \delta_Y^{{\rm pu}_a}  \Delta_{Y,i}^{a} + \sum_{Y} \delta^{\rm thu}_Y \Delta_{Y,i} \ ,
\label{DBstat}
\end{eqnarray}
where in the last line one introduces a compact notation for the error magnitudes.
The sum over $Y$ here must include all the individual Higgs decay channels (not only the ones effectively detected at colliders), 
namely $Y \equiv b \bar b, c \bar c$, WW, ZZ, $\tau \bar \tau$, $\gamma\gamma$, gg~\dots  

We stress that the parametric error $\delta^{{\rm pu}_a}_Y\Delta^{{\rm pu}_a}_Y$ on various decay rates $Y$ arises from the same source (namely, varying the fundamental parameter $a$). The parametric errors on the various decays  are thus fully correlated. Therefore, one could in principle drop the $Y$ index on  $\delta^{{\rm pu}_a}_Y$.   There is however a subtlety, because these errors can be either $100\%$ correlated or $100\%$ anti-correlated. The use of parameters $\delta^{{\rm pu}_a}$  would render the full correlation manifest, but minus signs would have to be included in certain $\Delta^{{\rm pu}_a}_Y$. Here instead, we chose positive $\Delta$'s by convention. We have thus to keep the $Y$ index on $\delta^{{\rm pu}_a}_Y$, bearing in mind that this $Y$ labels only $100\%$ correlation or anti-correlation. A second subtlety is that these signs are actually not clearly given in the literature. Rather, only the absolute values of the $\Delta^{{\rm pu}_a}_Y|_0$ are provided. We adopt a conservative choice by assuming that all these errors are $100\%$ correlated.

We can now apply the leading moment approximation on the combination of Eqs.~(\ref{DmuB})-(\ref{DBstat}), where the leading uncertainty is $\delta^P_{X_i}\Delta^P_i$ and the perturbation is  $\delta^B_{Y_i}\Delta^B_{Y_i}$, \ie~$\Delta^P_i\gg\Delta_{Y,i},\Delta_{Y,i}^a$.
The $1\sigma$-width of the global theoretical uncertainty in a channel $i$ is given by 
\begin{eqnarray}
(\Delta_{i}^{\mu})^2=( \Delta_{i}^{P} )^2
+ \sum_{a} \left [ \sum_{Y}  \Delta_{Y,{i}}^{a} \right ]^2
+ \sum_{Y} ( \Delta_{Y,{i}} )^2  \ , \  \  \  \ 
\label{VarianCombFull}
\end{eqnarray}
with $\Delta_{i}^{P}$ given by Eq.~(\ref{VarianCombGen}).
Regarding the prior distribution of the $\delta^\mu_X$, the discussion is exactly the same as the one in Section~\ref{CTerr}. That is, following the leading moment approximation, the joint distribution of the $\delta^\mu_X$ corresponds to the one of the leading uncertainties $\delta^P_X$, so that
\be
\pi^\mu\approx\pi^P \ . \label{eq:pimupiP}
\ee
This implies in particular that the $\delta^\mu_X$ inherit the correlations from the $\delta^P_X$, that is $\rho^\mu_{XX'}\approx\rho^P_{XX'}$.

Let us discuss the correlations used to derive Eq.~(\ref{VarianCombFull}), which are drawn from Ref.~\cite{LHCHWGweb,LHCHWG1,LHCHWG3}. 
First, a given parametric uncertainty associated to $\delta^{{\rm pu}_a}_Y$ introduces $100\%$ correlated errors among the various decay modes $Y$, so that the sum over $Y$ of the $\Delta^{a}_{Y,i}$ is linear. Recall the parametric correlations are taken to be all positive.
There is also a slight correlation between $\delta^P_{X_i}\Delta^P_{i}$ and $\delta^{{\rm pu}_{\alpha_s}}_Y\Delta^{\alpha_s}_{Y,i}$, because $\delta^P_X$ also contains a contribution from the  $\alpha_s$ error. The $\alpha_s$ contribution being subleading in $\delta^P_X$, its correlation with $\delta^{{\rm pu}_{\alpha_s}}_Y$ is expected to be small, so that we can neglect it. All the other sources of uncertainties are independent due to their different origins, so that summations in quadrature appear everywhere else in Eq.~(\ref{VarianCombFull}).
\\
Using the  definitions of the reduced $\Delta$'s in Eq.~(\ref{DBstat}), we finally write explicitly the total theoretical uncertainty on the signal strength of a Higgs detection channel $i$,
\begin{eqnarray}
(\Delta_i^{\mu})^2 = ( \Delta_i^{P} )^2 +
\sum_{a} \bigg [ \sum_{Y} \Delta^{{\rm pu}_a}_Y \left( B^{\rm SM}_Y -  \delta_{{Y_i}Y} \right) \bigg ]^2
+ \sum_{Y} \bigg[ \Delta^{\rm thu}_Y  \left( B^{\rm SM}_Y -  \delta_{{Y_i}Y} \right) \bigg ]^2  .  \  \  \  \ 
\label{FINDelta}
\end{eqnarray}

\subsection{Summary} \label{se:comb_summary}

In this section we have assembled step by step all the theoretical uncertainties on the Higgs signal strengths, starting from the Higgs likelihood Eq.~\eqref{GenLprior}. 
 This combination is made possible by the statistical analysis of Section~\ref{se:Combinations}, whose results have been extensively used here. 
The final Higgs likelihood involving the combined errors reads 
\begin{eqnarray} \label{LpriorH}  
L(c_V,c_f)  & & =  \ \int \bigg(\prod_{ X}  d \delta_X^\mu \bigg)\,\pi^\mu(\delta_X^\mu)\, \times
\\
 \exp  & & \left[{-\frac{1}{2}\sum_{i,j}\bigg(\mu_{i}^{\rm th}[c_V,c_f]-\mu_{i}^{\rm ex}(1+\delta^\mu_{X_i}\Delta^\mu_i)\bigg)\, 
\mathcal{C}^{{\rm ex}\,-1}_{ij}\, \bigg(\mu_{j}^{\rm th}[c_V,c_f]-\mu^{\rm ex}_{j}(1+\delta^\mu_{X_j}\Delta^\mu_j)\bigg) }\right] . 
\nonumber 
\end{eqnarray}
The only label for the combined nuisance parameters $\delta^\mu_{X_
i}$ is $X_i$, the dominant production mode for a given channel $i$ (see for instance Eq.~(\ref{eq:ExdmuXi})).
The prior $\pi^\mu$ is approximately equal to the prior of the production mode uncertainties $\pi^\sigma$, through Eq.~\eqref{eq:piPpisig} and Eq.~\eqref{eq:pimupiP}. 
In Section~\ref{se:Corr}, the assumptions on the correlations among the production modes will allow us to express $\pi^\sigma$ in terms of the priors of individual production mode uncertainties $\pi^\sigma_X$ (see Eqs.~\eqref{eq:LMA_comb_12}--\eqref{eq:LMA_comb_3}).

One of the outcome of the combination procedure followed throughout this section is that the shape of the combined priors $\pi^\sigma_X$ appears to be almost Gaussian. 
This comes partly because some of the priors for the individual sources of uncertainty are Gaussian. However, the main reason is actually that a substantial number of the individual sources of uncertainty are independent and of same order of magnitude. 
These conditions resemble to the ones of the
\textit{central limit theorem}, which predicts that the combination would converge towards a Gaussian distribution. 
 Besides, the small errors from contamination and partial decay widths  do not affect either the final prior shape under the leading moment approximation. It follows that the $\pi^\mu$ distribution is close to a multivariate Gaussian distribution.

Finally, we stress again that the famous question of the linear versus quadratic summation of individual errors (as the ones used in this section to derive $\Delta^\mu_i$ in Eq.~(\ref{FINDelta})) relies uniquely on  
the correlations among the errors, and is therefore independent of the shapes of the priors. 
This general feature holds when uncertainties are combined using Bayesian statistics.

\section{Marginalising the Higgs likelihood}  \label{se:MargBF}

\subsection{Correlations of the detection channels}  \label{se:Corr}
  
In this subsection we focus on the correlations among Higgs detection channels induced by the theoretical uncertainties. These correlations appear whenever a source of uncertainty contributes simultaneously to various channels. 

As a preliminary observation, let us recall that these correlations are sometimes not taken into account in the literature. What is typically done in such case is that some amount of error, typically  from Refs.~\cite{LHCHWGweb,LHCHWG1,LHCHWG2,LHCHWG3}, is added independently to the statistical error of each detection channel. Such combination typically reads $(\Delta\mu^{{\rm ex}}_i)^2+(\Delta \mu^{\rm th}_i)^2$ if done in quadrature. From the point of view of nuisance parameters, this combination would correspond to associating one independent $\delta_i^\mu\Delta_i^\mu$ to each detection channel, and thus performing one integration per channel in the marginal likelihood.

The issue with such approach is that the correlations among channels induced by the theoretical uncertainties are lost. As stated in Section~\ref{sec:BiasPrin}, these correlations are crucial because they potentially change the tension among the various channel measurements, which in turn can modify the best-fit regions. As slight modifications of the best-fit regions are expected in presence of new physics, treating correctly the theoretical uncertainties is fundamental.


Taking into account the correlations among channels amounts to consistently propagate the theoretical errors into the different detection channels. This is precisely what is done through the combination procedure of Section~\ref{se:marg}. 
Combining the errors together and using the leading moment approximation to treat subdominant errors, only five nuisance parameters $\delta_{\rm ggF}^\mu$,  $\delta_{\rm VBF}^\mu$,  $\delta_{\rm ZH}^\mu$,  
$\delta_{\rm WH}^\mu$ and $\delta_{\rm ttH}^\mu$ arise (see Eqs.~\eqref{FINDelta}--\eqref{LpriorH}). The uncertainty on each channel is described by only one of these $\delta^\mu_X$, where the $X$ corresponds to the dominant production mode in this channel. That is, all channels dominated by the same production mode $X$ have the same nuisance parameter $\delta_X^\mu$. This implies that these channels are $100\%$ correlated. 

In principle, the combination procedure of Section~\ref{se:marg} describes the complete distribution for the $\delta_X^\mu$, $\pi^\mu$, including the correlations $\rho^\mu_{XX'}$ \textit{among} the different $\delta_X^\mu$. In practice, a complete knowledge of the correlations among the individual  sources of uncertainties is needed to obtain $\rho^\mu_{XX'}$. Here we consider the determination of $\rho^\mu_{XX'}$ as beyond the scope of this paper, since for example one would have to work out clearly the correlations among the Higgs production modes induced by the PDF data uncertainties ($\delta^{\rm data}_X$). Using the information available in the literature we will rather consider some characteristic cases for $\rho^\mu_{XX'}$ .
\\

Let us first discuss the typical correlations induced by the PDF uncertainties (origi\-nating from the PDF data fit)  and the scale uncertainties ({\it c.f.} Section~\ref{EFTsca}) on the production cross sections. From now on, the $\delta^\mu_X$ are denoted as $\delta_X$ for simplicity,
\be
\delta_X^\mu\,\hat=\,\delta_X\,.
\ee
First, we will set $\delta_{\rm ggF}=-\delta_{\rm ttH}$ since an anti-correlation between the corresponding PDF errors is reported in Ref.~\cite{CMS-NOTE-2011-005}~\footnote{It is not clear from this reference whether the correlations include as well the whole error   
from $\alpha_s$ which is  $100\%$ correlated between the production modes. Nevertheless this source of error is minor compared to
the other ones.}. Note that in reality, this anti-correlation is not total (its value is -0.6 in Ref.~\cite{CMS-NOTE-2011-005}) and furthermore the other 
source of error, the scale uncertainty, does not correlate the ggF and ttH cross sections as these come from independent QCD calculations.
\\
The correlation coefficients of the PDF errors~-- between $\sim$~0.63 and 0.93~\cite{CMS-NOTE-2011-005}~-- for the three other production modes motivate us to take 
$\delta_{\rm VBF}=\delta_{\rm ZH}=\delta_{\rm WH}$. This assumption is further justified by the fact that the PDF error is larger than the scale error (particularly for VBF) and that the scale error most probably correlates the ZH and WH modes. 
\\
The correlation coefficients of the PDF errors between ggF and WH (-0.23), ZH (-0.14) or VBF (-0.57)  suggest to consider the two  extreme cases of vanishing correlation and $100\%$ anti-correlation.  The scale uncertainties tend to decorrelate these modes. It is thus coherent to consider
the cases of vanishing correlation and $100\%$ anti-correlation as  the two extreme cases to study.
All these assumptions are summarized as the two following configurations on the nuisance parameters,~\footnote{For consistency,
these two configurations are used as well to determine the $\rho^\sigma_{XX'}$ correlation matrix of Eq.~(\ref{VarianCombGen}).}
\begin{eqnarray}
\delta_{\rm ggF}=-\delta_{\rm ttH} \ , \ \delta_{\rm VBF}=\delta_{\rm ZH}=\delta_{\rm WH} \ , 
\label{Corr1} \\ 
-\delta_{\rm ggF}=\delta_{\rm ttH}=\delta_{\rm VBF}=\delta_{\rm ZH}=\delta_{\rm WH}  \ ,
\label{Corr2}
\end{eqnarray}
keeping in mind that the realistic situation lies in between these extreme cases.

Regarding the PDF set error, the individual uncertainties giving rise to  this error are not available in the literature. 
Rather, only the global PDF set error is estimated by changing various assumptions at a time.
One can at least notice that the PDF set errors can be potentially correlated either negatively or positively, respectively, for the ggF and VBF 
reactions or the VBF and VH processes, as observed from the relative signs of rate variations in Fig.~(57) of Ref.~\cite{LHCHWG3} when changing the PDF set.~\footnote{Recall that the Fig.~(57) of Ref.~\cite{LHCHWG3}  is for the $8$~TeV LHC.} 
These correlations are roughly consistent with the ones in Eq.~(\ref{Corr2}).

Let us describe how the correlation configurations of Eq.~(\ref{Corr1})-(\ref{Corr2}) are related to the $\pi^\mu$ appearing in the marginal likelihood~(\ref{LpriorH}). 
The prior $\pi^\mu$ is approximately equal to the prior of the production mode uncertainties $\pi^\sigma$ (Eq.~\eqref{eq:piPpisig} and Eq.~\eqref{eq:pimupiP}) which can
itself be expressed (according to \eqref{eq:LMA_comb_3a}--\eqref{eq:LMA_comb_3}) in terms of the $\pi^\sigma_X$ under the assumptions~(\ref{Corr1})-(\ref{Corr2}). 
One ends up with the two final priors, associated respectively to the correlation configurations of Eqs.~(\ref{Corr1})-(\ref{Corr2}), 
\begin{eqnarray}
\pi^\mu (\delta_X) = \pi^\sigma_{\rm ggF}(\delta_{\rm ggF})\ \delta(\delta_{\rm ggF}+\delta_{\rm ttH})
\ \pi^\sigma_{\rm VBF}(\delta_{\rm VBF})\ \delta(\delta_{\rm VBF}-\delta_{\rm ZH})\ \delta(\delta_{\rm VBF}-\delta_{\rm WH}) \  , 
\label{Corr1Cons} \\ 
\pi^\mu  (\delta_X) = \pi^\sigma_{\rm ggF}(\delta_{\rm ggF})\ \delta(\delta_{\rm ggF}+\delta_{\rm ttH})\ \delta(\delta_{\rm ggF}+\delta_{\rm VBF})
\ \delta(\delta_{\rm ggF}+\delta_{\rm ZH})\ \delta(\delta_{\rm ggF}+\delta_{\rm WH}) \  ,
\label{Corr2Cons}
\end{eqnarray}
where $\delta()$ denotes the Dirac distribution.

 \subsection{The Bayesian analytical likelihood} \label{GandQ}

The $\pi_{X}^\sigma$ priors deduced from the combination of all the cross section errors, in Section~\ref{PDFsca}, have been found to be nearly Gaussian distributions.
These Gaussian shapes are obtained by choosing flat shapes for all the unknown priors for theoretical uncertainties. As mentioned in Section~\ref{se:comb_summary}, one expects this result to hold approximatively for other choices of initial priors. 
Nevertheless,  in order to take into account in our numerical results the possibility of non-flat initial shapes, we also consider a totally different form of the final 
prior: we take it as a flat distribution. The choice of these two shapes (Gaussian and flat)  provides an estimate of the impact of the prior shape on the final results.
The distributions  $\pi^\sigma_X$ appearing in  Eqs.~\eqref{Corr1Cons}-\eqref{Corr2Cons} are hence defined as \be
\pi_{X}^\sigma(\delta_{X}^\sigma) = \frac{1}{\sqrt{2\pi}}e^{-(\delta_{X}^\sigma)^2/2}\, ,
\label{eq:GSpr}
\ee
\be
\pi_{X}^\sigma(\delta_{X}^\sigma)=
\begin{cases}
1/2\sqrt{3}~~\textrm{if}~~\delta_{X}^\sigma\in[-\sqrt{3},\sqrt{3}]\,, \\
0~~\textrm{otherwise}\,
\end{cases}
\label{eq:GFlat}
\ee
for the Gaussian and flat cases respectively. Recall that the variance of all the $\delta$'s, including $\delta^\sigma_X$, are chosen to be equal to one for any prior shape. This appears clearly in Eq.~\eqref{eq:GSpr} and implies the $[-\sqrt{3},\sqrt{3}]$ interval in Eq.~\eqref{eq:GFlat}.

For analytical integrations of the final likelihood~\eqref{LpriorH},
it is convenient to denote by ${\cal X}$ a subset of fully correlated production modes, $\{X,X',\ldots\}$. 
We then denote by $\Omega_{\cal X}$ the subset of channels (labelled by $i$) dominated by the production modes contained in ${\cal X}$. 
In presence of anti-correlations, one further divides $\Omega_{\cal X}$ into two anti-correlated subsets  $\Omega_{\cal X}^+$, $\Omega_{\cal X}^-$. Finally, the set of all channels is written $\Omega$. 
Assuming the correlations among production modes follow Eq.~\eqref{Corr1}, the set of detection channels is splitted into $\Omega_{\{\rm ggF,ttH\}}$ and $\Omega_{\{\rm VBF,WH,ZH\}}$. $\Omega_{\{\rm ggF,ttH\}}$ is then splitted into the anti-correlated subsets $\Omega_{\{\rm ggF,ttH\}}^+=\Omega_{\rm ggF}$, $\Omega_{\{\rm ggF,ttH\}}^-=\Omega_{\rm ttH}$.  
Assuming  the correlations of Eq.~\eqref{Corr2}, there is instead a unique set $\Omega=\Omega_{\{\rm ggF,ttH,VBF,WH,ZH\}}$. It is splitted into the anti-correlated subsets 
 $\Omega_{\{\rm ggF,ttH,VBF,WH,ZH\}}^+=\Omega_{\rm ggF}$,  $\Omega_{\{\rm ggF,ttH,VBF,WH,ZH\}}^-=\Omega_{\{\rm ttH,VBF,WH,ZH\}}$.

At that point it is also convenient to introduce the following quantities $\zeta_{\cal X}$ and $\eta_{{\cal X}{\cal X}'}$  defined as
\begin{eqnarray}
\zeta_{{\cal X}} \ &=& \sum_{i\in \Omega_{\cal X},\,j\in \Omega} \ \kappa_i \, \Delta_{i}^{\mu} \  (\mu^{\rm th}_{i}-\mu^{\rm ex}_{i})\ {\cal C}^{{\rm ex}\, -1}_{ij} \ \mu^{\rm ex}_{j} \,,\quad 
\kappa_i=\begin{cases}1\textrm{ if } i\in\Omega^+_{\cal X}\\
-1\textrm{ if } i\in\Omega^-_{\cal X}
\end{cases}
\nonumber \\
\eta_{{\cal X}{\cal X}'} \ &=& \sum_{i\in \Omega_{\cal X},\,j\in \Omega_{{\cal X}'}} \ \kappa_{i} \, \Delta_{i}^{\mu} \ \mu^{\rm ex}_{i} \ {\cal C}^{{\rm ex}\, -1}_{ij} \ \kappa_j \, \Delta_{j}^{\mu} \ \mu^{\rm ex}_{j}
\, .
\label{LpriorFC}
\end{eqnarray}
The overall sign of $\zeta_{\cal X}$ is irrelevant. Note also  that if ${\cal X} \neq {\cal X}'$ (as may occur in  the $\eta_{{\cal X}{\cal X}'}$ function),  there are no theoretical correlations at all between the channels belonging to $\Omega_{\cal X}$ and $\Omega_{{\cal X}'}$.

In the case of a Gaussian prior (Eq.~\eqref{eq:GSpr}), it is noticeable that the most general likelihood~\eqref{LpriorH} can be integrated analytically
 and results in the  
simple analytical expression~\footnote{A similar expression can also be obtained for an  arbitrary correlation matrix $\rho^\mu_{XX'}$. Note one dropped an overall factor, as the likelihood is defined up to a normalisation constant.}
\be
\boxed{ L_{\rm B}^{\rm Gauss} = L_\mu \,  \exp\left[ \frac{1}{2} \sum_{{\cal X}{\cal X}'} \zeta_{{\cal X}}( \delta_{{\cal X}{\cal X}'}+ \eta_{{\cal X}{\cal X}'})^{-1}\zeta_{{\cal X}'}\right]}\,.
\label{eq:LpriorGAUSS}
\ee
Here $\delta_{{\cal X}{\cal X}'}$ is the Kronecker symbol.
  $L_\mu$ is the base likelihood defined in Eq.~\eqref{pdfGAUSS}, \ie~the likelihood before introducing nuisance parameters. One observes that the marginal likelihood takes the form of a product of the base likelihood with a term generated by the theoretical uncertainties. 
     This term, which depends on $c_V,c_f$ through $\zeta_{\cal X}$, as well as on all theoretical and experimental uncertainties,  implements all the deformations and correlations induced by the theoretical uncertainties.  
\\ 
For the case of no experimental correlations between different  group of channels  of dominant production modes, including the case considered without experimental correlations at all (see Section~\ref{se:BaseLike}), one has $\eta_{{\cal X}{\cal X}'}=0$ for ${\cal X}\neq {\cal X}'$ and
 \be\eta_{{\cal X}{\cal X}}\equiv \eta_{\cal X} = \sum_{i,j \in \Omega_{\cal X}} \kappa_{i} \Delta_{i}^{\mu} \mu^{\rm ex}_{i} {\cal C}^{{\rm ex}\, -1}_{ij}\kappa_j  \Delta_{j}^{\mu} \mu^{\rm ex}_{j}\,. \label{eq:etachi} \ee The marginal likelihood \eqref{eq:LpriorGAUSS} then reduces to, 
\be
 L_{\rm B}^{\rm Gauss} = L_\mu \prod_{{\cal X}} e^{\zeta_{{\cal X}}^2/2(\eta_{{\cal X}}+1)} \,.
\label{eq:LpriorGAUSS_simp}
\ee
Note that this product is over different ${\cal X}$ subsets \ie~there are no theoretical correlations among the channels belonging to
the different $\Omega_{\cal X}$ groups. 

Note that if one assumes a single independent nuisance parameter per channel, there is no sum in Eqs.~\eqref{LpriorFC}, 
meaning that  no correlation among channels is induced.~\footnote{We 
recall that such a combination should be avoided as it is not realistic.} 
One can directly verify that in the purely de-correlated case (neither experimental nor theoretical correlations), 
Eq.(\ref{eq:LpriorGAUSS_simp}) gives back the primary likelihood~(\ref{pdfGAUSS})
with a summation in quadrature between the absolute experimental and theoretical errors, 
$\Delta\mu^{\rm ex}_i$ and $\mu^{\rm ex}_i\Delta_{i}^{\mu}$.

In the case of the flat prior of Eq.~\eqref{eq:GFlat}, there is no simple general form such as Eq.~\eqref{eq:LpriorGAUSS}. However, assuming no  experimental correlations among  various $\Omega_{\cal X}$ subsets,  the marginal likelihood takes  a simple form, 
\be
 L_{\rm B}^{\rm flat} = L_\mu \prod_{{\cal X}} e^{\zeta_{{\cal X}}^2/2\eta_{{\cal X}}}  \left[\rm Erf \left(\frac{\sqrt{3}\,\sqrt{\eta_{{\cal X}}}}{\sqrt{2}} +\frac{\zeta_{{\cal X}}}{\sqrt{2\eta_{{\cal X}}}}\right) -
\rm Erf \left(\frac{\sqrt{3}\,\sqrt{\eta_{{\cal X}}}}{\sqrt{2}} -\frac{\zeta_{{\cal X}}}{\sqrt{2\eta_{{\cal X}}}}\right)  \right]  \,,
\label{LpriorFLAT} 
\ee
where $\rm Erf$ is the standard error function.

\subsection{The frequentist treatment}  \label{se:F2}

\subsubsection{The marginal likelihood}  \label{FMmethod}

In classical frequentist statistics, hypotheses are not associated with probabilities, so that  there is no such thing as a prior  distribution for a nuisance  parameter. 
In the hybrid frequentist framework however, one can associate a parameter with a ``prior'' distribution that can be seen as an extra likelihood constraining the nuisance parameter.
Pushing forward the analogy with the Bayesian case, we worked out the way to combine uncertainties within frequentist statistics in Section~\ref{se:combTUfreq}. 
One may find however that the Bayesian combination of uncertainties are better defined than the frequentist one. 

More pragmatically, frequentist combinations are also more complicated, as the combination of the magnitude of the errors (the $\Delta$'s) depends on the shape of the frequentist ``priors'', contrary to the Bayesian case.
These drawbacks can constitute motivations to rather follow the Bayesian approach developed in previous sections. Nevertheless, for completeness we describe here the final part of the frequentist method for the Higgs fit. For that purpose we 
consider in the following, a generic prior, $\pi^\mu(\delta^{\mu}_{X})$, of width $\Delta_{i}^{\mu}$, obtained after a first phase of frequentist combination.

Recall that  the frequentist marginalisation procedure, also called profiling, consists in maximizing over $\delta^{\mu}_{X}$, 
instead of integrating as done in Eq.~(\ref{LpriorH}). Hence the frequentist marginal Higgs likelihood reads
\begin{eqnarray} \label{LpriorFrq}  
L(c_V,c_f)  & = & \max_{\delta_X^\mu}\Bigg[ \,\pi^\mu(\delta_X^\mu)\, \times
\\
& \exp & \bigg[{-\frac{1}{2}\sum_{i,j}(\mu_{i}^{\rm th}[c_V,c_f]-\mu_{i}^{\rm ex}(1+\delta^\mu_{X_i}\Delta^\mu_i))\, 
\mathcal{C}^{{\rm ex}\,-1}_{ij}\, (\mu_{j}^{\rm th}[c_V,c_f]-\mu^{\rm ex}_{j}(1+\delta^\mu_{X_j}\Delta^\mu_j)) }\bigg]\Bigg] . 
\nonumber 
\end{eqnarray}

As often done in practice for the frequentist treatment, one can equivalently minimize the $\chi^2$ distribution,   $\chi^2=-2\log L$, instead of the maximisation in Eq.~(\ref{LpriorFrq}),
\begin{eqnarray} \label{eq:chi2A}  
\chi^2(c_V,c_f)  & = & \min_{\delta_X^\mu}\Bigg[ \,-2\log\pi^\mu(\delta_X^\mu)\, +
\\
&  &{\sum_{i,j}\left(\mu_{i}^{\rm th}[c_V,c_f]-\mu_{i}^{\rm ex}(1+\delta^\mu_{X_i}\Delta^\mu_i)\right)\, 
\mathcal{C}^{{\rm ex}\,-1}_{ij}\, \left(\mu_{j}^{\rm th}[c_V,c_f]-\mu^{\rm ex}_{j}(1+\delta^\mu_{X_j}\Delta^\mu_j)\right) }\Bigg] . 
\nonumber 
\end{eqnarray}
The best-fit point given by the $\chi^2$ minimum in the $(c_f,c_V)$ parameter space is noted $(\hat c_f,\hat c_V)$ 
and the best-fit regions are obtained by drawing contour levels of the difference ({\it c.f.}  Section~\ref{se:STATbasics})
\be \Delta \chi^2 (c_f,c_V)= \chi^2 (c_f,c_V) - \chi^2 (\hat c_f,\hat c_V)\,\ee
at the values given by Eq.~\eqref{eq:Chi2val}.

\subsubsection{The frequentist analytical likelihood}  \label{GandQII}

Assuming that the Bayesian and frequentist combinations of the errors lead to analogous shapes for the final priors, we consider both a Gaussian and a flat shape for each $\pi^\sigma_X$ prior, as in Eqs.~\eqref{eq:GSpr}--\eqref{eq:GFlat}.
 In the Gaussian case, the marginal likelihood~\eqref{LpriorFrq} can be computed analytically,
\be
\boxed{ L_{\rm F}^{\rm Gauss} = L_\mu \,  \exp\left[ \frac{1}{2} \sum_{{\cal X}{\cal X}'} \zeta_{{\cal X}}( \delta_{{\cal X}{\cal X}'}+ \eta_{{\cal X}{\cal X}'})^{-1}\zeta_{{\cal X}'}\right]}\,,
\label{eq:LpriorGSfq}
\ee
where the $\zeta_{\cal X}$, $\eta_{\cal X X'}$ are defined as in Section~\ref{GandQ}.
This is precisely the same result as  for the Bayesian likelihood of Eq.~\eqref{eq:LpriorGAUSS},  $L_{\rm B}^{\rm Gauss}$. 

For the case of no experimental correlations between the $\Omega_{\cal X}$'s, the marginal likelihood with Gaussian prior thus simplifies just like in 
Eq.~\eqref{eq:LpriorGAUSS_simp}.~\footnote{Hence the same likelihood (with a sum in quadrature) as in the Bayesian framework arises, in the case of
neither experimental nor theoretical correlations.} 
In this case, the marginal likelihood with a flat prior also gets an analytical expression,
\begin{eqnarray}
& L_{\rm F}^{\rm flat} & =  \prod_{\cal X}    \label{eq:L_marg_F_flat}
\\ & \exp & \left[{-\frac{1}{2}\sum_{i,j}\left(\mu_{i}^{\rm th}[c_V,c_f]-\mu_{i}^{\rm ex}(1+\xi_{\cal X}\kappa_i\Delta^\mu_i)\right)\, 
\mathcal{C}^{{\rm ex}\,-1}_{ij} \,  
\left(\mu_{j}^{\rm th}[c_V,c_f]-\mu^{\rm ex}_{j}(1+\xi_{\cal X}\kappa_j\Delta^\mu_j)\right) }\right]\,
\nonumber
\end{eqnarray}
with
\be\xi_{\cal X}=\begin{cases}
\zeta_{\cal X} / \eta_{\cal X}&\textrm{if } \zeta_{\cal X}/\eta_{\cal X} \in[-\sqrt{3} ,\sqrt{3} ]\\
\;\,\sqrt{3}  & \textrm{if } \zeta_{\cal X}/\eta_{\cal X} >\sqrt{3}\\
\;\, -\sqrt{3} &\textrm{if } \zeta_{\cal X}/\eta_{\cal X} <-\sqrt{3}
\end{cases}\,,
\ee
where $\zeta_{\cal X}$, $\eta_{\cal X}$ are defined as in Eq.~(\ref{LpriorFC}),~\eqref{eq:etachi}.

\subsection{Numerical results}  \label{PlotsMargFreq}

The frequentist marginalisation (likelihood~(\ref{eq:LpriorGSfq}) for the Gaussian prior or (\ref{eq:L_marg_F_flat}) for the flat one) is 
not illustrated here because the frequentist framework may seem slightly less consistent than the Bayesian one and the error combinations are more delicate. For these reasons,  we rather recommend to use the  Bayesian marginalisation technics for the Higgs fits. 
In any case, the Bayesian and frequentist approaches 
are expected to converge  as the experimental uncertainties become small  relatively to the theoretical ones. This situation will gradually occur in the next LHC Runs
due to the decrease of the statistical uncertainties and the expected improvement in the knowledge of the experimental systematic errors. We have described  this feature in Ref.~\cite{ProceedHouches}.  

Now as a general remark allowing a better comprehension of the following subsections,
let us try to explain in simple words the reason why the  presence of nuisance parameters can indeed modify the size and the location  of the best-fit domains in $c_V-c_f$. 
\\
For the sake of understanding the impact on the size, it is easier to focus on frequentist marginalisation. Frequentist marginalisation can be seen as an approximation of Bayesian marginalisation, so that the same explanation holds for both. 
The frequentist marginalisation consists  of a maximisation of the nuisance parameter (say $\delta^\mu_X$) at any point in the space of the parameters of interest. This means that the value of $\delta^\mu_X$ at a given point is chosen in order to maximise goodness-of-fit. Now, this improvement of goodness-of-fit is typically larger for the points far away from the best-fit point than for those close by the best-fit point. When this fact is true (which is usually the case), the operation of marginalising  tends to enlarge the best-fit regions.
\\
The effect of the nuisance parameters on the location of the best-fit regions in $c_V-c_f$ can be understood as follows. Recall that the nuisance parameters enter  in the likelihood as $\mu_{i}^{\rm ex} ( 1+\delta^{\mu}_{X_i} \Delta_{i}^{\mu} )$ (see Eq.~(\ref{LpriorH})), so that  they  shift the central experimental value of the signal strength. This in turn  can induce a change in the location of the 
best-fit point in $c_V-c_f$. Such a shift actually occurs  if a non-zero value of $\delta^\mu_X$ is preferred. This happens when a non-zero value for $\delta^\mu_X$ helps relaxing the tensions (\ie~different preferred values of $c_V$, $c_f$) 
among various signal strengths $\mu_{i}^{\rm ex}$. Notice that this means that the likelihood itself favours a non-zero value for $\delta^\mu_X$, even though the prior of  
$\delta^\mu_X$ is centered on zero.

\subsubsection{The forbidden case: no correlations}

\begin{figure}[t]
\begin{picture}(400,200)
\put(100,0){\includegraphics[width=7.5cm]{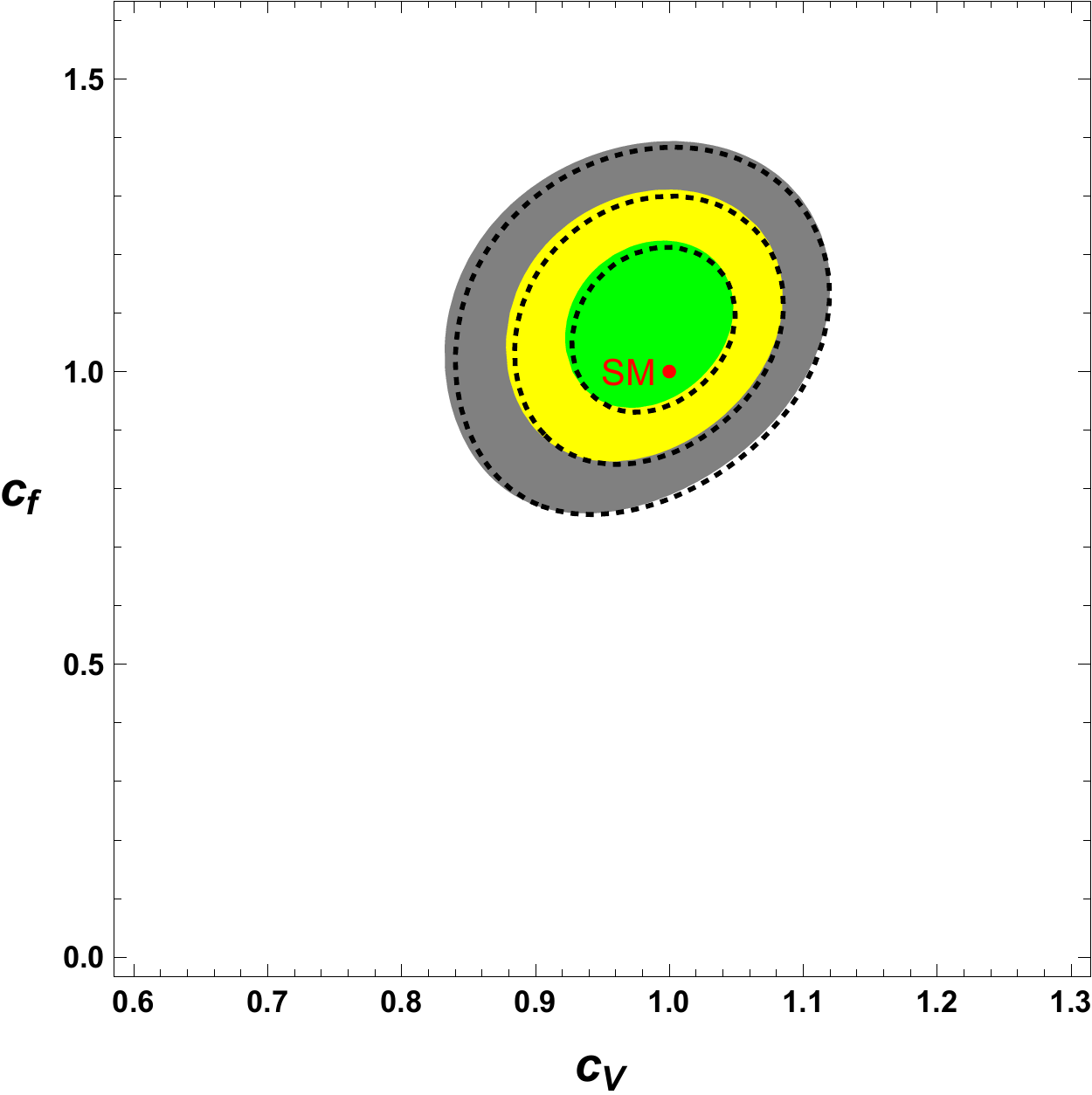}}
\put(133,60){{\color{blue} Bayesian marginalisation}}
\put(133,40){{\color{blue} -- {\it Gaussian prior, no correlations } -- } }
\end{picture}
\caption{
The best-fit regions in the $c_V-c_f$ plane obtained from Bayesian marginalisation
and Gaussian priors for the theoretical uncertainties. 
The $68\%$, $95\%$ and $99\%$ credible regions are represented respectively  by the green, yellow and grey domains. 
No theoretical error correlations between the Higgs detection channels are taken into account in this figure.
The dashed contours illustrate the case without theoretical uncertainties. The SM prediction is shown by the red point.
}\label{fig:Combqd}
\end{figure}
 
 

Following our overview approach, let us start with the simplest case: the Bayesian margina\-lisation in the absence of correlations 
between the theoretical errors of the different Higgs channels. Let us take for instance a Gaussian prior (taking a flat one would not change our conclusions).
This case was described in more details in the beginning of 
Section~\ref{se:Corr} as well as in Section~\ref{GandQ}. In this  ``de-correlated'' case, the likelihood is simply the primary likelihood~(\ref{pdfGAUSS}) 
with a summation in quadrature of the absolute experimental and theoretical errors, $(\Delta\mu^{\rm ex}_i)^2+(\mu^{\rm ex}_i\Delta_{i}^{\mu})^2$. 
The best-fit domains in the $c_V-c_f$ plane are derived following the 
standard procedure described in Section~\ref{se:STAT}, and are shown  in Fig.~(\ref{fig:Combqd}). Here and throughout Section~\ref{PlotsMargFreq}, the priors for $c_V, c_f$ are taken flat, $\pi(c_{V,f})\propto 1$.
\\
We see on this figure that the theoretical SM prediction ($c_V=c_f=1$) 
lies well within the $68\%$~C.L.~\footnote{The acronym C.L. will stand for Credible Level within the Bayesian framework and for
Confidence Level in the frequentist framework.} region. Physically, this implies that, with such a fit, no physics  beyond the SM is required to interpret the $8$~TeV LHC 
measurements of the Higgs rates. The increase of the best-fit domain sizes induced by the existence of theoretical errors is relatively weak, due to the sum 
in quadrature, as observed when comparing to the best-fit regions obtained with vanishing theoretical errors. The latter regions are  
superimposed on Fig.~(\ref{fig:Combqd}) for illustration purpose (as the dashed contours) and to ease the comparison with  next plots. 
\\
However let us recall that the likelihood used here (and leading to the colored regions of Fig.~(\ref{fig:Combqd})) is not realistic as the correlations among  the Higgs channels should not be neglected. We thus do not recommend the use of this likelihood.

\subsubsection{Flat prior}

\begin{figure}[t]
\begin{picture}(400,200)
\put(0,0){\includegraphics[width=7.5cm]{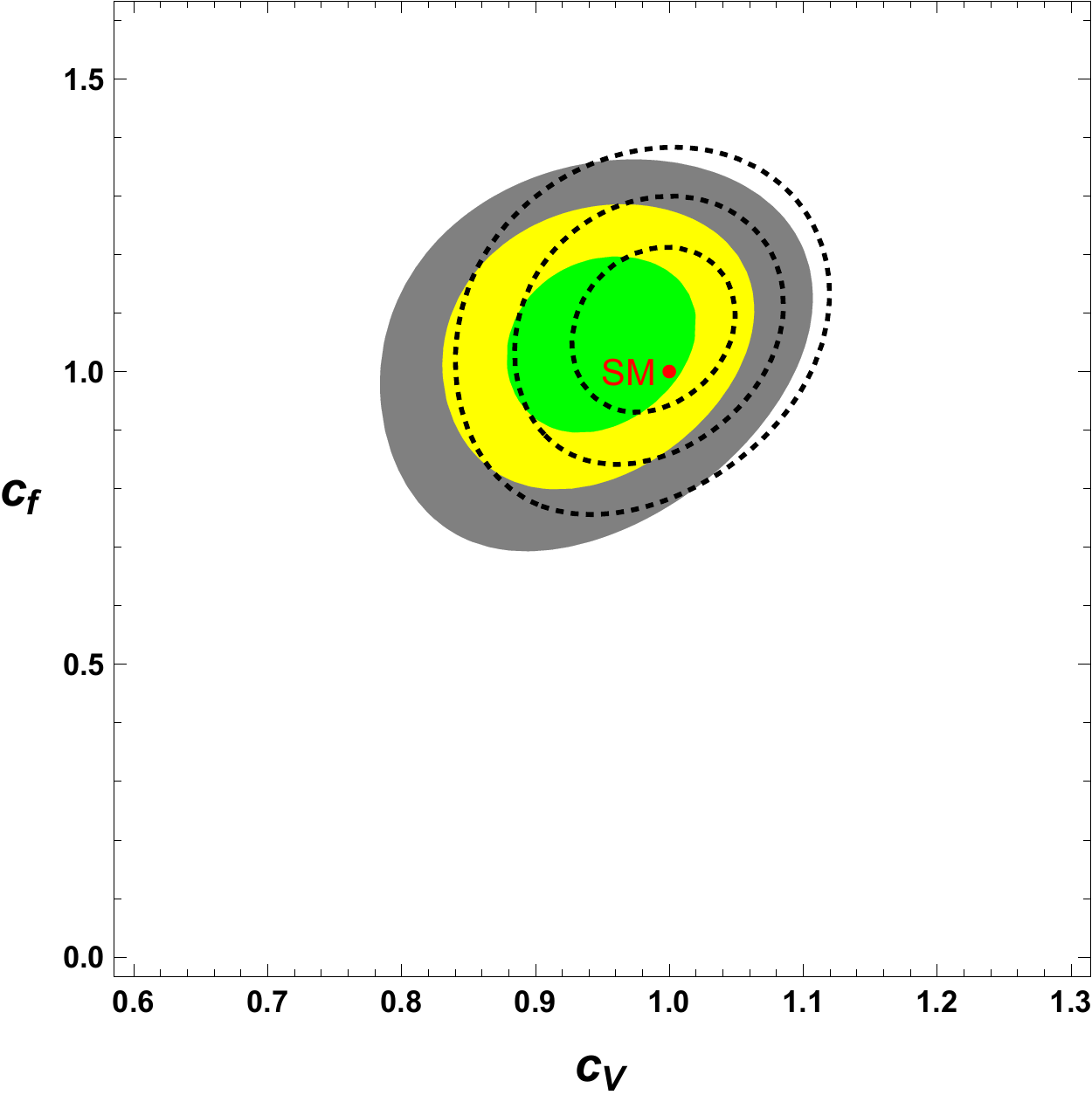}}
\put(230,0){\includegraphics[width=7.5cm]{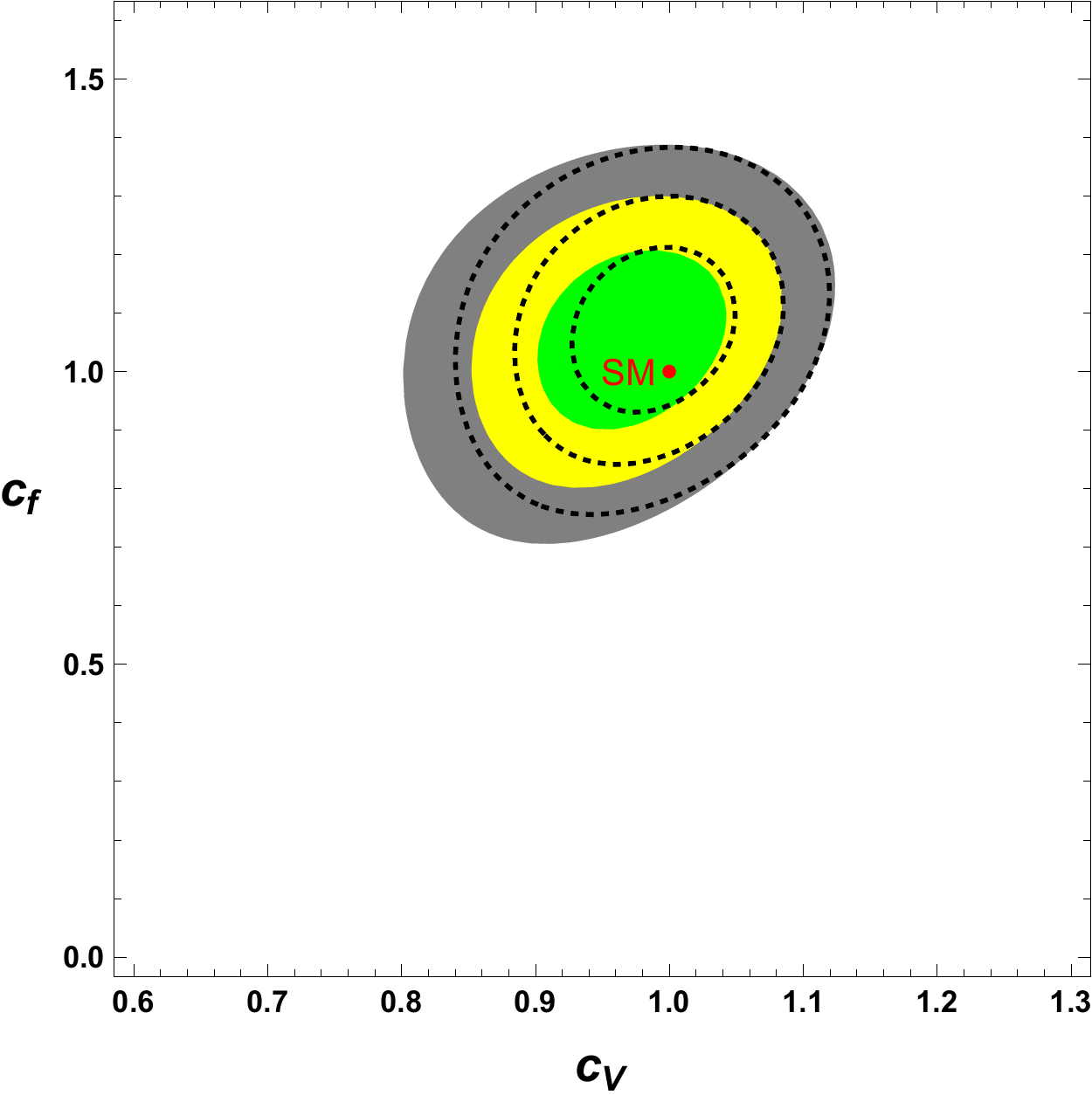}}
\put(40,60){{\color{blue} Bayesian marginalisation}}
\put(40,40){{\color{blue}  -- {\it flat prior} --  [a] }}
\put(275,60){{\color{blue} Bayesian marginalisation}}
\put(275,40){{\color{blue} -- {\it flat prior} -- [b] }}
\end{picture}
\caption{
The best-fit regions in the $c_V-c_f$ plane obtained from Bayesian marginalisation  and  flat priors for the theoretical uncertainties. 
The $68\%$, $95\%$ and $99\%$ credible regions are represented respectively  by the green, yellow and grey domains. 
The [a] and [b] plots correspond, respectively, to the two characteristic correlation configurations described in Eq.~\eqref{Corr1Cons} and Eq.~\eqref{Corr2Cons}. 
The dashed contours illustrate the case without theoretical uncertainties.  The SM prediction is shown by the red point.
}\label{fig:CombMarg}
\end{figure}

From now on we  consider the more realistic likelihoods obtained in Section~\ref{GandQ}. 
These likelihoods contain all the correlations between Higgs channels induced by the theore\-tical uncertainties. 
First, we consider the configuration with two independent nuisance parameters (see Eq.~(\ref{Corr1}) and Eq.~(\ref{Corr1Cons})). 
The Bayesian marginalisation over these two nuisance parameters leads to the analytical likelihood~(\ref{LpriorFLAT}) for flat final 
priors. Applying the standard Bayesian procedure, described in Section~\ref{se:STAT}, we find the best-fit regions 
of Fig.~(\ref{fig:CombMarg})[left].

By compa\-ring the colored plots in Fig.~(\ref{fig:Combqd}) and 
Fig.~(\ref{fig:CombMarg})[left], one observes clearly a shift of the best-fit regions.
This shift originates from the theoretical correlations that are taken into account in Fig.~(\ref{fig:CombMarg})[left]. This shift occurs because the relaxation of the tensions between the individual signal strength measurements (see discussion in the introduction of Section~\ref{PlotsMargFreq})  is different in the correlated case and in the ``de-correlated'' one. We emphasize that this shift is a consequence of taking into account the theoretical  correlations. Indeed we will see in next subsection that the same effect occurs for a different prior shape. Concerning the region size, a slight increase occurs relatively to Fig.~(\ref{fig:Combqd}). This comparison can be done by looking at
the reference case (dashed contours) without theoretical errors at all, which is once more superimposed on Fig.~(\ref{fig:CombMarg})[left].

The plot on the right hand side of Fig.~(\ref{fig:CombMarg}) is the same as the left plot but for the second correlation configuration, involving a single nuisance parameter (discussed 
in Eq.~(\ref{Corr2}) and Eq.~\eqref{Corr2Cons}). The effect of the theoretical correlations  (relatively to Fig.~(\ref{fig:Combqd})) appears to be softer than for the left plot: the shift is smaller. This difference between the two colored regions of Fig.~(\ref{fig:CombMarg}) makes clear that the theoretical correlations have an important impact on the fits, and should thus be carefully  taken into account.

As described below Eq.~(\ref{Corr2}), the 
most realistic correlation configuration is most proba\-bly an intermediate configuration between those adopted in the two plots of Fig.~(\ref{fig:CombMarg}). 
We thus conclude that, with the statistical treatment adopted here, the SM
prediction remains in a good agreement ($1\sigma$ level) with the $8$~TeV LHC Higgs data, even once realistic theoretical correlations are taken into account.

\subsubsection{Gaussian prior}

\begin{figure}[t]
\begin{picture}(400,200)
\put(0,0){\includegraphics[width=7.5cm]{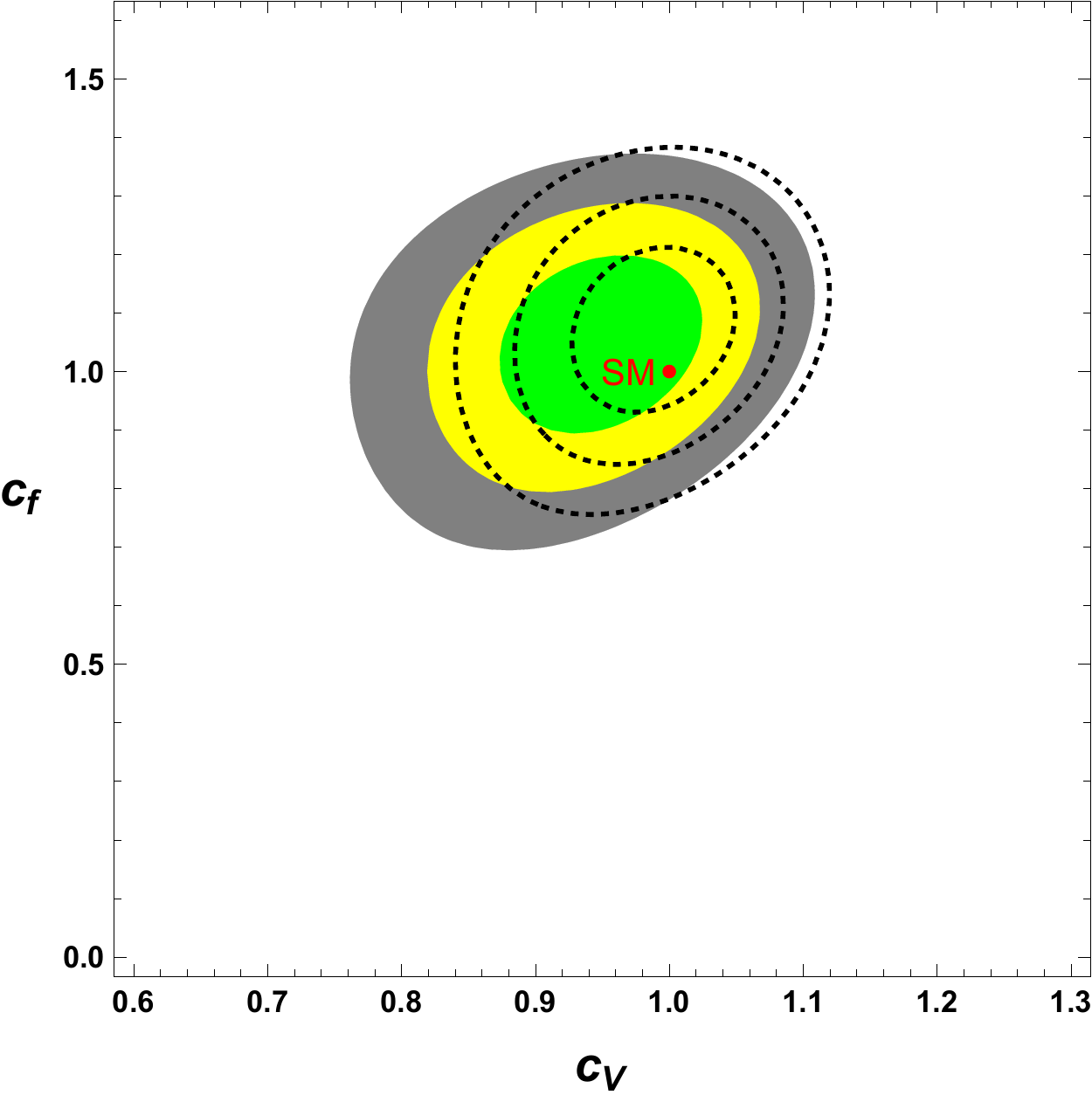}}
\put(230,0){\includegraphics[width=7.5cm]{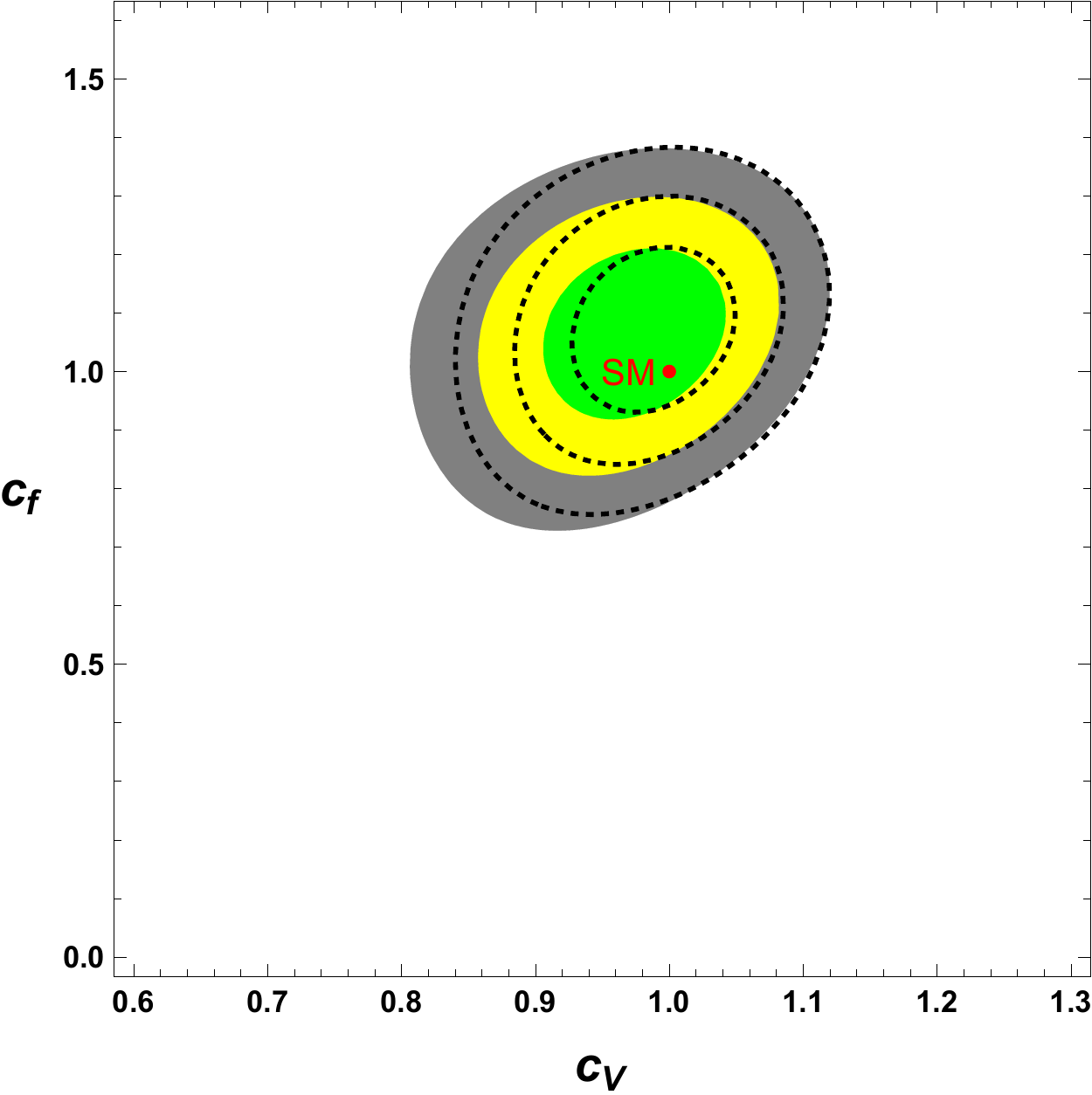}}
\put(40,60){{\color{blue} Bayesian marginalisation}}
\put(40,40){{\color{blue}  -- {\it Gaussian prior} --  [a] }}
\put(275,60){{\color{blue} Bayesian marginalisation}}
\put(275,40){{\color{blue} -- {\it Gaussian prior} -- [b] }}
\end{picture}
\caption{
The best-fit regions in the $c_V-c_f$ plane obtained from Bayesian marginalisation  and  Gaussian priors for the theoretical uncertainties. 
The $68\%$, $95\%$ and $99\%$ credible regions are represented respectively  by the green, yellow and grey domains. 
The [a] and [b] plots correspond, respectively, to the two characteristic correlation configurations described in Eq.~\eqref{Corr1Cons} and Eq.~\eqref{Corr2Cons}. 
The dashed contours illustrate the case without theoretical uncertainties.  The SM prediction is shown by the red point.
}\label{fig:CombMargG}
\end{figure}

Fig.~(\ref{fig:CombMargG}) illustrates the same case as in Fig.~(\ref{fig:CombMarg}) except that the final priors are now Gaussian,~\footnote{At this stage, we 
recall that the Gaussian priors are obtained from a combination of all the individual priors, 
while the  flat priors have just been chosen `by hand' to illustrate what happens for completely 
different distributions.} which leads to the marginalised 
Bayesian likelihood of Eq.~(\ref{eq:LpriorGAUSS}) and Eq.(\ref{eq:LpriorGAUSS_simp}). 
It appears that there is no substantial difference (neither in location, size nor shape of the best-fit regions) between these 
two figures. 
This illustrates the mild impact of the choice of the shape for the prior of the theoretical uncertainties.
We conclude that, with the present statistical uncertainties on Higgs data, 
the recurring question of the exact shape of the prior,~\footnote{Including 
the details of the form at the boundaries in case {\it e.g.} of a flat distribution.} in particular for the errors due to truncated perturbative expansions in QCD, is nearly irrelevant.  

However we should stress that this insensitivity to the  prior shape occurs because the experimental uncertainties of the current data are typically larger 
or of the same order as the theoretical ones. This situation is expected to change with the upcoming LHC runs, as the statistical uncertainties will decrease with the integrated luminosity.

\subsubsection{The nuisance parameters favoured by the data}

\begin{figure}[t]
\begin{picture}(400,180)
\put(100,20){\includegraphics[width=7.5cm]{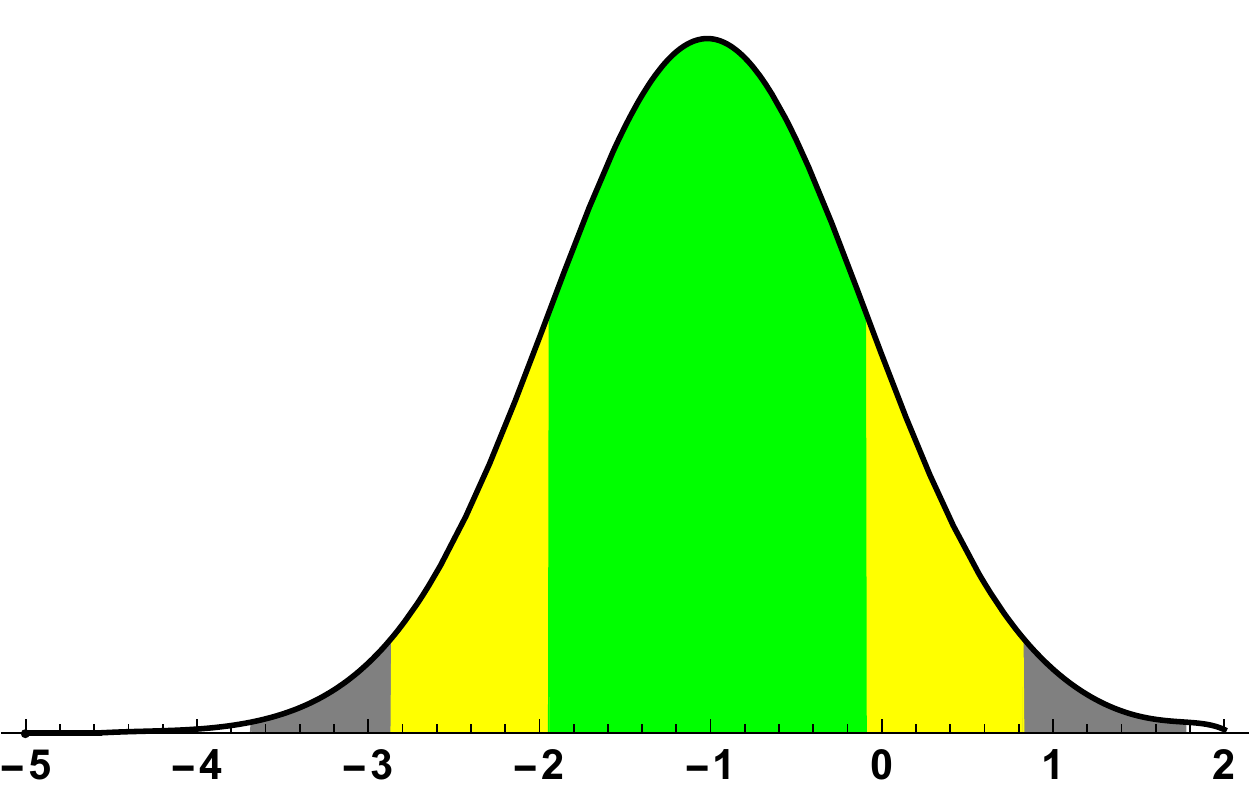}}
\put(125,120){{\color{blue} $p(\delta_{\rm ggF}|\mu_i^{\rm ex})$ 
}}
\put(201,10){$\delta_{\rm ggF}$}
\put(199,70){{\color{black} $68\%$~C.L.}}
\end{picture}
\caption{
The data-dominated  posterior $p (\delta_{\rm ggF}|\mu_i^{\rm ex})$ (Eq.~\eqref{eq:post_delta}).
The $68\%$, $95\%$ and $99\%$ credible domains are indicated respectively  by the green, yellow and grey  areas. 
}\label{fig:Lshape}
\end{figure}
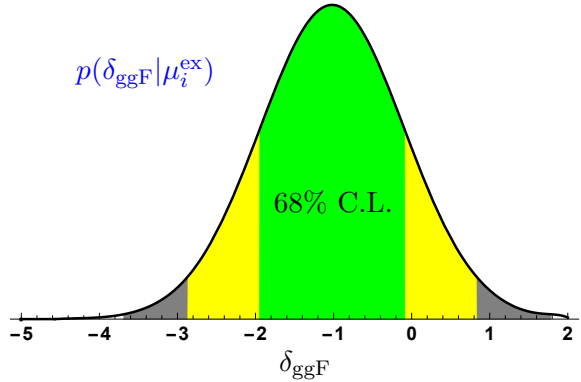

Let us now consider the posterior distribution for the theoretical uncertainties themselves, instead of the posterior for the parameters of interest.
Here we shall take the  priors associated with the theoretical uncertainties ($\pi^\sigma_X$) as flat and with an \textit{infinite} range.  For such choice of prior,  the information of the posterior is fully contained in the likelihood (second line in Eq.~\eqref{eq:post_delta}).
The interest of this data-dominated posterior is that it allows us to study exclusively the 
information that the sole Higgs data provide about the theoretical uncertainties, $\Delta_{i}^{\mu}$. 

We first consider the case with 
a single nuisance parameter $\delta_{\rm ggF}$ (\ie~the fully correlated case), given in Eq.~(\ref{Corr2}), and we present in Fig.~(\ref{fig:Lshape}) the data-dominated posterior for $\delta_{\rm ggF}$,   
\begin{eqnarray} \label{eq:post_delta}
&&  p(\delta_{\rm ggF}|\mu_i^{\rm ex})   =  \  \int dc_V \, dc_f \ \pi(c_V,c_f) \ \pi^\sigma_{\rm ggF}(\delta_{\rm ggF})  \ \times
\\
&& \exp   \left[{-\frac{1}{2}\sum_{i,j}\bigg(\mu_{i}^{\rm th}[c_V,c_f]-\mu_{i}^{\rm ex}(1\pm \delta_{\rm ggF}\Delta^\mu_i)\bigg)\, 
\mathcal{C}^{{\rm ex}\,-1}_{ij}\, \bigg(\mu_{j}^{\rm th}[c_V,c_f]-\mu^{\rm ex}_{j}(1\pm \delta_{\rm ggF} \Delta^\mu_j)\bigg) }\right] . 
\nonumber 
\end{eqnarray}
This posterior is obtained by integrating the likelihood of Eq.~(\ref{LpriorH}) (with $\pi^\mu$ given by Eq.~(\ref{Corr1Cons}))
 over all $\delta$'s but one, chosen to be $\delta_{\rm ggF}$, and marginalising with respect  to the  $c_V,c_f$ parameters  with $\pi(c_V,c_f)\propto 1$.

It appears in Fig.~(\ref{fig:Lshape}) that the posterior for $\delta_{\rm ggF}$ is centred on $\delta_{\rm ggF} \simeq -1$.~\footnote{For comparison, the 
maximum of $p (\delta_{\rm ggF},c_V=c_f=1|\mu_i^{\rm ex})$ is reached for $\delta_{\rm ggF} \simeq -0.7$.}
This means that for each  signal strength, the data typically favour a value   falling at $\pm1\sigma$  (\ie~at $\pm\Delta_{i}^{\mu}$)
from the nominal value $\mu_{i}^{\rm ex}$. 
In other words, for the correlation configuration of Eq.~(\ref{Corr2})
the  Higgs data provide a non-trivial indication that the magnitudes  of the theoretical errors  are reasonably well estimated. Indeed, the theoretical estimations predict the  $\mu_i^{\rm ex}$ 
 to lie  typically within the $1\sigma$ interval $ \pm \Delta_i^\mu$.

This compatibility suggests that the $\Delta^\mu_i$ uncertainties, whose estimations rely on quite ad hoc QCD scale variations  and on the arbitrariness in the choice  of PDF sets, are nevertheless quite robust.
On the other hand,  one  also notices in Fig.~(\ref{fig:Lshape}) that the credible intervals for $p(\delta_{\rm ggF}|\mu_{i}^{\rm ex})$ go beyond $-1$. This could be taken as an argument for slightly increasing the overall magnitude of the theoretical uncertainties (see next subsection).

The correlation configuration with two nuisance parameters, given by Eq.~(\ref{Corr1}), leads to larger preferred values for the nuisance parameters $\delta_{\rm ggF}\simeq-2$, $\delta_{\rm VBF}\simeq-5$.
We interpret these very large values as the fact that neglecting totally the correlation between the two nuisance parameters is an unrealistic hypothesis (as already described in Section~\ref{se:Corr}).  
As a matter of fact, if one restored the usual prior for the $\delta$'s (\ie~a prior with unit variance, $V[\delta]=1$), a hypothesis testing would show that the data favour the correlation configuration of Eq.~(\ref{Corr2})   with respect to the configuration of Eq.~(\ref{Corr1}).

\subsubsection{More conservative theoretical errors}

Throughout this paper, we have been observing that, among the various origins of theoretical uncertainty involved in the Higgs fit, some are of a nature (see Section~\ref{PDFsole}~-~\ref{BRerr}) which renders difficult
the {\it exact determination} of the associated $1\sigma$ interval. These are the truncation of the perturbative expansion for the QCD calculation of Higgs
rates translated into an arbitrary error range for the renormalisation/factorisation scale $\mu = \mu_{\rm R} = \mu_{\rm F}$ (affecting the production and
decay amplitudes as well as the $\alpha_s$ coupling constant), the choices made (on the statistical method, the number of
free parameters\dots) in the different PDF sets, and finally the  $m_b$ renormalisation scheme and EFT assumptions for the ggF mechanism.
These considerations can be taken as a motivation to adopt more conservative theoretical errors.

 Moreover, we have seen in the previous subsection (see Fig.~(\ref{fig:Lshape})) that  the data tend to prefer  theoretical uncertainties that are somewhat larger than the combined $1\sigma$ width $\Delta_i^\mu$  obtained in Section~\ref{se:marg}, see \textit{e.g.} the $68\%$~C.L. interval in Fig.~(\ref{fig:Lshape}).
Taking seriously this fact, it makes sense to perform the fits with a slight overall increase of the uncertainties. We suggest a rescaling \be\Delta_{i}^{\mu}\rightarrow 1.5\,\Delta_{i}^{\mu} \label{eq:Delta_rescale} \ee
as a reasonable estimation for a most conservative choice of theoretical uncertainties.
Notice that the rescaling of Eq.~\eqref{eq:Delta_rescale} 
is equivalent ({\it c.f.} Eq.~(\ref{LpriorH})) to rescale by $1.5$ the axis on Fig.~(\ref{fig:Lshape}). For example, the point $\delta_{\rm ggF}=-1$ becomes $\delta_{\rm ggF}=-1.5$.

\begin{figure}[t]
\begin{picture}(400,200)
\put(0,0){\includegraphics[width=7.5cm]{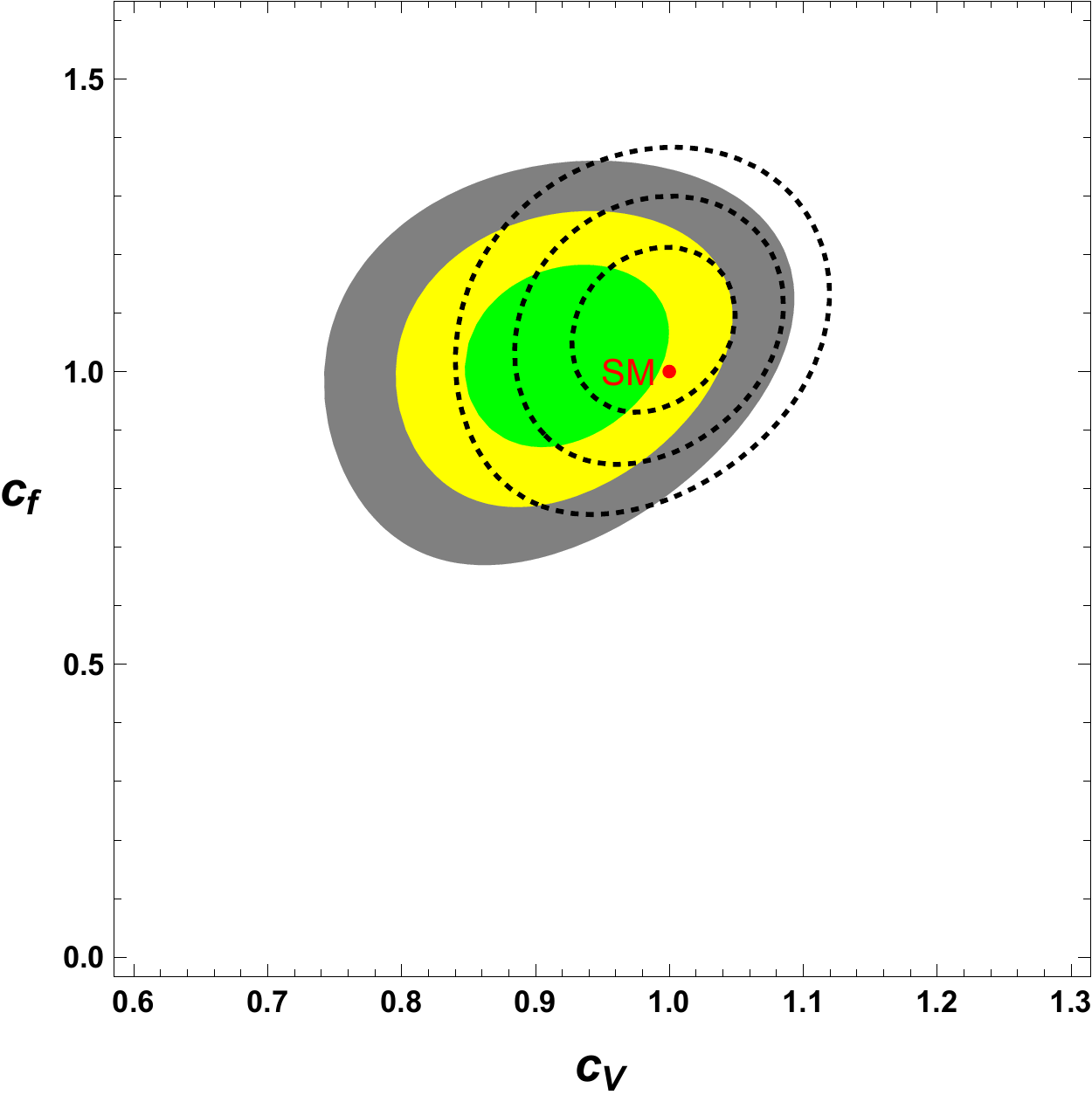}}
\put(230,0){\includegraphics[width=7.5cm]{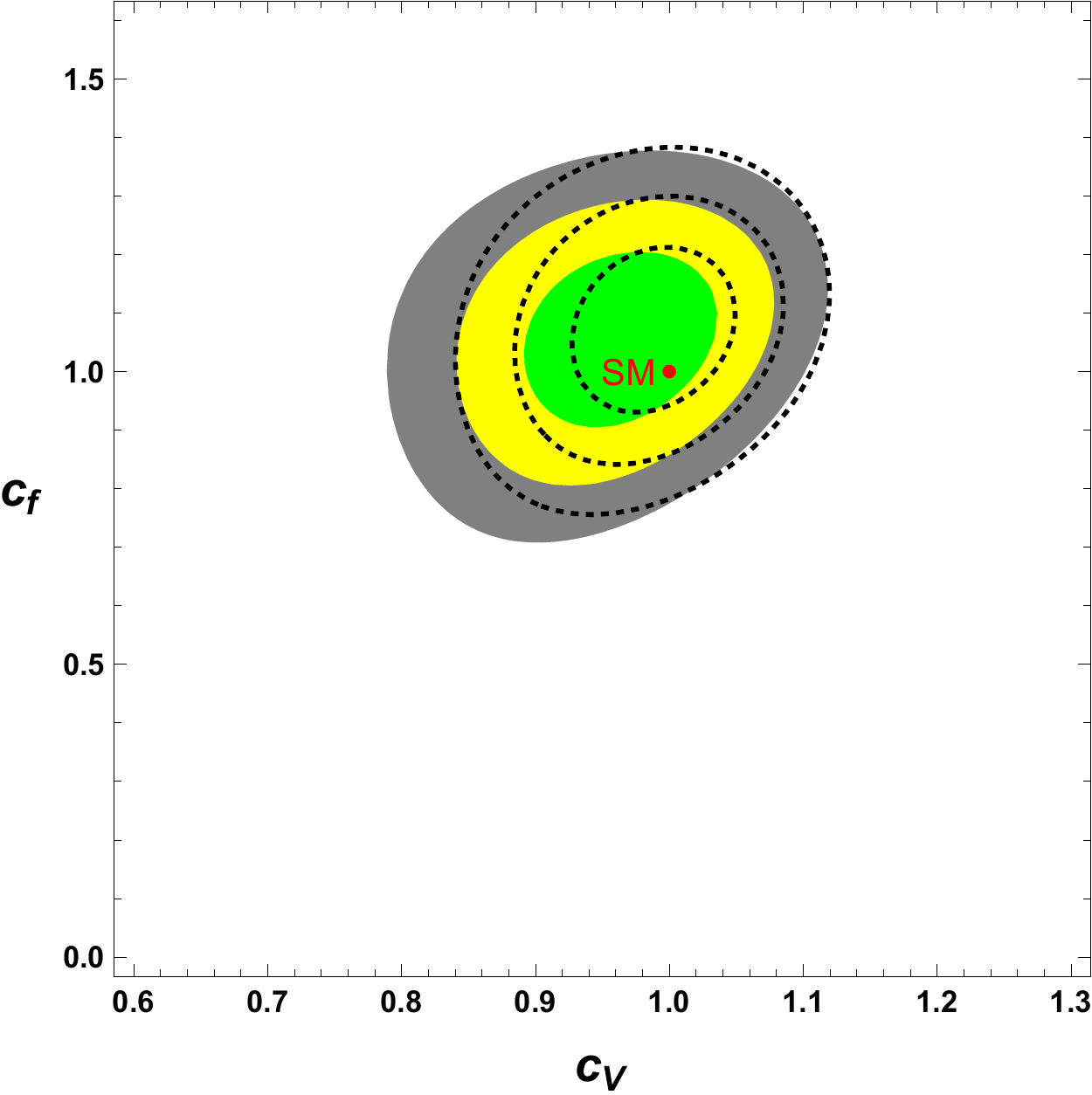}}
\put(40,60){{\color{blue} Bayesian marginalisation}}
\put(40,40){{\color{blue}  -- {\it flat prior} --  [a] }}
\put(150,40){{\color{blue}  $\Delta_{i}^{\mu} \times 1.5$ }}
\put(275,60){{\color{blue} Bayesian marginalisation}}
\put(275,40){{\color{blue} -- {\it flat prior} -- [b] }}
\put(380,40){{\color{blue}  $\Delta_{i}^{\mu} \times 1.5$ }}
\end{picture}
\caption{
The best-fit regions in the $c_V-c_f$ plane obtained from Bayesian marginalisation  and flat priors for the theoretical uncertainties. 
The $68\%$, $95\%$ and $99\%$ credible regions are represented respectively  by the green, yellow and grey domains. 
The [a] and [b] plots correspond, respectively, to the characteristic correlation configurations  described in Eq.~\eqref{Corr1Cons} and Eq.~\eqref{Corr2Cons}. 
The dashed contours illustrate the case without theoretical uncertainties.  The SM prediction is shown by the red point.
 The difference with Fig.~(\ref{fig:CombMarg}) is the enhancement of the uncertainties, accordingly to $\Delta_{i}^{\mu}\rightarrow\Delta_{i}^{\mu} \times 1.5$. 
}\label{fig:Conserv}
\end{figure}
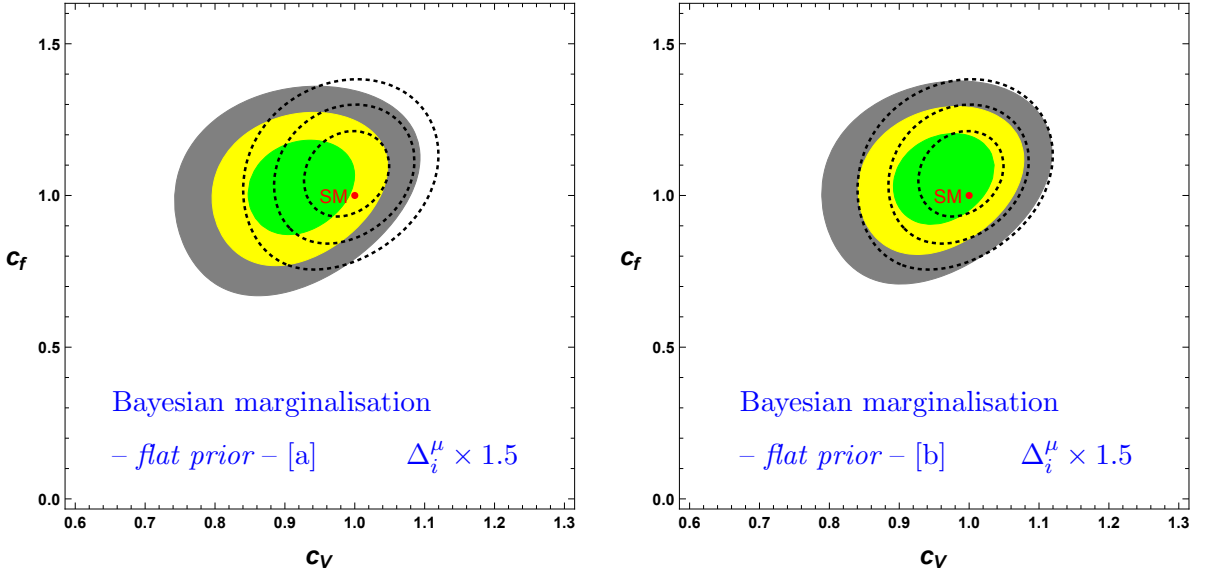

The best-fit regions with $\Delta_{i}^{\mu} \times 1.5$ are shown in Fig.~(\ref{fig:Conserv}) for the two correlation
configu\-rations and considering the flat prior case (Eq.~\eqref{LpriorFLAT}), keeping in mind that with the current Higgs data, the final prior shape does not affect significantly those best-fit domains.  
 The impact of the increase of the theoretical uncertainties (Eq.~\eqref{eq:Delta_rescale}) on the fit of the current Higgs data can be seen by comparing Fig.~(\ref{fig:CombMarg}) and Fig.~(\ref{fig:Conserv}). It turns out that the shift of the preferred regions  with respect to the case without theoretical errors   gets slightly accentuated. In the correlation configuration of Eq.~\eqref{Corr1Cons}, \ie~with two independent $\delta_X$, it even appears (see Fig.~(\ref{fig:Conserv})[left]) that the SM point moves just outside the $68\%$ C.L. region. 
 
The increase of this shift can be understood by recalling that rescaling the $\Delta^\mu_i$ is equivalent to increase the width of the $\delta_X$ prior. It is then clear that 
 more possibilities are opened for the preferred values of $\delta_X$. 
It turns out that these preferred values move further away from zero, which  induces a more pronounced shift of the best-fit regions.

Even though these effects are not statistically significant for the current Higgs data, we stress that the impact of the theoretical errors will increase  while more data will be accumulated at the LHC. 
The ambiguity existing in the theoretical errors estimation deserves thus to be taken into account. For future LHC phenomenological studies, we suggest to take into account, in the same way as proposed in this subsection, the impact on the fits from the lack of knowledge in theoretical errors.

\section{Biasing the Higgs likelihood}  \label{se:bias}

The principle of bias has been presented in Section~\ref{sec:BiasPrin}. To have a self-consistent section, we recall here the basics of a ``biasing'' procedure. We distinguish two realisations of the bias principle: the \textit{extremal bias} and the \textit{envelope method}.

The method of extremal biasing consists in drawing the best-fit regions for the  para\-meters of interest  for extreme fixed values of the theoretical errors. By the word `extreme', we mean that we set the nuisance parameters $\delta$ at $\pm 1$ (corresponding to one-standard 
deviations with our conventions) in order to obtain a strong impact on the fit. 
In our Higgs fit, the theoretical uncertainties affect the signal strengths $\mu_{i}^{\rm ex}$, which in turn modify the preferred value of $\mu^{\rm th}_{i}(c_f,c_V)$ and thus the best-fit regions of $c_V,c_f$.
Note that the choice of extreme values $\delta\pm 1$ can be seen as natural, and for that reason will be used in our numerical results, but 
strictly speaking remains only a choice with a certain degree of arbitrariness.

The envelope method corresponds formally to the continuous version of this extremal biasing. Loosely speaking, this is what one obtains if one does the fit for each fixed value of the nuisance parameters between the extreme values $\delta=\pm1$. One expects typically a deformed contour somehow interpolating between the regions of extremal biasing.
For a more formal and unified description of these biasing methods, see Section~\ref{sec:BiasPrin}.

What are the motivations for choosing the marginalisation or the  bias approaches (extremal bias or envelope method) in the Higgs fits? 
The lack of knowledge on the shape of the prior associated to the main QCD uncertainties discussed in Section~\ref{PDFsca} encourages one to apply a bias method,  which does not rely on the prior shape -- in contrast with the marginalisation. \\
Besides, the bias is  more conservative. 
Indeed, while in the marginalisation the best-fit domain corresponds roughly to nuisance parameters centered around a preferred $\delta_X$ value, 
in the bias methods $\delta_X$ rather spans  by construction its  $[-1,1]$ interval without favouring any value. Hence, generally speaking (and this is the case for the Higgs fit), the best-fit regions in the space of the parameters of interest 
 obtained through the bias methods are wider than the ones from marginalising.
\\
In addition, the envelope method allows one to see at a glance the whole best-fit domain in the $c_{V}-c_f$ plane  spanned by varying the nuisance parameters 
inside their entire $[-1,1]$ intervals. The price to pay here is maybe a heavier technical approach than in the margina\-lisation
procedure: compare  the margina\-lisation definitions in Eqs.~\eqref{eq:marg},\eqref{eq:marg_freq} with the biasing definitions in Eqs.~\eqref{eq:bias_bayes},\eqref{eq:bias_freq} (see for example  Eq.~(\ref{LpriorH}) and Eq.~(\ref{eq:bias_bayesAPP}) 
for the application to the Higgs likelihood). It is clear that more operations (either integrations or maximisations) are needed for the envelope method.

\subsection{Combining the uncertainties}  \label{se:BiasComb}

The starting point is the likelihood~(\ref{pdfGAUSS}), and then (\ref{eq:L0}).
Applying the Eqs.~(\ref{eq:comb_bias_0})-(\ref{eq:comb_bias_+1})-(\ref{eq:comb_bias_-1})-(\ref{eq:comb_bias_3})
together with the definition of Eq.~(\ref{DBstat}) and,
\begin{eqnarray}  
\Delta^{\rm P}_{X,i}   
\ \hat = \ 
\frac{\epsilon_{X}^{i} \sigma_X^\textrm{SM}}{\sum_{X'}\epsilon^{i}_{\rm X'} \sigma^{\rm SM}_{\rm X'}} 
\ \left ( \Delta_X^{\rm amp}+\Delta_X^{\rm PDF+\alpha_s} \right )
   \,,  \label{DeltaBS}
\end{eqnarray}
which is a new compact notation comparable to Eq.~(\ref{DmuCt}), we obtain the likelihood depending on a unique nuisance parameter, $\delta_b$,
\be
L_{\rm bias}  (\delta_b)
= \exp  \left[{-\frac{1}{2}\sum_{i,j}\bigg(\mu_{i}^{\rm th}[c_V,c_f]-\mu_{i}^{\rm ex}(1+\delta_b\Delta^b_i)\bigg)\, 
\mathcal{C}^{{\rm ex}\,-1}_{ij}\, \bigg(\mu_{j}^{\rm th}[c_V,c_f]-\mu^{\rm ex}_{j}(1+\delta_b\Delta^b_j)\bigg) }\right]
\label{eq:pdfBS}
\ee
relying on the combined error,
\begin{eqnarray}
\Delta_i^{b} & =  &  \big \vert  \Delta^{\rm P}_{\rm ggF,i} -  \Delta^{\rm P}_{\rm ttH,i}  \big \vert  +  \Delta^{\rm P}_{\rm VBF,i} +  \Delta^{\rm P}_{\rm WH,i} +  \Delta^{\rm P}_{\rm ZH,i}  
+ \sum_{Y,a} (  \Delta_{Y,i}^{a} +  \Delta_{Y,i} )  \ ,
\label{DeltaBS1}
\end{eqnarray}
or,
\begin{eqnarray}
\Delta_i^{b} & =  &  \big \vert  \Delta^{\rm P}_{\rm ggF,i} - ( \Delta^{\rm P}_{\rm ttH,i}   +  \Delta^{\rm P}_{\rm VBF,i} +  \Delta^{\rm P}_{\rm WH,i} +  \Delta^{\rm P}_{\rm ZH,i} ) \big \vert 
+ \sum_{Y,a} (  \Delta_{Y,i}^{a} +  \Delta_{Y,i} )  \ ,
\label{DeltaBS2}
\end{eqnarray}
for the two  configurations of correlations defined in Eq.~(\ref{Corr1})-(\ref{Corr2}), respectively.

The combinations of the errors on the partial decay widths are dictated by the fact that their nuisance parameters are either independent (among them and from the nuisance parameters at the
production level) or taken $100\%$ correlated to each other, as discussed in Section~\ref{BRerr}.
\\
In Eq.~(\ref{DeltaBS}), $\Delta_X^{\rm amp}$ is either equal to $\Delta_X^{\rm scale}$ (see Section~\ref{EFTsca}) or taken as
$\Delta_{\rm ggF}^{\rm amp} \ = \ \Delta^{\rm scale}_{\rm ggF} + \Delta_{\rm ggF}^{\rm Q,V}$, for the ggF channel (instead of Eq.~(\ref{eq:ggFamp}))
with now, $\Delta_{\rm ggF}^{\rm Q,V}\simeq 9\%$, from the linear sum of the three errors originating from EFT assumptions and $m_b$ scheme 
dependence~\cite{bottomEFT}.  These linear summations are all motivated by the fact that these errors are independent.
 \\
The $\Delta^{\rm PDF+\alpha_s}_X$ uncertainty entering Eq.~(\ref{DeltaBS}) is obtained from Ref.~\cite{LHCHWG3,0905.3531} using an ``envelope method'', which corresponds exactly to the combinations in the bias approach presented in Section~\ref{se:comb_bias}. Indeed, this combination is equivalent to a linear sum of the individual errors $\Delta^{\rm set}_X$, $\Delta^{\rm data}_X$ and $\Delta^{\alpha_s}_X$, which are independent (\textit{c.f.} Section~\ref{PDFsole}).
Finally, the linear sum in Eq.~(\ref{DeltaBS}) is justified by the independence of the errors $\Delta_X^{\rm amp}$ and $\Delta^{\rm PDF+\alpha_s}_X$. 

The 1$\sigma$-errors  ($\Delta$'s) are taken to be exactly the symmetrized errors provided by the LHCHWG~\cite{LHCHWGweb,LHCHWG1,LHCHWG3}
in order to be conservative (similar discussion as in Sections~\ref{PDFsca} and \ref{BRerr}).
These errors are consistent with the previous marginalisation framework, so that the results from bias and marginalisation can readily be compared.

\subsection{The Bayesian approach}  \label{se:B3}

\subsubsection{Extremal bias}   \label{se:BayExtr}

According to Section~\ref{sec:BiasPrin}, the extremal bias within the Bayesian framework consists in deriving the best-fit regions in the $c_V-c_f$ plane 
for two fixed values of the nuisance parameters, $\delta_b= \pm 1$, using the likelihood $L_{\rm bias} (\delta_b)$ of Eq.~(\ref{eq:pdfBS}). 
Recall that in the Bayesian case, the best-fit regions are computed by integrating the posterior density probability, according to Eqs.~(\ref{eq:PostBay})-(\ref{eq:bayes_contours})-(\ref{eq:CLBay}).
 The priors ($\pi(\theta)$) for the parameters of interest 
(here $\theta \equiv c_V,c_f$) entering Eq.~(\ref{eq:PostBay}) are taken flat, \ie~$\pi(c_{V,f})\propto 1$. 

Note that, if the two extreme regions have an overlap, one cannot display them together consistently. Instead, one  has to follow the rigorous definition of Eq.~\eqref{eq:bias_bayes}, using a discrete domain $\mathcal{D}=\{-1,1\}$. This equation dictates to use the sum of the posteriors at $\delta_b=-1$ and $\delta_b=1$, with each posterior  separately  normalised
 by its integral over the
$c_V-c_f$ plane.

\subsubsection{Envelope method}  \label{sec:DEMbis}

The envelope method corresponds to letting vary continuously $\delta_b$ within $[-1,1]$, \ie~this is the continuous version of the extremal bias, as discussed in Section~\ref{sec:BiasPrin}.
The corresponding likelihood is
\be
\bar L_{\rm B} (c_f,c_V)=  \int^{1}_{-1} d\delta_b \left[ \frac{L_{\rm bias}(c_f,c_V,\delta_b)}{ \int dc_f  \int dc_V  \  L_{\rm bias}(c_f,c_V,\delta_b) }\right] \ . \label{eq:bias_bayesAPP}
\ee
This likelihood is derived by applying Eq.~(\ref{eq:bias_bayes}) with the likelihood $L_{\rm bias}  (c_V,c_f,\delta_b)$ from Eq.~(\ref{eq:pdfBS}).
The best-fit regions are obtained through  the standard procedure of Eqs.~(\ref{eq:PostBay})-(\ref{eq:bayes_contours})-(\ref{eq:CLBay}).  Again, we  take the priors for the parameter of interest to be flat, $\pi(c_{V,f})\propto 1$.

\subsection{The frequentist approach}  \label{se:F3}

\subsubsection{Extremal bias}  \label{se:FreqExtr}

For the extremal bias in the frequentist framework (see Section~\ref{sec:BiasPrin}), one uses again the likelihood $L_{\rm bias} (\delta_b)$ (Eq.~\eqref{eq:pdfBS}), with $\delta_b$ fixed at the two extreme values $\delta_b=\pm1$.
 In practice, in order to draw the best-fit regions in $c_V-c_f$, one can define a $\chi$-squared function difference
\be
\Delta  \chi^2 (c_f,c_V,\delta_b)= \chi^2(c_f,c_V,\delta_b) - \chi^2  (\hat c_f,\hat c_V,\delta_b)
\ , \  \ 
\chi^2 (c_f,c_V,\delta_b)  = - 2 \log [L_{\rm bias}  (\delta_b)]  \,, 
\label{eq:FreExt}
\ee
as follows from Eq.~(\ref{eq:freq_contours}).
Remind that $\chi^2  (\hat c_f,\hat c_V,\delta_b)$ stands for the minimum of $\chi^2$
with respect to $c_f, c_V$ for a given $\delta_b$. The best-fit regions are obtained by drawing the contour levels of $\Delta  \chi^2$ set at the values given in  Eq.~(\ref{eq:Chi2val}).
Once more, the prior for the parameters of interest entering in Eq.~(\ref{eq:freq_contours}) are taken flat, $\pi(c_{V,f})\propto 1$.

If the two extreme regions  overlap, the same remark as in the Bayesian case holds.
To display consistently the two regions together,  one  has to follow the rigorous definition of Eq.~\eqref{eq:bias_freq}, using a discrete domain $\mathcal{D}=\{-1,1\}$. 
This equation dictates to use the minimum of the two $\Delta \chi^2$,  
 \ie~$\textrm{min}_{\delta_b\in\{-1,1\}}[\Delta  \chi^2 (c_f,c_V,\delta_b)]$.

\subsubsection{Envelope method}  \label{sec:DEM}

For the envelope method in the frequentist case, one can proceed with the $\chi^2$ introduced in Eq.~\eqref{eq:FreExt}  and define
\be
\bar \chi^2(c_f,c_V) = \min\limits_{\delta_b\in[-1,1]}      \bigg[ \chi^2 (c_f,c_V,\delta_b) 
- \chi^2  (\hat c_f,\hat c_V,\delta_b) \bigg] 
\ , 
\label{DEMmin}
\ee
 according to the general definition  of Eq.~(\ref{eq:bias_freq}). This equation is the frequentist analog of Eq.~\eqref{eq:bias_bayesAPP}.
 In order to draw the best-fit regions in the $c_V-c_f$ plane, one should then define
\begin{eqnarray}
\Delta \bar  \chi^2(c_f,c_V) =  \bar \chi^2 (c_f,c_V) 
- \bar \chi^2  (\hat c_f,\hat c_V)  \ .
\label{eq:barDelta}
\end{eqnarray}
The best-fit regions are obtained by drawing the contour levels of $\Delta \bar \chi^2$ set at the values given in  Eq.~(\ref{eq:Chi2val}).
Again, the prior for the $c_V,c_f$ parameters entering in Eq.~(\ref{eq:freq_contours}) are taken flat, $\pi(c_{V,f})\propto 1$.

Let us finally recall the parallel between Eq.~(\ref{eq:bias_bayesAPP}) and Eq.~(\ref{DEMmin}).
 As first explained in Section~\ref{sec:BiasPrin}, the subtracted term in Eq.~(\ref{DEMmin}) is the frequentist analogy of the ratio  over $\int dc_f  dc_V\,  L_{\rm bias}(c_f,c_V,\delta_b)$ in Eq.~(\ref{eq:bias_bayesAPP}). In both cases, the effect of this term is to remove the contribution of $\delta_b$ to goodness-of-fit (which avoids favouring specific values of $\delta_b$).
Both formulas are analog up to exchanging integration over $\delta_b$  with minimisation over $\delta_b$.
 The fact that the integration/minimisation over $\delta_b$ is performed on the whole range $[-1,1]$,  rather than on the discrete domain $\{-1,1\}$, leads to an envelope  in the $c_f-c_V$ plane, instead of two distinct domains as in the extremal bias.

\subsection{Numerical results}  \label{PlotsBias}

In this section, we apply both the frequentist and Bayesian versions of the bias method  to the Higgs likelihood. We stress that the Higgs likelihood $L_{\rm bias}  (\delta_b)$ is exactly the same in the two statistical frameworks, so that the discrepancies observed among the plots originate solely from the different statistical treatments. 
These two treatments differ in their definition of the best-fit regions (see Section~\ref{se:STATbasics}) and their realisation of the bias principle (see Eqs.~\eqref{eq:bias_bayes},~\eqref{eq:bias_freq}).

\subsubsection{Extremal bias}

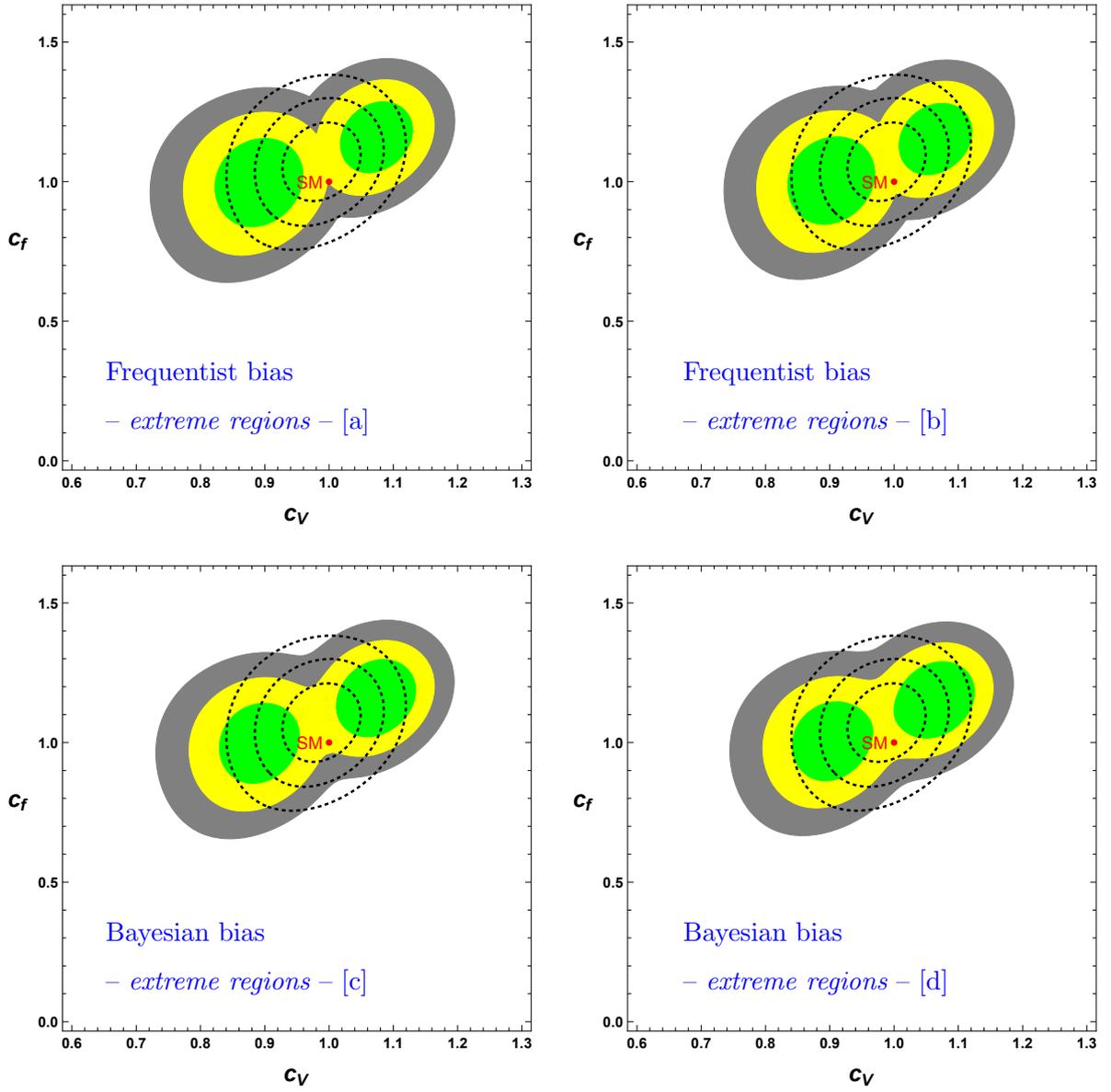
\begin{figure}[t]
\begin{picture}(400,450)
\put(0,0){\includegraphics[width=7.5cm]{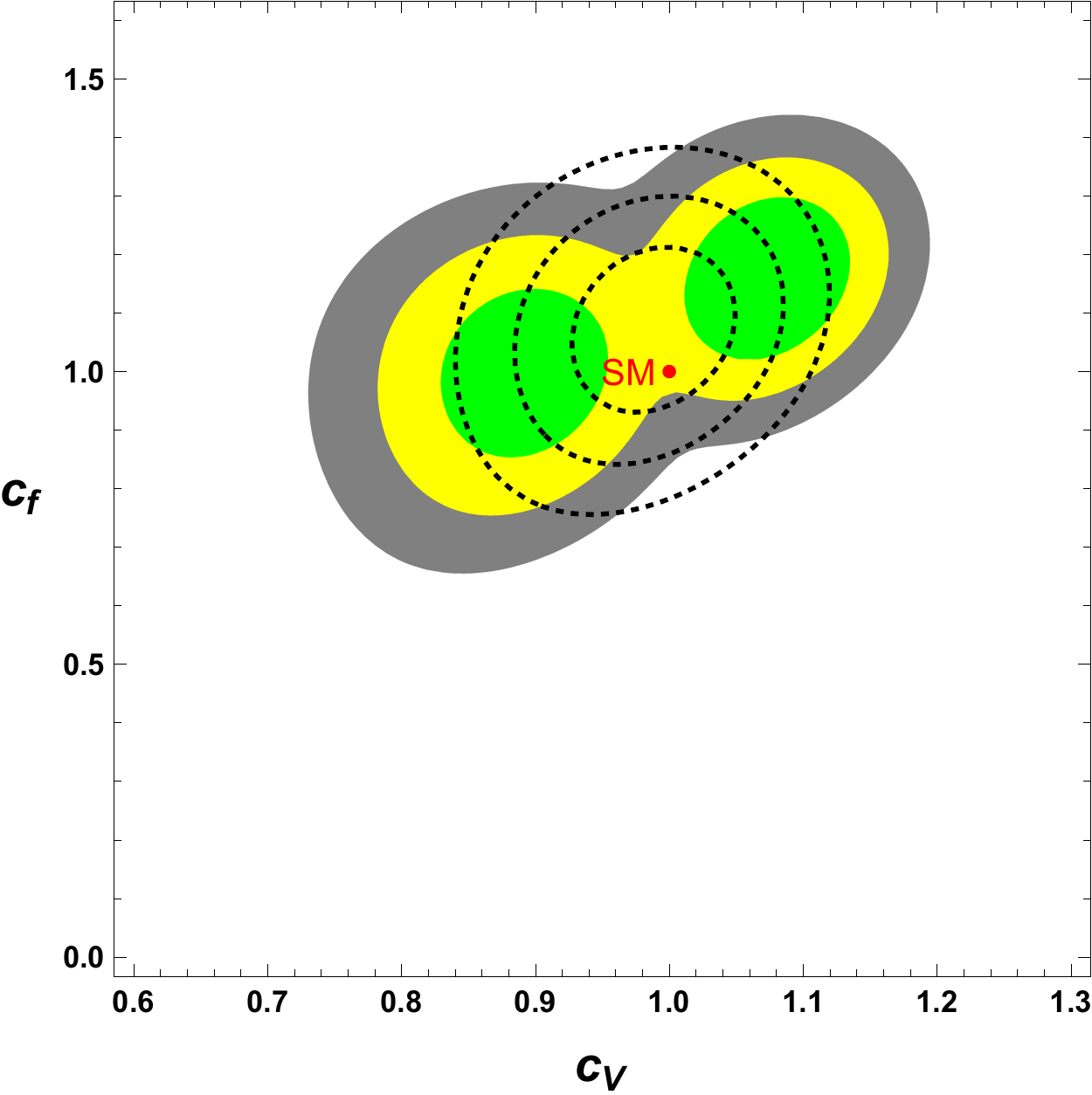}}
\put(230,0){\includegraphics[width=7.5cm]{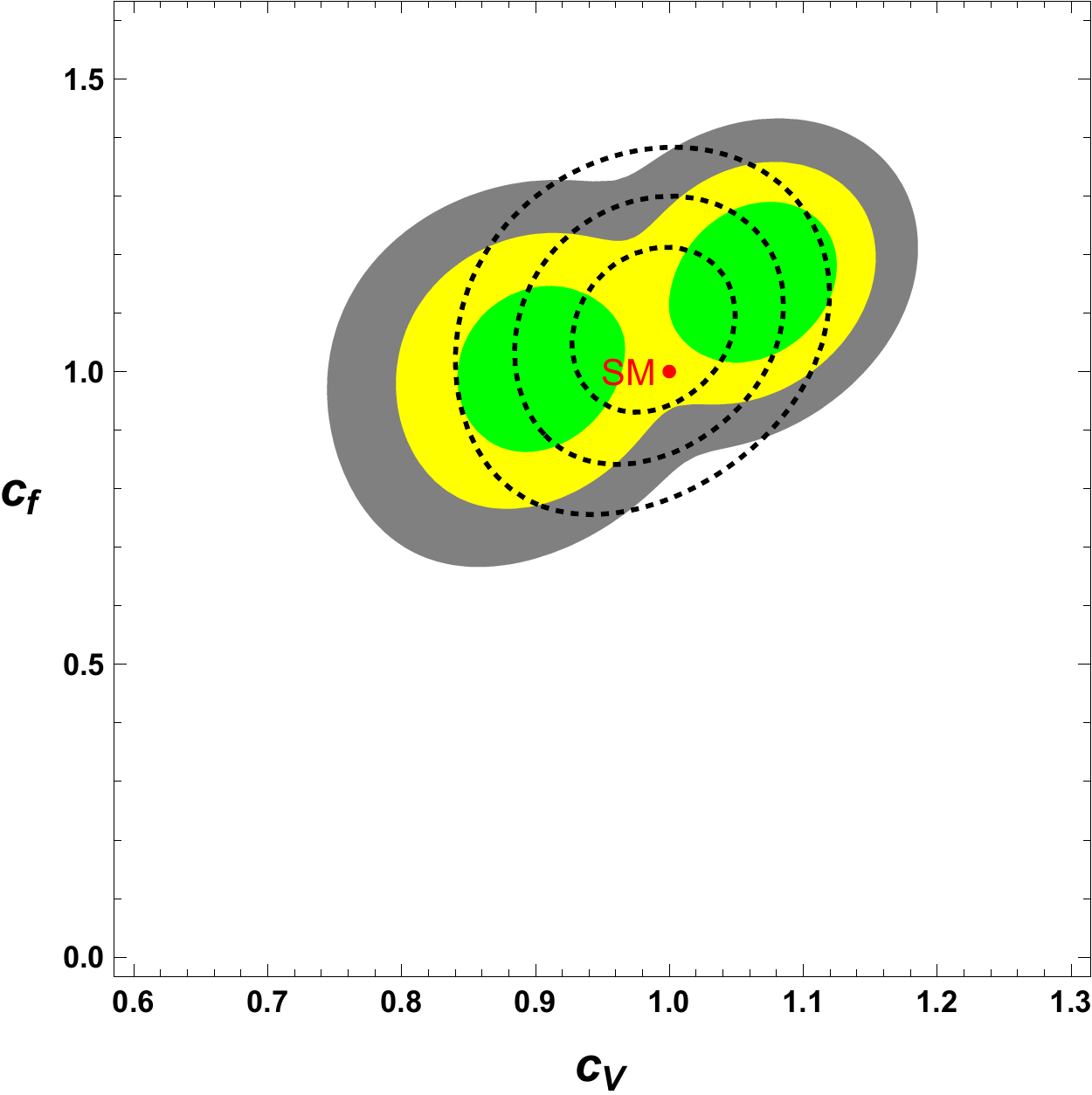}}
\put(0,230){\includegraphics[width=7.5cm]{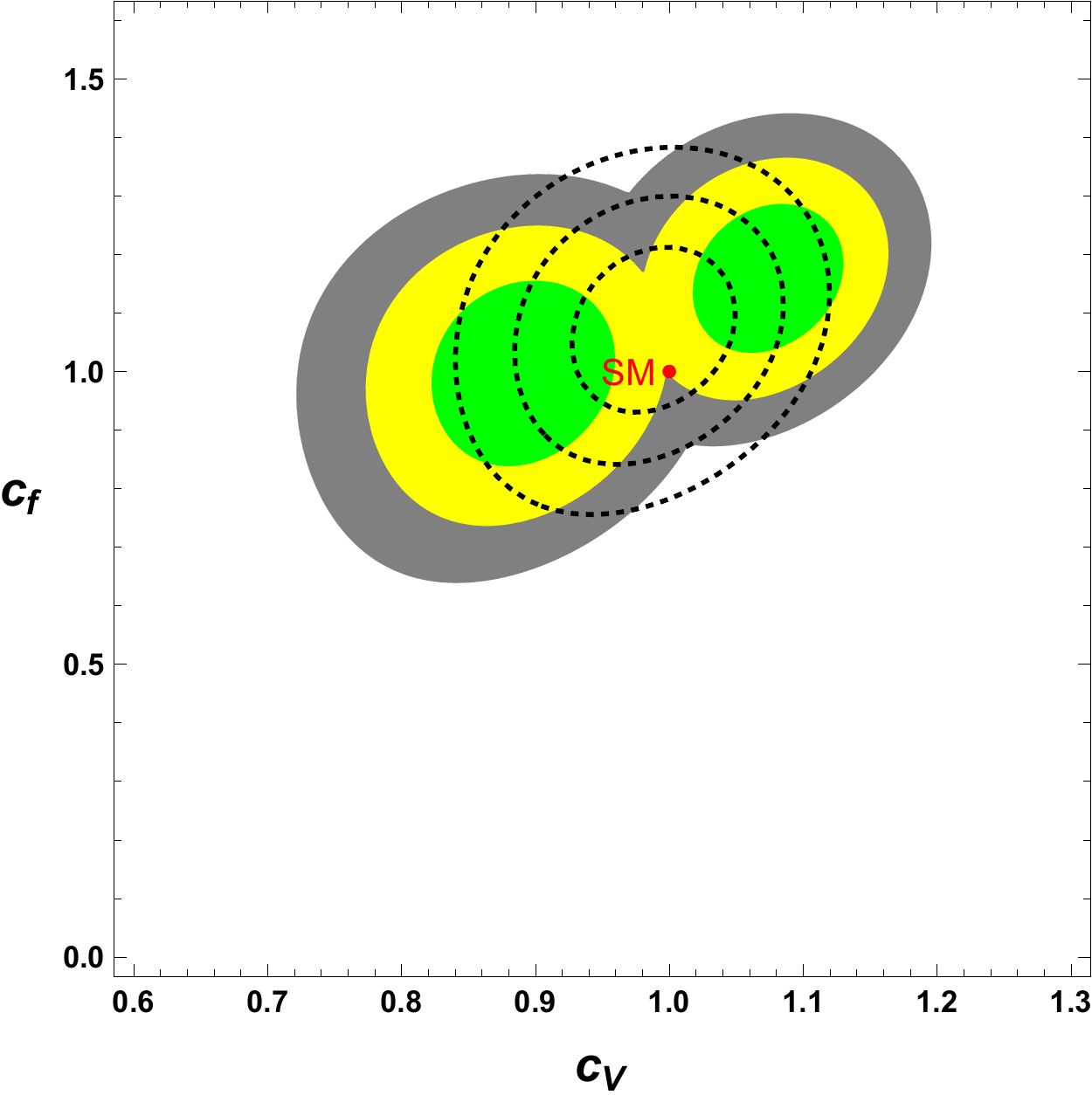}}
\put(230,230){\includegraphics[width=7.5cm]{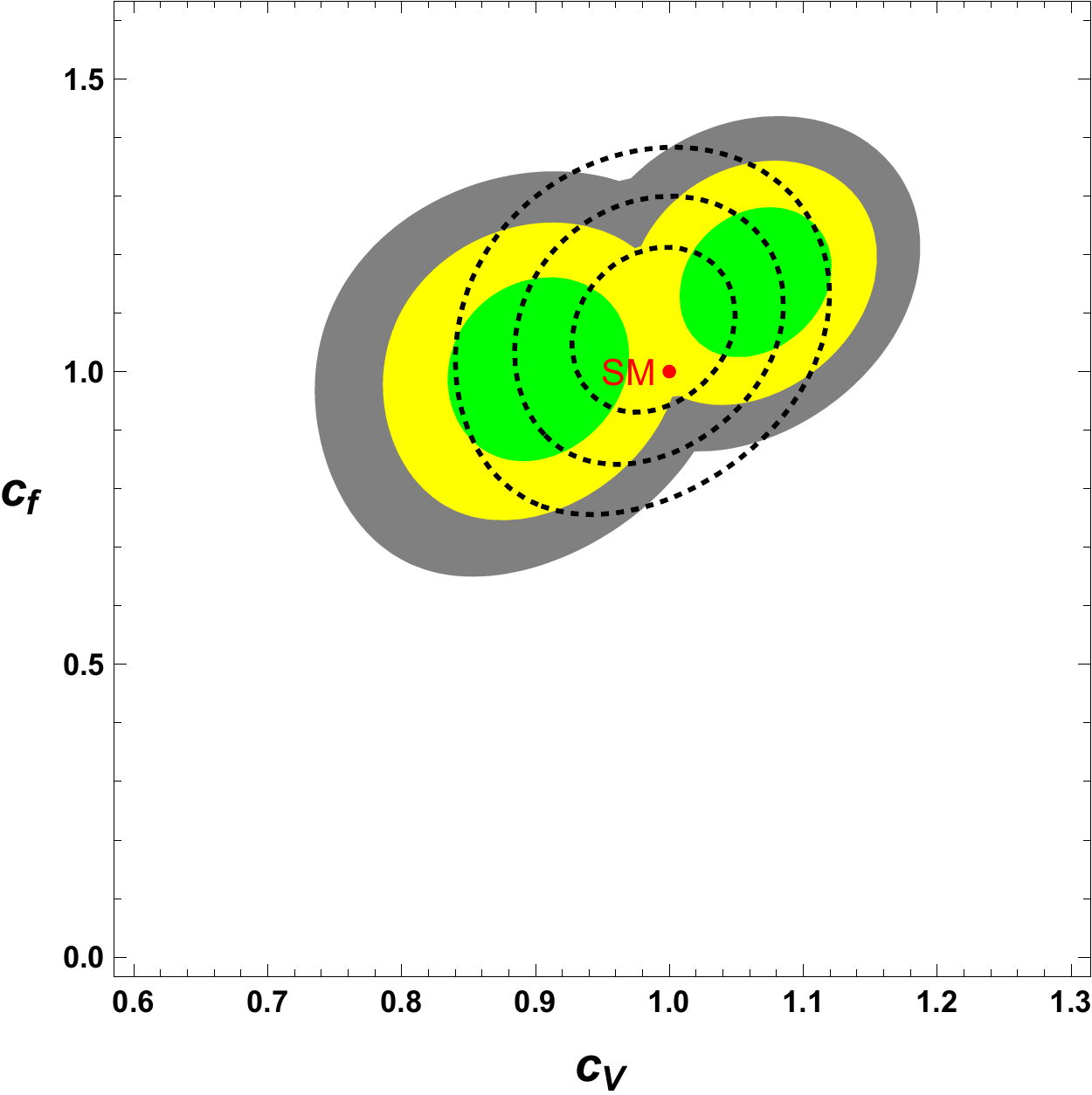}}
\put(40,60){{\color{blue} Bayesian bias}}
\put(40,40){{\color{blue}  -- {\it extreme regions} --  [c] }}
\put(40,290){{\color{blue} Frequentist bias}}
\put(40,270){{\color{blue}  -- {\it extreme regions} --  [a] }}
\put(275,60){{\color{blue} Bayesian bias}}
\put(275,40){{\color{blue} -- {\it extreme regions} -- [d] }}
\put(275,290){{\color{blue} Frequentist bias}}
\put(275,270){{\color{blue} -- {\it extreme regions} -- [b] }}
\end{picture} 
\caption{
The best-fit regions in the $c_V-c_f$ plane obtained through an extremal bias. 
The $68\%$, $95\%$ and $99\%$ confidence regions are represented respectively by the green, yellow and grey domains. 
The upper plots illustrate the frequentist approach whereas the two lower ones show the Bayesian approach. 
The [a], [c] and [b], [d] plots correspond, respectively, to the characteristic correlation configurations described in 
Eq.~\eqref{DeltaBS1} and Eq.~\eqref{DeltaBS2}.
The dashed contours illustrate the case without theoretical uncertainties.  The SM prediction is shown by the red point.
}\label{fig:CombBias}
\end{figure}

In Fig.~(\ref{fig:CombBias}), we present the best-fit regions obtained through the Bayesian and frequentist bias methods, respectively described in Sections~\ref{se:BayExtr} and~\ref{se:FreqExtr}. The likelihood, $L_{\rm bias}  (\delta_b)$ of Eq.~(\ref{eq:pdfBS}), is used together with one of the two combined 
errors~(\ref{DeltaBS1})-(\ref{DeltaBS2}) depending on which correlation configuration is considered (Eq.~(\ref{Corr1}) or Eq.(\ref{Corr2}) respectively). 
\\
The left and right pannels of Fig.~(\ref{fig:CombBias}) correspond  to the two correlation configurations surrounding the case with realistic correlations. It turns out that the best-fit regions obtained in
these two extreme correlation configurations have  only mild differences.
\\
Now, compare the  two upper plots and lower plots of Fig.~(\ref{fig:CombBias}), corresponding respectively to the  frequentist and Bayesian treatments.  A small difference appears at the junction of the two set of regions, coming from the different realisation of the bias principle in the two statistical frameworks. Besides, the frequentist best-fit regions are slightly larger than the Bayesian ones, due to the non-equivalent  definitions of the Bayesian  and frequentist contours.
Overall, there is a strong resemblance between  the Bayesian and frequentist results. This 
reflects the weak impact of choosing the Bayesian or frequentist procedure for the extremal bias.

Let us now compare the lower plots of Fig.~(\ref{fig:CombBias}) with the previous Bayesian margina\-lisation plots obtained in Fig.~(\ref{fig:CombMarg}) -- considering of course respectively
the two correlation configurations used in the left and right plots. 
One  can  clearly see that the best-fit regions~\footnote{Notice that these best-fit regions 
 include essentially the two extreme sub-domains corresponding to $\delta_b=\pm1$. 
} obtained from the extremal bias are larger than the ones obtained through marginalisation. This is because the regions in Fig.~(\ref{fig:CombMarg}), derived by marginalising, 
correspond somehow to fix the nuisance parameters to their values favoured by the fit. For the present Higgs fits, it turns out that these preferred values  are close to $\delta \approx -1$. 
Hence, the regions from the extremal bias (Fig.~(\ref{fig:CombBias}))
 being obtained for $\delta_b = \pm 1$ (lower left set is for $\delta_b = - 1$~\footnote{The dependence of the best-fit region location 
on the nuisance parameter is induced by the dependence of the likelihood~(\ref{eq:pdfBS}) on, $\mu_{i}^{\rm ex} [1+\delta_b \Delta_{i}^{b}]$.}), they
clearly cover more space in the $c_V - c_f$ plane than the domains in Fig.~(\ref{fig:CombMarg}).

\subsubsection{Envelope method} 

\begin{figure}[t]
\begin{picture}(400,450)
\put(0,0){\includegraphics[width=7.5cm]{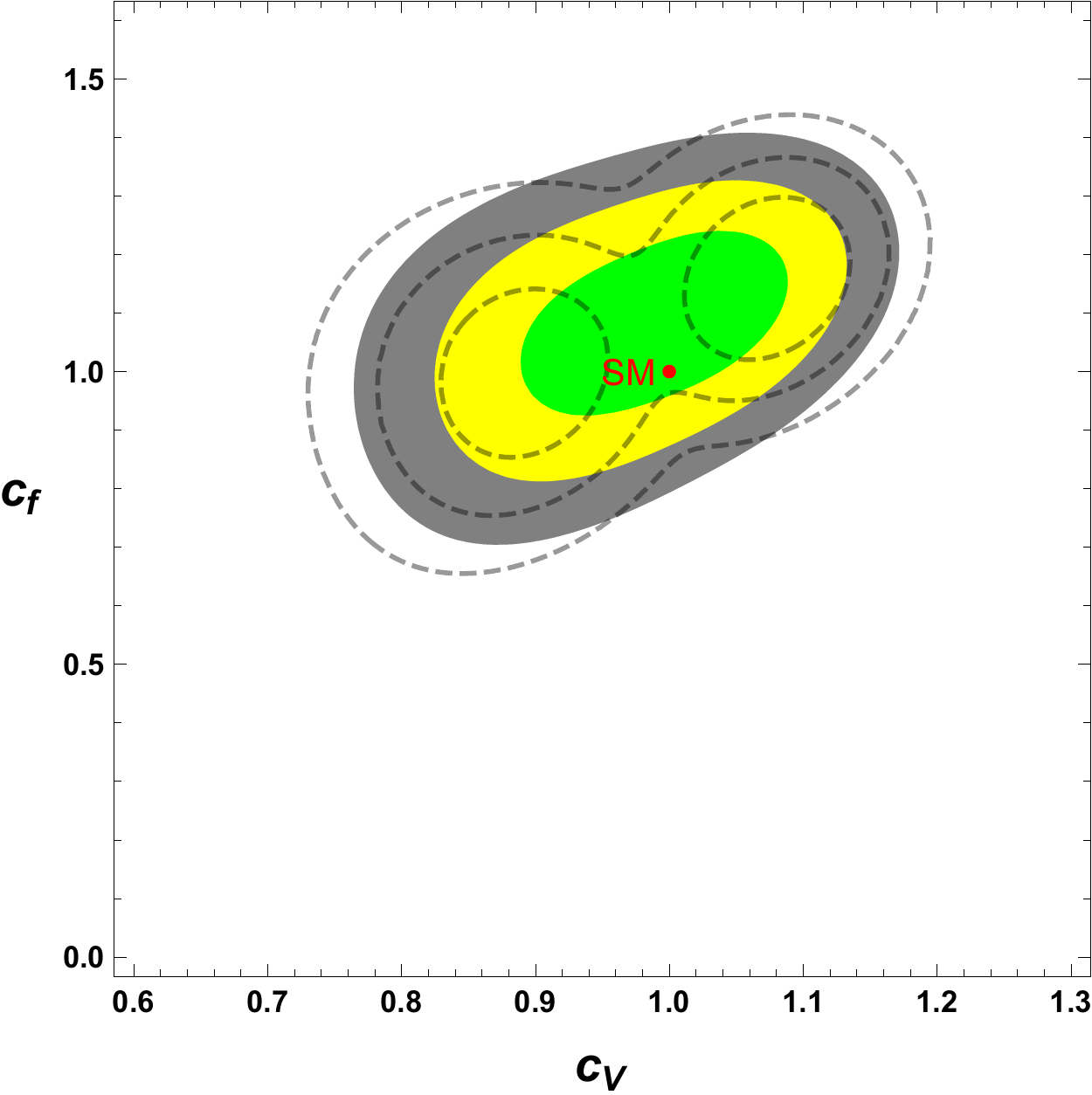}}
\put(230,0){\includegraphics[width=7.5cm]{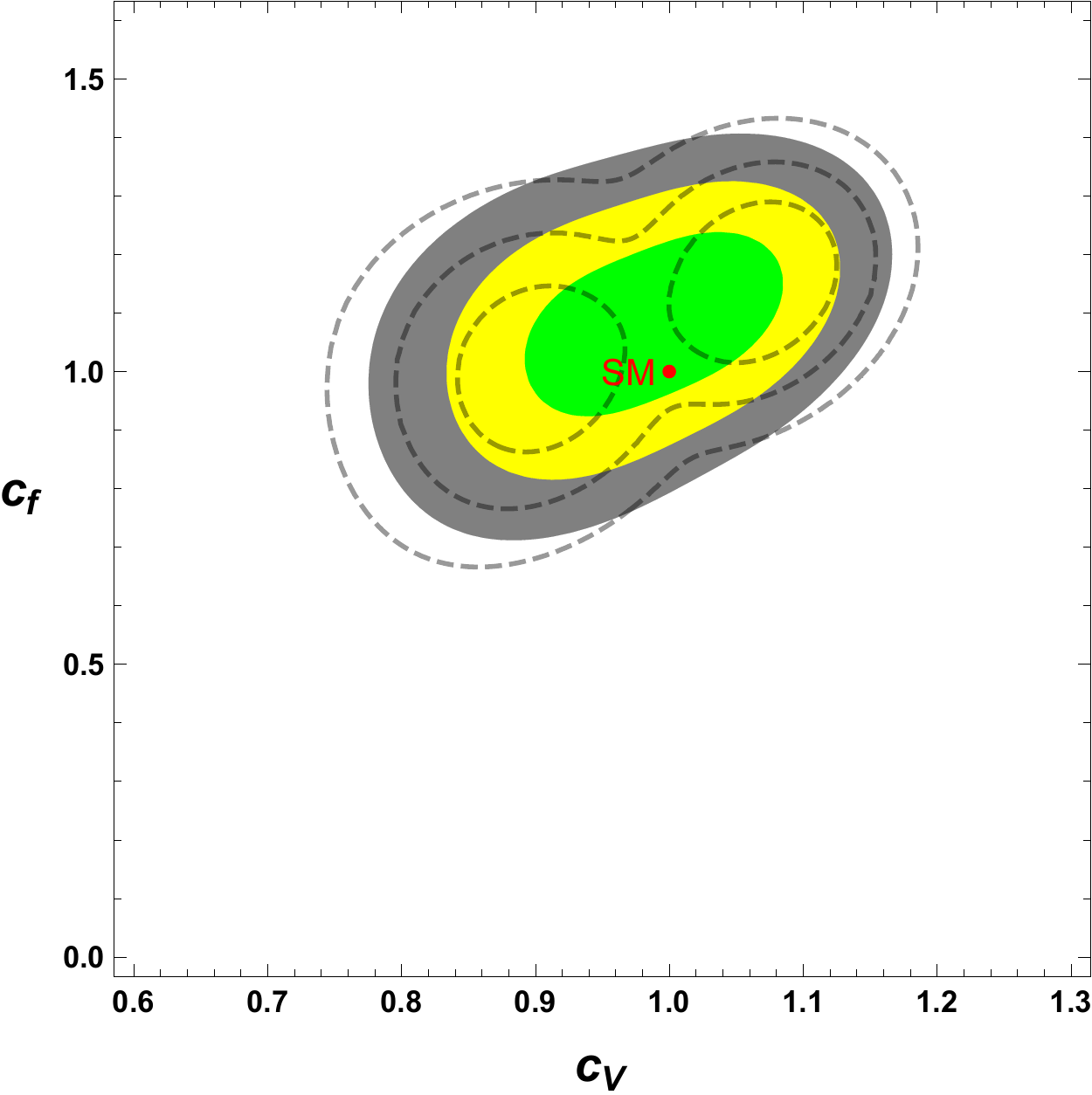}}
\put(0,230){\includegraphics[width=7.5cm]{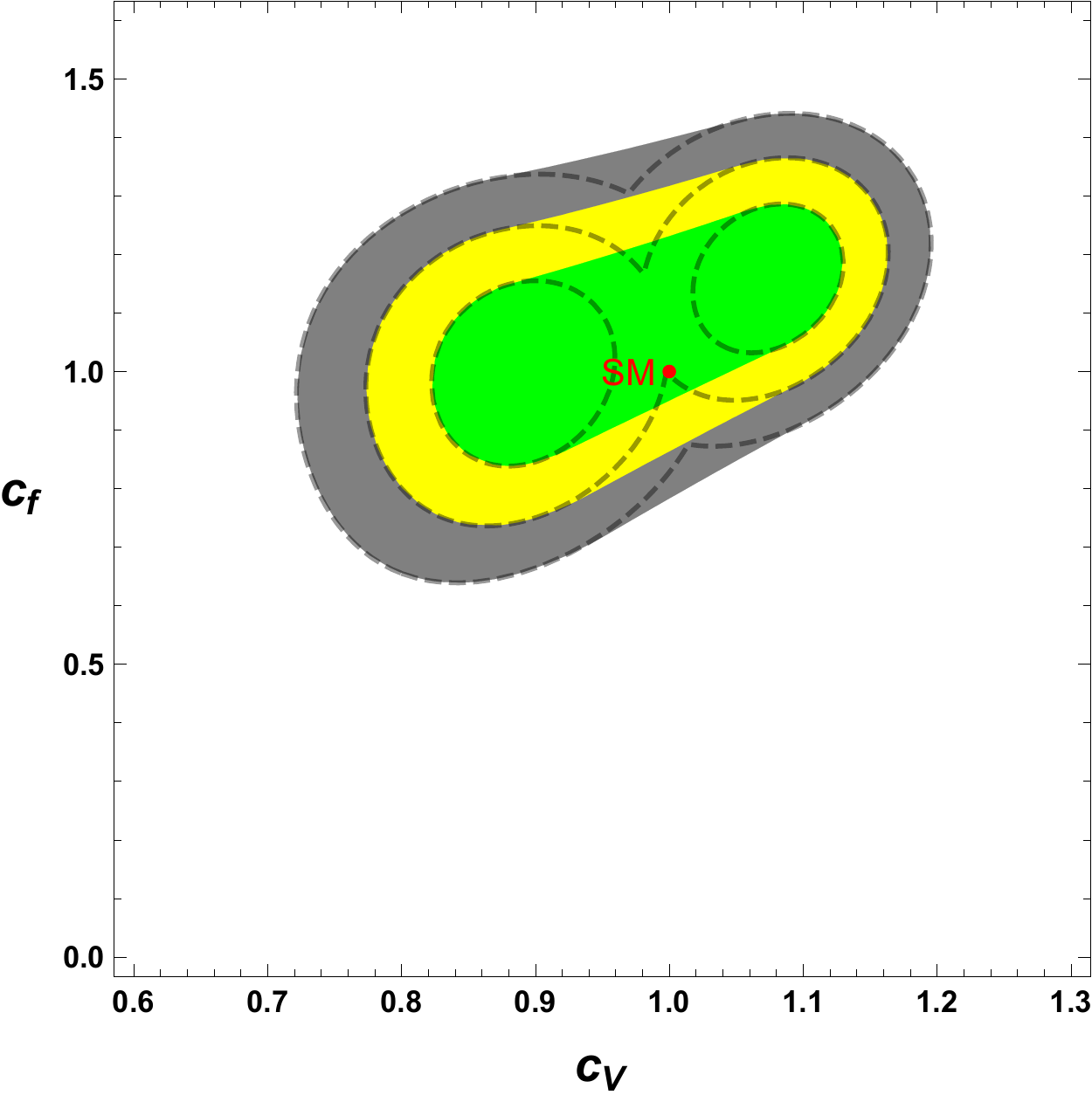}}
\put(230,230){\includegraphics[width=7.5cm]{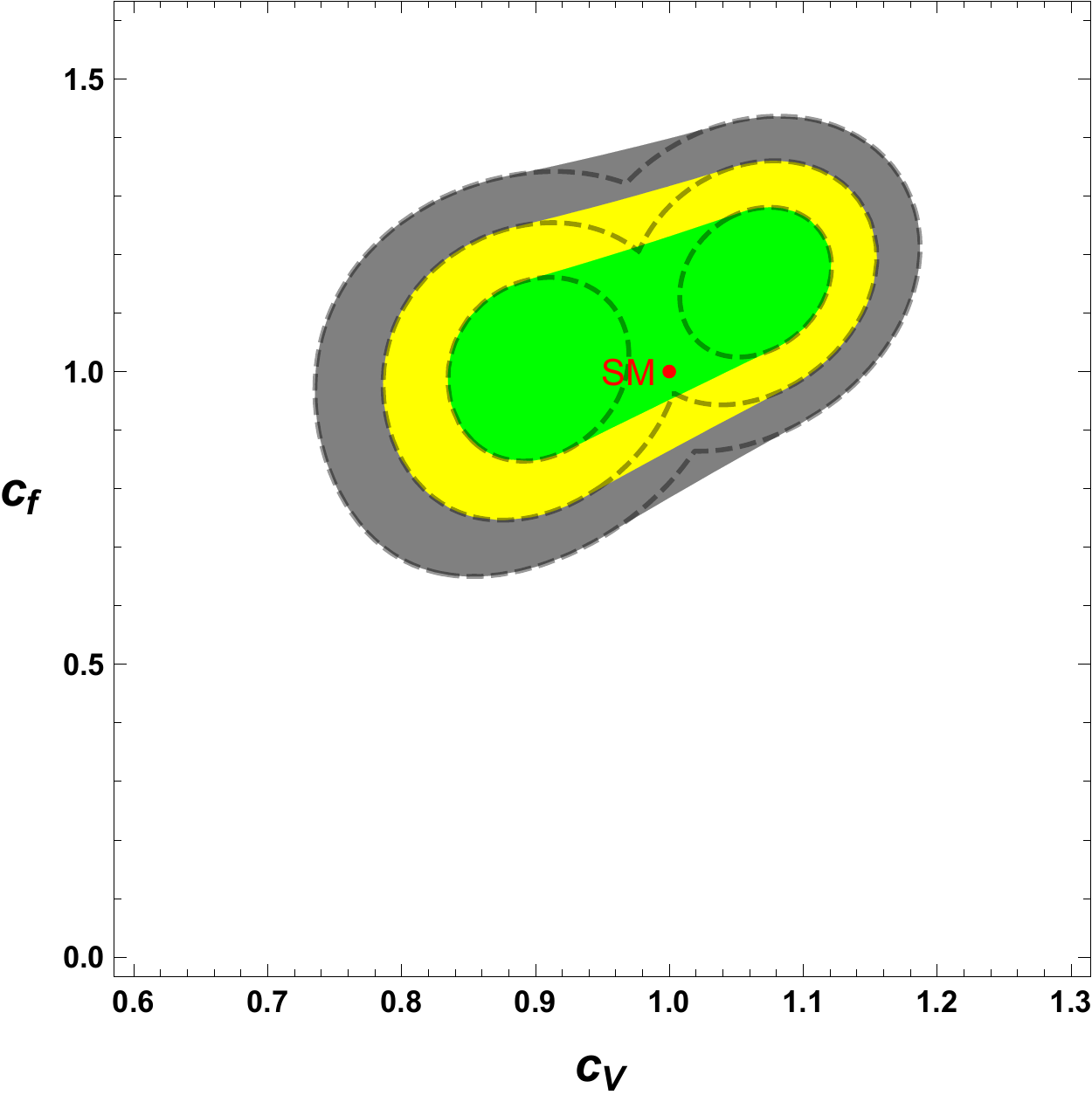}}
\put(40,60){{\color{blue} Bayesian bias}}
\put(40,40){{\color{blue}  -- {\it envelope} --  [c] }}
\put(40,290){{\color{blue} Frequentist bias}}
\put(40,270){{\color{blue}  -- {\it envelope} --  [a] }}
\put(275,60){{\color{blue} Bayesian bias}}
\put(275,40){{\color{blue} -- {\it envelope} -- [d] }}
\put(275,290){{\color{blue} Frequentist bias}}
\put(275,270){{\color{blue} -- {\it envelope} -- [b] }}
\end{picture}
\caption{
The best-fit regions in the $c_V-c_f$ plane obtained through the envelope method. 
The $68\%$, $95\%$ and $99\%$ confidence regions are represented respectively by the green, yellow and grey domains. 
The upper plots illustrate the frequentist approach whereas the two lower ones show the Bayesian approach. 
The [a], [c] and [b], [d] plots correspond, respectively, to the characteristic correlation configurations described in 
Eq.~\eqref{DeltaBS1} and Eq.~\eqref{DeltaBS2}. 
The dashed grey contours illustrate the  best-fit regions at $68\%$~C.L., $95\%$~C.L. and $99\%$~C.L.,  
obtained in  Fig.~(\ref{fig:CombBias}).  
The SM prediction is shown by the red point.
}\label{fig:CombBiasEnv}
\end{figure}
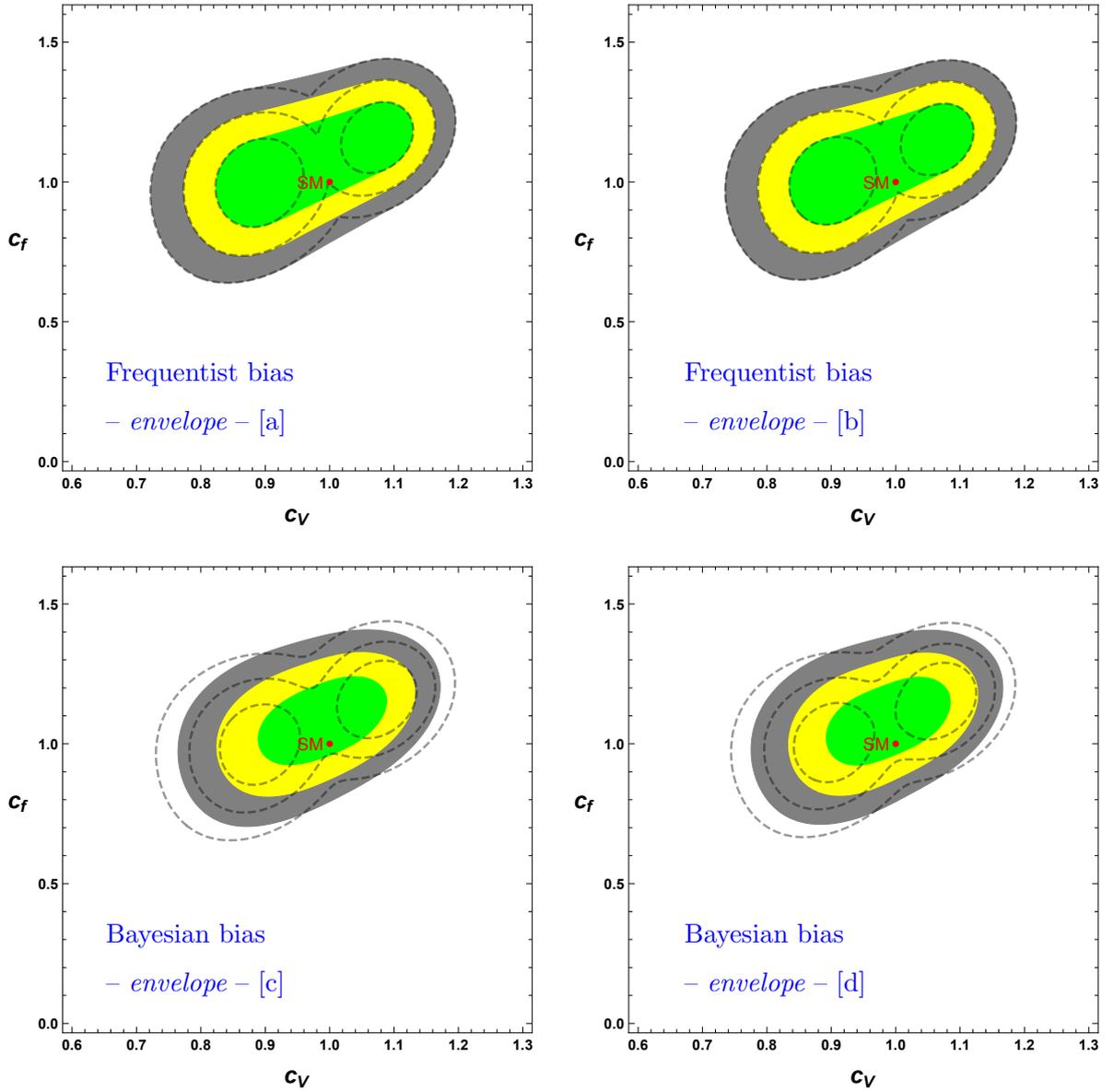

The four plots of Fig.~(\ref{fig:CombBiasEnv}) illustrate the Bayesian and frequentist envelope methods performed accordingly to Sections~\ref{sec:DEMbis} and \ref{sec:DEM}.
Again, both correlation configurations, giving rise to the combined errors of Eq.~(\ref{DeltaBS1})-(\ref{DeltaBS2}), are studied numerically. 
The two upper and lower plots of Fig.~(\ref{fig:CombBiasEnv}) differ due to the direct envelope method being not equivalent within the Bayesian and frequentist 
cases. 
\\
The sets of frequentist envelopes represent the best-fit areas that would be obtained by superimposing the best-fit regions of the extremal bias, but for $\delta_b$ 
spanning continuously the  interval $[-1,1]$. This correspondence between the envelope method and extremal bias appears clearly when one realises (\textit{c.f.} end of Section~\ref{sec:BiasPrin}) that the former is based on the Eqs.~(\ref{DEMmin})-(\ref{eq:barDelta}) while the latter can be obtained through the same equations just with a minimisation over the discrete 
domain $\delta_b\in {\cal D}= \{-1,1\}$ in Eq.~(\ref{DEMmin}), instead of the continuous range $[-1,1]$. The correspondence is visible when comparing the envelopes with the extreme sets of best-fit domains
at $\delta_b = \pm 1$, obtained previously from the frequentist bias method and also superimposed on upper plots of Fig.~(\ref{fig:CombBiasEnv}), 
as dashed contours: these contours draw exactly the extreme limits of the envelopes.
\\
The two sets of Bayesian envelopes obtained in the two lower plots of Fig.~(\ref{fig:CombBiasEnv})  represent less conservative regions with respect to the frequentist envelope.
Besides, the envelopes of these plots cover smaller regions than the best-fit domains that would be obtained by superimposing the best-fit regions of the extremal bias, but for $\delta_b$ 
spanning continuously the  interval $[-1,1]$.
This appears clearly when comparing those envelopes to the extreme sets of best-fit 
regions at, $\delta_b = \pm 1$, obtained previously from the Bayesian bias method (once more superimposed on the lower plots of 
Fig.~(\ref{fig:CombBiasEnv}), as dashed contours).

Finally, we mention that the SM point belongs to all the $68\%$~C.L. regions of Fig.~(\ref{fig:CombBiasEnv}). 
At this level, we can illustrate one of the interests of the bias.
Let us consider an hypothetical but plausible situation. 
For example, suppose that 
with future LHC data, the SM point would fall outside the $3\sigma$ region obtained by marginalising. Such a discrepancy could be interpreted either
as an indirect effect of physics underlying the SM on the Higgs sector, or as a shift of the best-fit regions induced by values of the nuisance parameters favoured statistically by the fit.  
 This shift induced by the nuisance parameters would come from the fact that the nuisance parameters and the parameters of interest  are determined simultaneously.  
In contrast, in the envelope method, a SM prediction falling beyond the $3\sigma$ region  would indicate the presence of new physics without any alternative explanation relying on the statistical treatment (the entire interval of the nuisance parameters being covered).  
This example provides a motivation to apply both  bias and marginalisation methods, which are somehow complementary.

\section{Conclusions}  \label{se:conclu}

The main goal of this analysis was to work out a consistent statistical treatment of the theoretical uncertainties in the fits of the Higgs boson rates. We have analysed in a unified formalism both the Bayesian and frequentist approaches to theoretical uncertainties.  We systematically analysed how to perform error combinations in a given  statistical context and we have introduced a framework to use the bias principle on firm ground.

This analysis has been the opportunity to update the Higgs rate fit based on the latest LHC data at $7$ and $8$~TeV. In the case of Bayesian marginalisation, we have found that the SM prediction for the Higgs couplings still falls into
the $68\%$~C.L. region of the $c_V-c_f$ plane. 
 Bayesian marginalisation  benefits from well-defined distributions for the nuisance parameters  and from an easier convolution of these error distributions compared to frequentist marginalisation.

We have reviewed all the fundamental sources of the individual theoretical errors involved in the SM Higgs cross sections and branching ratios. Then those errors have been 
combined in a careful `step-by-step' approach following the Bayesian rules. In this task of combining a significant number  of uncertainties (various Higgs production modes, 
decay channels\dots), we were helped by the leading moment approximation -- which has been deduced from considerations on the moment-generating function.

This has allowed us to show that the prior of the total uncertainty resulting from the  combination of all the theoretical
errors (using flat priors for the unknown ones) converges to a nearly Gaussian shape.
 Besides, it also came out from the numerical results that
the precise form of this final theoretical prior is not crucial with respect to the determination of the best-fit regions.
This conclusion holds only for the present data, which still have large  experimental errors  with respect to the theoretical ones.  

In contrast, our analysis has shown  that  the correlations of the theoretical uncertainties among the Higgs detection channels induce a significant shift of the best-fit domains
in the space of the parameters of interest. These correlations appear thus to be an unavoidable ingredient of the fits.
The Higgs fits were performed in two extreme configurations of theoretical correlations between the various detection channels. The most realistic correlation setup is an intermediate configuration between those two. Such an approach is thus conservative. Besides, considering  characteristic configurations
has allowed us to derive   simple analytic expressions  for the marginal likelihood functions.

For future Higgs fits, given the ambiguities inherent to the estimation of the theoretical error magnitudes, 
 we recommend to present an additional analysis with $1\sigma$ errors enhanced by a typical factor of $1.5$ as a conservative benchmark.
Such a factor is consistent with the $1\sigma$ theoretical errors preferred by the data. 
Of course the present degree of arbitrariness in 
the theoretical error magnitudes could be improved for instance with future higher order QCD calculations or new methods to determine the PDFs.

Finally, we have provided a rigorous statistical framework for 
the bias principle, which constitutes an alternative to marginalisation.
This framework has lead us to define two complementary bias treatments:
 the extremal bias and the envelope method.  
The bias principle is  more conservative than marginalisation by construction, and does not depend on  the shape of the priors of the nuisance parameters, which are not always known.
Therefore, a reasonable advice is to apply both the marginalisation and bias methods to the Higgs data. Using the envelope method, we find  that the SM prediction belongs to the $68\%$ C.L. region of the $c_V-c_f$ plane.

\section*{Acknowledgements} 

The authors gratefully  acknowledge Glen Cowan for his involvement and valuable inputs. The authors also thank  Henri Bachacou, Damir Be\v{c}irevi\'c, Abdelhak Djouadi, Guillaume Drieu La Rochelle,  
Gero von Gersdorff, Andrea Massironi, Sezen Sekmen, Michael Spira
for stimulating and useful discussions,  and especially 
 Marumi Kado for helpful exchanges about the  bias methods. It is finally a pleasure to thank the organizers 
of the ``Physics at TeV Colliders'' Workshop at {\it Les Houches} (2013) where this work was initiated~\cite{ProceedHouches}. 
SF acknowledges the {\it Brazilian Ministry of Science, Techno\-logy and Innovation} for financial support as well as the LPT at Orsay University 
for hospitality during a part of this work. The work of GM is partly supported by the {\it Institut Universitaire de France}, the European 
ERC Grant ``Higgs@LHC'' and the European Union FP7 ITN INVISIBLES (Marie Curie Actions, PITN-GA-2011-289442).

\vspace*{2cm}

\noindent{\Large\bf Appendix}

\appendix

\section{The leading moment approximation}
\label{app:approx}

 Consider a linear combination $\delta_C$ of random variables $\delta_A$, $\delta_B$, given by
 \be \delta_C \Delta_C=\delta_A \Delta_A+\delta_B \Delta_B,\ee with $\Delta_b\ll \Delta_a$ and ${\rm E}[\delta]=0$, ${\rm V}[\delta]=1$ by convention.  The \textit{pdf} of  $\delta_{A}$, $\delta_{B}$, $\delta_{C}$  are noted respectively $\pi_A$, $\pi_B$, $\pi_C$.

 We  mainly work in Laplace space, using the moment-generating function \be \phi_Z(t)={\rm E}[e^{Z\,t}]=\int dz\, e^{z\,t} \pi_Z(z)\,. \label{eq:phi_c}
 \ee
  If all moments are finite, $\phi_Z(t)=\sum_{n=0}^{\infty} \frac{m^Z_n}{n!}t^n$, where $m^Z_n$ denotes the $n$-th moment of $Z$. $m_1$ being the mean, we have $m_1^A=0=m_1^B$.  $m_2$ being the variance, we have $m_2^A=1=m_2^B$.

   Let us assume in a first place that $\delta_A$, $\delta_B$ are uncorrelated. This implies that $\pi_{A,B}=\pi_A \pi_B$, that the \textit{pdf} of $\delta_C$ is given by a convolution product, and that the moment generating function of $\delta_C$ is given by the product  \be\phi_C(\Delta_C t)=\phi_A(\Delta_A t)\, \phi_B(\Delta_B t)\,.\ee Having $\Delta_B\ll \Delta_A$ by assumption, we can use $\Delta_B/\Delta_A$ has an expansion parameter.
At leading order, neglecting the contribution from $\delta_B$  to the combination amounts to appro\--ximate \be \phi_B (\Delta_B t)=1+O(\Delta_B^2 \, t^2)\ee in the  product~\ref{eq:phi_c}. This corresponds to  approximating $\pi_B$ as a Dirac distribution centred on zero.

Going one order further in the expansion leads to keep 
\be \phi_B (\Delta_B t)=1+  \Delta_B^2 \frac{t^2}{2!} + O(\Delta_B^3 \, t^3)\,.\ee
This subleading term induces $O(\Delta_B^2/\Delta_A^2)$ corrections to the moments of $\delta_C$. 
Explicitly one finds 
\be
\Delta^n_C\, m_n^C=\Delta^n_A\left( m_n^A+\frac{\Delta_B^2}{\Delta_A^2}\, m_{n-2}^A \, N_n \right)\,.
\ee
with $N_n=n!/(2(n-2)!)$. 
At that point, the corrections to all moments $m_n^C$ should in principle be kept.

We then take a second step in our approximation, by considering that the amount of information relevant for our problem somehow decreases with the order of the moment. As a consequence, the corrections to the first moments  are the more relevant. 
Keeping the next-to-leading corrections up to order $p$, our approximation scheme thus reads
\be
\Delta^n_C\,m_n^C =
\begin{cases}
\Delta^n_A\left( m_n^A+\frac{\Delta_B^2}{\Delta_A^2}\, m_{n-2}^A \, N_n +O(\frac{\Delta_B^3}{\Delta_A^3})\right) & \textrm{if}\quad 1\leq n\leq p \\
\Delta^n_A\left( m_n^A+O(\frac{\Delta_B^2}{\Delta_A^2}) \right) & \textrm{if}\quad p < n\,.
\end{cases}
\ee

In particular, truncating the corrections at $p=2$ amounts to take into account only the correction to the variance,  
\be
\Delta_C^2=\Delta_A^2+\Delta_B^2\,.
\ee
 The other details of  the shape remaining unperturbed, it follows that
\be
\pi_C=\pi_A+O\left(\frac{\Delta_B^3}{\Delta_A^3}\delta^{(2)},\frac{\Delta_B^2}{\Delta_A^2}\delta^{(3)}\right)\,. \label{eq:LMA_piC}
\ee
Here $\delta^{(n)}$ is the $n$-th derivative of the Dirac distribution. It comes from the Laplace transform of the $t^n$ term of the moment-generating function 
(see also Ref.~\cite{SylvainShape}). These $\delta^{(n)}$ should be understood as the leading functional deformation to $\pi_A$.
In practice, it appears that keeping only the first leading moment is appropriate when $\pi_A$ is a one-parameter \textit{pdf}. 
In that case, the parameter characterising $\pi_C$ is identified through the combination of variances. 
For example,  taking the normal distribution $\pi_A=\mathcal{N}(0,\sigma^2_A)$ gives $\sigma^2_C=\sigma_A^2+\Delta^2_B$ and $\pi_C=\mathcal{N}(0,\sigma^2_C)$. 
~\footnote{It is worth noticing that in the Gaussian case, this identification reproduces exactly the correction to the  $m_n^C$ at any order. This is not true for other distributions.}

The approach above also extends to correlated variables. The difference with respect to the uncorrelated case is that the moment-generating functions do not factorise, as $\delta_A$, $\delta_B$ now share common moments. 
For example, truncating the corrections at $p=2$ gives the correction
\be
m_2^C=\Delta_A^2+\Delta_B^2+2\Delta_A\Delta_B \rho\,, \label{eq:comb_corr_app}
\ee
where $\rho$ ($=m_1^{AB}$) is the covariance of $(\delta_A,\delta_B)$.   In the limit of full correlation, one has $\rho=1$, so that $ \Delta_C^2=(\Delta_A+\Delta_B)^2$.
Note that when $\rho> \Delta_B/\Delta_A$ in Eq.~\eqref{eq:comb_corr_app}, the contribution from the correlation term becomes larger than the contribution from the square  term $\Delta_B^2$.

Finally, the leading moment approximation also extends to the case of several li\-ne\-ar combinations of variables. Here we consider the case with two linear combinations of two variables $\delta_A$, $\delta_B$ with correlation $\rho$. The combinations are defined as 
\be
\delta_{C_1}\Delta_{C_1}=\delta_{A}\Delta_{A_1}+\delta_{B}\Delta_{B_1}\,,
\ee
\be
\delta_{C_2}\Delta_{C_2}=\delta_{A}\Delta_{A_2}+\delta_{B}\Delta_{B_2}\,.
\ee
The variances are found to be 
\be
\Delta_{C_1}^2=\Delta_{A_1}^2+\Delta_{B_1}^2+2\rho\Delta_{A_1}\Delta_{B_1}\,,
\ee
\be
\Delta_{C_2}^2=\Delta_{A_2}^2+\Delta_{B_2}^2+2\rho\Delta_{A_2}\Delta_{B_2}\,,
\ee
like in the one-combination case described above. 
In the case $\Delta_{A_1}\gg \Delta_{B_1}$, $\Delta_{A_2}\gg \Delta_{B_2}$, the correlation coefficient  $\rho_{12}$ between $\delta_{C_1}$ and $\delta_{C_2}$ reads
\be
\rho_{12}=
1-\frac{1}{2}\left(\frac{\Delta_{B_1}}{\Delta_{A_1}}-\frac{\Delta_{B_2}}{\Delta_{A_2}}\right)^2-\rho^2\left(\frac{\Delta_{B_1}}{\Delta_{A_1}}+\frac{\Delta_{B_2}}{\Delta_{A_2}}\right)^2+O\left(\frac{\Delta^3_{B_{1,2}}}{ \Delta^3_{A_{1,2}}}\right)
\label{eq:LMA_comb_app1}
\,.
\ee
In the case $\Delta_{A_1}\gg \Delta_{B_1}$, $\Delta_{A_2}\ll \Delta_{B_2}$, the correlation coefficient is instead
\be
\rho_{12}=
\rho+\left(\frac{\Delta_{B_1}}{\Delta_{A_1}}+\frac{\Delta_{A_2}}{\Delta_{B_2}}\right)(1-\rho^2)
+O\left(\frac{\Delta^2_{B_{1,2}}}{ \Delta^2_{A_{1,2}}}\right)
\,.
\label{eq:LMA_comb_app2}
\ee

\end{document}